\documentclass[12pt]{article}
\pdfoutput=1

\usepackage{jheppub}
\usepackage{amsmath,amssymb,euscript,array,mathrsfs}
\usepackage{graphicx,amsbsy,bm,amsmath}
\usepackage{dcolumn}
\usepackage{siunitx} 
\usepackage{textcomp} 
\usepackage{color,soul} 

\usepackage{natbib}

\usepackage{savesym}

\newcommand{\isum}%
{\mathop{\hbox{$\displaystyle\sum\kern-13.2pt\int\kern1.5pt$}}}

\usepackage{mathtools}

\newcommand{\PTzero}{\mathrlap{T}{P}}

\def\a{\alpha}
\def\b{\beta}
\def\c{\gamma}
\def\d{\delta}
\def\e{\epsilon}

\def\k{\kappa}
\def\l{\lambda}
\def\m{\mu}
\def\n{\nu}

\def\r{\rho}
\def\s{\sigma}
\def\t{\tau}

\def\w{\omega}

\def\D{\Delta}
\def\L{\Lambda}

\def\tr{{\rm tr}}

\def\Odanish{\,\,{\raise.15ex\hbox{/}\mkern-12mu {\rm O}}}
\def\odanish{\,\,{\raise.15ex\hbox{/}\mkern-12mu {\rm o}}}
\def\Dbarslash{\,\,{\raise.15ex\hbox{/}\mkern-12mu {\bar D}}}
\def\Dslash{\,\,{\raise.15ex\hbox{/}\mkern-12mu D}}
\def\delslash{\,\,{\raise.15ex\hbox{/}\mkern-9mu \partial}}
\def\delbarslash{\,\,{\raise.15ex\hbox{/}\mkern-9mu {\bar\partial}}}

\def\thalf{\tfrac{1}{2}}

\def\rta{\rightarrow}
\def\ltrta{\leftrightarrow}
\def\tr{{\rm tr}}

\def\Hbar{{\bar{\rm H}}}

\title{\begin{center}Testing Fundamental Physics in Antihydrogen Experiments\end{center}}

\author{\vskip0.3cm
M.~Charlton, S.~Eriksson and G.\,M.~Shore}
\emailAdd{m.charlton@swansea.ac.uk, s.j.eriksson@swansea.ac.uk, g.m.shore@swansea.ac.uk}

\affiliation{\vskip0.2cm 
Department of Physics, College of Science, Swansea University, Singleton Park, Swansea, SA2 8PP, UK}

\date{\today}

\abstract{

The recent advent of high precision antihydrogen ($\overline{\rm H}$) spectroscopy opens the way to 
stringent experimental tests of the fundamental principles
underlying particle physics and general relativity (GR), such as Lorentz and $\mathsf{CPT}$ invariance
and the Einstein Equivalence Principle (EEP), on pure antimatter systems.
In this paper, the nature and implications of these tests is investigated, with special reference
to the ALPHA antihydrogen programme at CERN. 
This is underpinned by a theoretical review of the role of antiparticles, causality and fundamental
symmetries in relativistic quantum field theory (QFT) and the theory of time measurement in GR. 
Low-energy effective theories which break Lorentz and $\mathsf{CPT}$ symmetry, or the 
Strong Equivalence Principle (SEP), are then introduced, together with a review of several
`fifth force' scenarios involving new long-range forces which would effectively violate the 
universality of free-fall, or Weak Equivalence Principle (WEPff). 
The possible role of $\mathsf{CPT}$ violation in determining the matter-antimatter asymmetry 
of the Universe is discussed.
Explicit calculations are given for the dependence on possible Lorentz and $\mathsf{CPT}$ violating
couplings of the transition frequencies amongst the $1S$, $2S$ and $2P$ hyperfine states 
measured in the magnetic field of the ALPHA trap and the resulting bounds are compared with existing limits.
An analysis of the implications for the EEP of current free-fall and spectroscopic measurements 
with antihydrogen is presented and existing and potential bounds on WEPff and the 
universality of clocks (WEPc) are derived, together with constraints on fifth forces.
Future prospects for high-precision antihydrogen spectroscopy, free-fall and gravitational
redshift experiments, and anti-atom matter-wave interferometry are described and experimental
possibilities involving other antimatter species are briefly outlined.       } 

\note{
\noindent {\it This is a preprint of the following work:  M.~Charlton, S.~Eriksson and G.\,M.~Shore,
``Antihydrogen and Fundamental Physics'', 2020, Springer,
reproduced with permission of Springer Nature Switzerland AG.
The final authenticated version is available online at} http://dx.doi.org/10.1007/978-3-030-51713-7.}

\begin{document}

{
\savesymbol{o}
\newcommand{\o}{\theta}

\savesymbol{O}
\newcommand{\O}{\Theta}

\maketitle

\setlength{\parskip}{10pt}

\section{Introduction}\label{sect 1}

The recent measurement by the ALPHA collaboration of the $1S$-$2S$ spectral line in antihydrogen
with a precision of a couple of parts in $10^{12}$ \cite{ALPHA1S2S,Ahmadi:2018eca} 
marks the beginning of a new era of precision anti-atomic physics.
Future experiments on antihydrogen and other antimatter species will enable exceptionally high-precision tests 
of many of the fundamental tenets of relativistic quantum field theory and general relativity, 
such as $\mathsf{CPT}$ invariance, Lorentz symmetry and the Equivalence Principle. It is therefore timely 
to examine critically what each of these  experiments may be said to test and what any violation 
from standard expectations would mean for fundamental physics.

Experiments on pure antimatter systems, whether elementary particles or bound states such as antihydrogen, are
especially interesting from this point of view since they are constrained so directly by the fundamental principles
underlying the standard model. For example, the discovery of a new $Z^\prime$ boson, right-handed neutrinos,
a supersymmetric dark matter candidate {\it etc.}~would be of immense interest but could readily be assimilated into
an extension of the standard model. In contrast, an anomalous result on the charge neutrality of antihydrogen,
or a difference in the $1S$-$2S$ transitions of hydrogen and antihydrogen, would impact directly on the
foundations of local relativistic QFT. In these theories, the existence of antiparticles with precisely 
the mass and spin, and opposite charge, of the corresponding particles is required by Lorentz invariance 
and causality. Moreover, for a {\it local} QFT, Lorentz invariance implies invariance under $\mathsf{CPT}$,
according to the celebrated theorem \cite{Pauli:1940, Bell, Luders, Pauli:1955}.
Antimatter experiments therefore directly test these principles.

The situation is not so clear when we consider gravity, where such experiments are often presented as tests 
of ``the equivalence principle''. The difficulty is that there are several versions of the equivalence principle in
the literature -- weak, strong, Einstein -- with definitions which are not always either unique or well-defined.
Indeed, as emphasised by Damour \cite{Damour:2012rc}, it should not really be considered as a `principle' of
GR in the more rigorous sense that Lorentz symmetry and causality are principles of QFT.
A more satisfactory approach is to recognise that we have a well-defined, and extraordinarily successful, theory
of gravity in GR, which makes clear and precise predictions for the gravitational interactions 
of all forms of matter. Like other experiments, those on antimatter should simply be viewed as tests of this
theory. 

General Relativity is based on the idea that gravitational interactions may be described in terms of a 
curved spacetime. As described more precisely in section \ref{sect 2.4}, this spacetime is taken to be a 
Riemannian manifold, since this has the property that at each point it locally resembles
Minkowski spacetime.  The global Lorentz symmetry of non-gravitational physics is reduced to 
{\it local Lorentz symmetry} in curved spacetime. This is the mathematical realisation of the physical 
requirement of the existence of local inertial frames ({\it i.e.} freely-falling frames) even in the presence of gravity.
Further to this, the standard formulation of GR makes a simplifying, though well-motivated, choice of dynamics 
for the interaction of matter and gravity,
which is encapsulated in the following statement of the Strong Equivalence Principle:

\noindent $\bullet$ {\it In a local inertial frame, the laws of physics take their special relativistic form} ~(SEP).

We will also discuss frequently two further expressions of the universality at the heart of GR. These are best
viewed as experimental predictions of GR, though we refer to them here as
versions of the Weak Equivalence Principle:

\noindent $\bullet$ {\it Universality of free-fall  -- all particles (or antiparticles) fall with the same 
acceleration in a gravitational field}~(WEPff).

\noindent $\bullet$ {\it Universality of clocks  -- all dynamical systems which can be viewed as clocks, 
{\it e.g.}~atomic or anti-atomic transition frequencies, measure the same gravitational
time dilation independently of their composition}~(WEPc).

Taken together, these three properties of GR are usually referred to as the
{\it Einstein Equivalence Principle}.\footnote{The Einstein Equivalence Principle may be stated in 
various essentially equivalent ways.  In \cite{Will:2014kxa}, the three principles are referred to as
{\it Local Lorentz Invariance} (LLI), which we have called SEP; the {\it Weak Equivalence Principle} (WEP), which
is simply our WEPff; and {\it Local Position Invariance} (LPI), which states that `the outcome of any local 
non-gravitational experiment is independent of where and when in the universe it is performed' \cite{Will:2014kxa}.
LPI implies the universality of gravitational redshift, or WEPc, and can also be tested through the 
space and time-independence of fundamental constants. Note that while GR implies WEPc, the latter is
a more general property of any metric theory of spacetime. Also note that, as described above, a metric theory
like GR on a Riemannian spacetime manifold exhibits LLI, but the dynamics need
not be the same as special relativity, or be independent of the local curvature, if SEP is violated.}

Apparent violations of these predictions, especially WEPff, can also arise not from the actual violation of any 
fundamental principle of QFT or GR but from the existence of new interactions not present in the standard model, 
so-called `fifth forces'. Low-energy precision experiments on antimatter, whether involving spectroscopy
or free-fall equivalence principle tests, may be sensitive to such new interactions and can place limits on their
range and coupling strength. Here, we consider two such possiblities, both well-motivated
by fundamental theory. The first is an extension of the standard model gauge group to include a new
$U(1)_{B-L}$ factor, with a corresponding gauge boson $Z'$ coupling to $B-L$ (baryon minus lepton number)
charge. The second involves the spin 1 `gravivector' boson which arises in some supergravity theories
with extended, ${\cal N}\ge 2$, supersymmetry. Both have the potential to modify gravitational free-fall
in violation of WEPff, distinguishing between matter and antimatter. We also consider a more general 
phenomenological approach to the possible existence of new, gravitational strength, vector or scalar 
interactions.

From an experimental perspective, the study of the fundamental properties of antiparticles and 
atomic systems constituted wholly, or partially, from them is a growing area of endeavour. 
In this paper, our main focus is on antihydrogen, and in particular the current experiments being
performed by the ALPHA collaboration at CERN and their implications as tests of fundamental physics.
Later, we briefly consider a range of other antimatter species which may offer complementary opportunities
for such tests.

We start by describing some of the practical aspects of current experiments with $\overline{\rm H}$ at low energy. 
Positrons ($e^+$) are available in the laboratory, typically via pair production and from radioactive materials 
(see {\it e.g.}~\cite{CharltonHumb} for a review). We concentrate on the latter, and the isotope $^{22}$Na 
(half-life around 2.6 years, $\beta^+$ fraction about 90\%) is the typical choice of source. Sealed capsules 
of around GBq activity can be held in vacuum and, using well-documented procedures 
(see {\it e.g.}~\cite{SchultzLynn}),
eV-energy beams can be produced with efficiencies of 0.1-1\% of the source strength. 
Such beams can be readily transported in vacuum to devices which enable their trapping and
 accumulation: the most common instrument to achieve this is the so-called buffer gas trap which, 
using a Penning-type trap and energy loss via inelastic positron-gas collisions, can accumulate 
around $10^8~e^+$ in a few minutes, if required. The positrons can then be transferred \cite{Jorgensen05} 
on demand for further experimentation, and of most relevance here for $\overline{\rm H}$ production 
and trapping.

Antiprotons ($\overline{p}$) are only available at laboratories such as CERN where high energy protons
 (typically 20-30\,GeV) produce the $\overline{p}$\,s in collision with fixed targets. 
CERN's unique Antiproton Decelerator 
(AD) \cite{Maury97,Eriksson09} syphons off $\overline{p}$\,s at a kinetic energy of about 3.5\,GeV and
 then decelerates and cools them in stages to reach 5.3\,MeV, whereupon they are ejected to experiments 
in bursts of 100\,ns duration containing around $10^7$ particles, about once every 100 s. The kinetic energy 
of the $\overline{p}$\,s is typically moderated using foils whose thickness is carefully adjusted to maximise 
the transmitted yield (of around $10^{-2}$ - $10^{-3}$ of the incident flux) below 5-10\,keV, and these are 
then captured in dynamically switched high field Penning traps \cite{Gabrielse86} where they can be 
efficiently electron cooled \cite{Gabrielse89,Feng97} to sub-eV energies. The $\overline{p}$\,s and electrons 
can then easily be separated, and the former then transferred to another apparatus or stored for further 
manipulation and experimentation.

The mixing of $\overline{p}$\,s and $e^+$s to form $\overline{\rm H}$ has been described in detail elsewhere 
\cite{TopRev,ALPHAAcc}, and a number of techniques have been developed to hold the antiparticle species 
in close proximity and manipulate the properties of the respective clouds ({\it e.g.}~number, density and temperature) 
in a system of Penning traps to facilitate anti-atom creation. Under the conditions of $e^+$  cloud density 
(around $10^{14}$ m$^{-3}$) and temperature (typically in the range 5-20\,K) commonly used in 
$\overline{\rm H}$ experiments, the dominant formation reaction is the three body process 
$e^+ + e^+ + \overline{p} \rightarrow \overline{\rm H} + e^+$. It is well-documented 
(see {\it e.g.}~\cite{MMM,Robicheaux08}) that this reaction produces highly excited 
$\overline{\rm H}$ states: thus, if experimentation on the ground state is required, the neutral should be held 
in a suitable trap, otherwise most of the nascent anti-atoms will annihilate on contact with the Penning trap walls, 
typically within a few $\mu$s.

The traps used for antihydrogen have, to date, been variants of the magnetic minimum neutral atom trap  
\cite{ALPHATrap1,ATRAPTrap}, as widely applied in cold atom physics (see {\it e.g.}~\cite{Foot}). The trapping force 
is due to a magnetic field gradient acting upon the anti-atom magnetic moment, such that the low-field
 seeking anti-atoms ({\it i.e.}~these whose positron spin is anti-parallel to the local magnetic field) are trapped. 
The practical details need not concern us here (see \cite{ALPHAApp}). However, even with advanced 
superconducting magnet technology the traps are shallow, at around 0.5 K deep, with respect to the 
aforementioned $\overline{p}/e^+$ temperatures on mixing. Thus, only a small fraction of the 
antihydrogen yield can be captured \cite{ALPHAAcc,ALPHATrap1,ALPHATrap2,ALPHATrap3}, and the 
recent state-of-the-art  is about 10-20 trapped anti-atoms from around 50,000 created during mixing 
of 90,000 $\overline{p}$\,s with 3 million $e^+$ at around 15-20 K \cite{ALPHA1S2P}. Nevertheless, 
the very long lifetime of the antihydrogen atoms in the cryogenic trap environment - in excess of 
60 hours \cite{Ahmadi:2018eca} - has allowed their accumulation over extended periods \cite{ALPHAAcc} 
with the record being over 1500 $\overline{\rm H}$s in ALPHA's 400 cm$^3$ magnetic trap. 
It is from this basis, of 1-$10^3$ stored $\overline{\rm H}$s that the experiments described 
below and in section \ref{sect 3} have been achieved.\footnote{The antihydrogen experiments are 
clearly $\overline{p}$ flux limited and in promotion of this field CERN 
is developing ELENA, an ultra-low energy add-on to the AD \cite{Maury14}. This will provide $\overline{p}$\,s 
at 100\,keV (rather than 5\,MeV), which will enhance capture efficiencies in most experiments by a factor of 
around 100. Furthermore, the low $\overline{p}$ kinetic energy  will allow delivery using electrostatic transport 
and switching technology, thus facilitating rapid changeover of beam between experiments. This will result 
in a $\overline{p}$ on-demand mode of operation, which together with the higher capture efficiencies will 
vastly enhance capabilities.}

So far, all precision experiments with antihydrogen in the ALPHA apparatus rely on detecting the by-products of 
annihilation when the anti-atom escapes from the magnetic trap and comes in contact with the (matter) 
walls of the electrodes used to confine the charged particle plasmas during antihydrogen synthesis. 
The high-energy particles (mostly pions) produce hits in silicon strips arranged axially in three 
cylindrically shaped layers surrounding the exterior of the trapping apparatus~\cite{ALPHADet}. 
The particle tracks are then traced back and the location of the annihilation event is found by 
searching for a vertex occurring on the electrode wall. Artificial learning is used to efficiently 
distinguish between background events and as a result, the tracking detector is able to spatially 
resolve single antihydrogen annihilation events. Antihydrogen detection events in this silicon 
vertex detector occur in two modes. In disappearance mode an experiment which produces loss 
of anti-atoms from the trap, {\it e.g.}~by resonantly inducing a transition to an untrapped state, is conducted, 
after which the trap is rapidly turned off and the remaining anti-atoms are detected. In appearance mode 
anti-atoms escaping due to the interaction are detected while the trap is energised. The modes of 
detection can be applied together, allowing consistent monitoring of the trapping rate.

The first precision experiment carried out by ALPHA to interrogate the properties of antihydrogen was the 
verification of charge neutrality \cite{Amole14ch, Ahmadi:2016ozp}. 
Interpreted as confirmation that the electron and positron, and the proton and 
antiproton, have equal and opposite charges, it is clearly a requirement of $\mathsf{CPT}$ invariance.
We emphasise, however, that this basic property of antiparticles is much more fundamental and is necessary
to preserve causality. The exact equality of the magnitudes of the proton and electron charges
is in turn necessary in the standard model to ensure unitarity.

Control over the dynamics of the antihydrogen atoms in the ALPHA magnetic trap \cite{Zhmoginov, Zhong}
allowed a preliminary gravity experiment \cite{Amole13} setting the loose bound $F\lesssim 110$
on the ratio $F = m_g/m_i$ of the `gravitational mass' to the `inertial mass'. These definitions are
discussed critically here in sections \ref{sect 2.4} and \ref{sect 3.3}. Far more precise bounds on 
WEPff violation for antimatter, at the percent level,   are expected with the dedicated 
ALPHA-g apparatus and other experiments in development: see section \ref{sect 3.3}. 

The spectroscopy programme, which we discuss in more detail in section \ref{sect 3}, 
began with a demonstration of microwave-induced spin flips in
antihydrogen \cite{Amole12}, and further development has resulted in the current high-precision
measurement of the hyperfine structure. A measurement of the hyperfine splitting for 
antihydrogen from the difference of the $1S_d$\,-\,$1S_a$ and $1S_c$\,-\,$1S_b$ transitions 
was presented in \cite{Ahmadi17} and found to be in agreement with hydrogen at the $10^{-4}$ level,
consistent with expectations.

The gold standard is of course the two-photon $1S$\,-\,$2S$ transition, which for hydrogen has been 
experimentally determined with a frequency precision of $4.2 \times 10^{-15}$ \cite{Parthey:2011lfa}.
In \cite{ALPHA1S2S}, ALPHA verified this result for antihydrogen with a precision of $10^{-10}$.
Note that the anti-atom measurement is made on the hyperfine state transitions
$1S_d$\,-\,$2S_d$ and $1S_c$\,-\,$2S_c$ in the 
$\sim 1\,{\rm T}$ magnetic field of the confining trap.
More recently \cite{Ahmadi:2018eca}, this precision has been improved still further with the 
measurement of the antihydrogen $1S$\,-\,$2S$ transition at the level of $2\times 10^{-12}$.
In terms of relative precision of the experimental measurement, this is the most precise test
of $\mathsf{CPT}$ symmetry achieved to date with the anti-atom.

So far, therefore, the ALPHA collaboration investigations of antihydrogen have proved consistent
with the fundamental principles outlined above, especially $\mathsf{CPT}$. This is, however, 
still only the early stages of an extensive programme of high precision spectroscopy \cite{ALPHA2019}.
Gravity tests, of both WEPff and WEPc, as well 
as explicit tests of Lorentz symmetry, are just beginning and will achieve competitive levels of 
precision in the coming years, exploiting the special nature of antihydrogen as a neutral,
pure antimatter state \cite{ALPHA2019}.  Atom interference experiments with antihydrogen are also
feasible, while ultra-high precision spectroscopy with molecular $\overline{\rm H}$ states can be
envisaged. 

\vskip0.2cm
The paper is organised as follows. Section \ref{sect 2} describes the main fundamental principles 
which will be tested in these experiments, especially Lorentz, $\mathsf{CPT}$, causality, unitarity 
and the equivalence principles embodied in GR. Modifications to standard theory
are considered, ranging from effective field theories incorporating Lorentz, $\mathsf{CPT}$ or 
SEP breaking to models with new `fifth forces' which would modify WEPff especially.

Section \ref{sect 3} is devoted to antihydrogen. We describe the ALPHA spectroscopy programme
and derive explicit formulae for the dependence on the Lorentz and $\mathsf{CPT}$ violating
couplings in the SME effective theory \cite{Colladay:1996iz, Colladay:1998fq} 
for the specific transitions between $1S$, $2S$ and $2P$ hyperfine states measured by ALPHA.
We then turn to gravity and, after a review of the planned experimental programme for 
antimatter equivalence principle tests with ALPHA-g, GBAR and AEgIS at the CERN AD, 
we discuss how these are interpreted in the framework of GR and possible
EP-violating phenomenological models. A careful discussion of physical time measurement in the 
context of metric theories such as GR is given in section \ref{sect 2} as background 
to the interpretation of these experiments.

Finally, in section \ref{sect 4}, we briefly describe a number of other antimatter bound states
which could be studied experimentally in future as complementary tests of fundamental physics
principles. A general summary and outlook is presented in section \ref{sect 5}.

\section{Fundamental Principles}\label{sect 2}

We begin by reviewing some of the fundamental principles of local, relativistic QFT 
and GR that may be tested in current and forthcoming low-energy experiments on antimatter,
especially with antihydrogen. The discussion is elementary and is intended simply to highlight how 
basic principles such as Lorentz invariance, $\mathsf{CPT}$, the equivalence principle, {\it etc}.~are 
embedded in the standard theories of 
particle physics and gravity.

\subsection{Antimatter and Causality}\label{sect 2.1}

The first key question to address is why antiparticles exist and why their properties (mass, spin, charge $\ldots$)
must {\it exactly} match those of the corresponding particles.  While this is often presented in historical terms,
going back to early and now superseded interpretations of the Dirac equation in relativistic quantum 
mechanics, the real reasons for the existence of antiparticles are much deeper and more general, and certainly not
specific to spin 1/2 particles. In fact, antiparticles are required to maintain causality in a Lorentz invariant 
quantum theory.

First, we recognise that single-particle relativistic quantum mechanics is not a consistent theory and has 
to be replaced by a quantum field theory, with the Dirac equation promoted to a field equation for a 
spinor quantum field $\psi(x)$. The corresponding action for QED, with $\psi(x)$ representing the electron and
the gauge field $A_\m(x)$ describing the photon, is
\begin{equation}  
S \,=\, \int d^4x\,{\cal L}_{\rm QED} 
~=~ \int d^4x\, \left(\bar\psi \left( i \c^\m D_\m -m\right)\psi  ~-~ \frac{1}{4} F_{\m\n} F^{\m\n} \right) \ , 
\label{b1}
\end{equation}
with $D_\m = \partial_\m - i e A_\m$. 

In this Lorentz-invariant QFT, the Dirac field $\psi(x)$ is expanded (in the Heisenberg picture) in terms of creation 
and annihilation operators for electrons and positrons as
\begin{equation}
\psi_\a(x) = \int \frac{d^3{\bf p}}{(2\pi)^3}\frac{1}{\sqrt{2E_{\bf p}}}  ~\sum_s\left(a^s_{\bf p} u^s_\a(p) e^{-ip.x} \,+\,
b^{s\dagger}_{\bf p} v^s_\a(p) e^{ip.x} \right) \ , 
\label{b2}
\end{equation}
where $p^0$ is identified as $E_{\bf p} = \sqrt{{\bf p}^2 + m^2}$, according to the standard relativistic dispersion relation,
and the momentum integral is over the Lorentz invariant measure.
The adjoint admits a similar decomposition with $a^s_{\bf p} \leftrightarrow b^s_{\bf p}$ and
$u^s(p) \leftrightarrow \bar v^s(p)$.
Note that these forms assume the full symmetries of Minkowski spacetime, including translation invariance.
The charge conjugate field $\psi^{\mathsf C}(x) = C \bar\psi^T(x)$, with $C = i \c^0 \c^2$, takes the form
\begin{equation}
\psi^{\mathsf C} (x) = \int \frac{d^3{\bf p}}{(2\pi)^3}\frac{1}{\sqrt{2E_{\bf p}}}  
~\sum_s\left(b^s_{\bf p} u^s_\a(p) e^{-ip.x} \,+\,
a^{s\dagger}_{\bf p} v^s_\a(p) e^{ip.x} \right) \ .
\label{b3}
\end{equation}
Here, $a^s_{\bf p}$ ($a^{s\dagger}_{\bf p}$) is the annihilation (creation) operator for an electron of spin $s$
and momentum $\bf p$, while $b^s_{\bf p}$  ($b^{s\dagger}_{\bf p}$) are the corresponding operators for positrons.
Under charge conjugation, $a^s_{\bf p} \xrightarrow{\mathsf C} b^s_{\bf p}$. The spinors $u^s_\a(p)$,
$v^s_\a(p)$ are the standard solutions of the Dirac equation,
satisfying notably\footnote{See ref.~\cite{Peskin:1995ev} 
for our conventions and the various identities amongst the spinor quantities used here.}
\begin{align}
\sum_s u^s_\a(p) \bar u^s_\b(p) &= (\c. p + m)_{\a\b} \ ,  \nonumber \\
\sum_s v^s_\a(p) \bar v^s_\b(p) &= (\c. p - m)_{\a\b} \ .
\label{b4}
\end{align}

The free Dirac theory has a global $U(1)$ symmetry, promoted to a local $U(1)$ with the inclusion of the
photon field $A_\m(x)$ in QED, with Noether current $J^\m = \bar\psi \c^\m \psi$. This implies the existence
of a conserved charge, the electric charge in QED, which is expressed in terms of the number operators for
electrons and positrons as
\begin{equation}
 Q = \int d^3x\, J^0 = \int \frac{d^3{\bf p}}{(2\pi)^3}
\sum_s\left(a^{s\dagger}_{\bf p} a^s_{\bf p} - b^{s\dagger}_{\bf p} b^s_{\bf p} \right) \ .
\label{b5}
\end{equation}
This shows that the antiparticles appearing in the Dirac fields $\psi$, $\psi^{\mathsf C}$ have exactly
the opposite charge to the corresponding particles.

We now illustrate why the existence of antiparticles with these properties is necessary to preserve causality
\cite{Peskin:1995ev,Weinberg:1995mt}.   The Wightman propagator $S_+(x,y)$ describing the propagation of an electron 
from $y$ to $x$ is 
\begin{align}
S_{+\a\b}(x,y) ~&=~ \langle 0|\psi_\a(x)\,\,\bar\psi_\b(y) |0\rangle \nonumber \\
 &= \int \frac{d^3{\bf p}}{(2\pi)^3}\frac{1}{\sqrt{2E_{\bf p}}} \int \frac{d^3{\bf q}}{(2\pi)^3}\frac{1}{\sqrt{2E_{\bf q}}} 
~\sum_{s,s'} \langle0|a_{\bf p}^s a_{\bf q}^{s'\dagger}|0\rangle \,u_\a^s(p) u_\b^{s'}(q) \,e^{-ip.x + i.q.y} \nonumber\\
&= \int \frac{d^3{\bf p}}{(2\pi)^3}\frac{1}{2E_{\bf p}} \,\left(\c.p + m\right)_{\a\b} e^{-ip.(x-y)} \ ,
\label{b6}
\end{align}
with $p^0$ as defined previously. 
The last line follows using the anticommutation relations
\begin{equation}
\{ a_{\bf p}^s, a_{\bf q}^{s'\dagger}\} = (2\pi)^3 \d^3({\bf p} - {\bf q}) \d^{ss'} \ ,
\label{b7}
\end{equation}
and the identity (\ref{b4}).  This may also be written in a form which makes clear it is a solution
of the homogeneous Dirac equation, {\it viz.}
\begin{equation}
S_+(x,y) = -i \int_{\cal C_+} \frac{d^4 p}{(2\pi)^4} \,\frac{\c.p +m}{p^2 - m^2}\, e^{-ip.(x-y)} \ ,
\label{b8}
\end{equation}
where the contour ${\cal C_+}$  in the $p^0$ plane wraps completely around the pole at $p^0 = E_{\bf p}$ on the 
positive real axis, as shown in Figure \ref{Fig 1}. 
\begin{figure}[h!]
\centering
\includegraphics[scale=0.6]{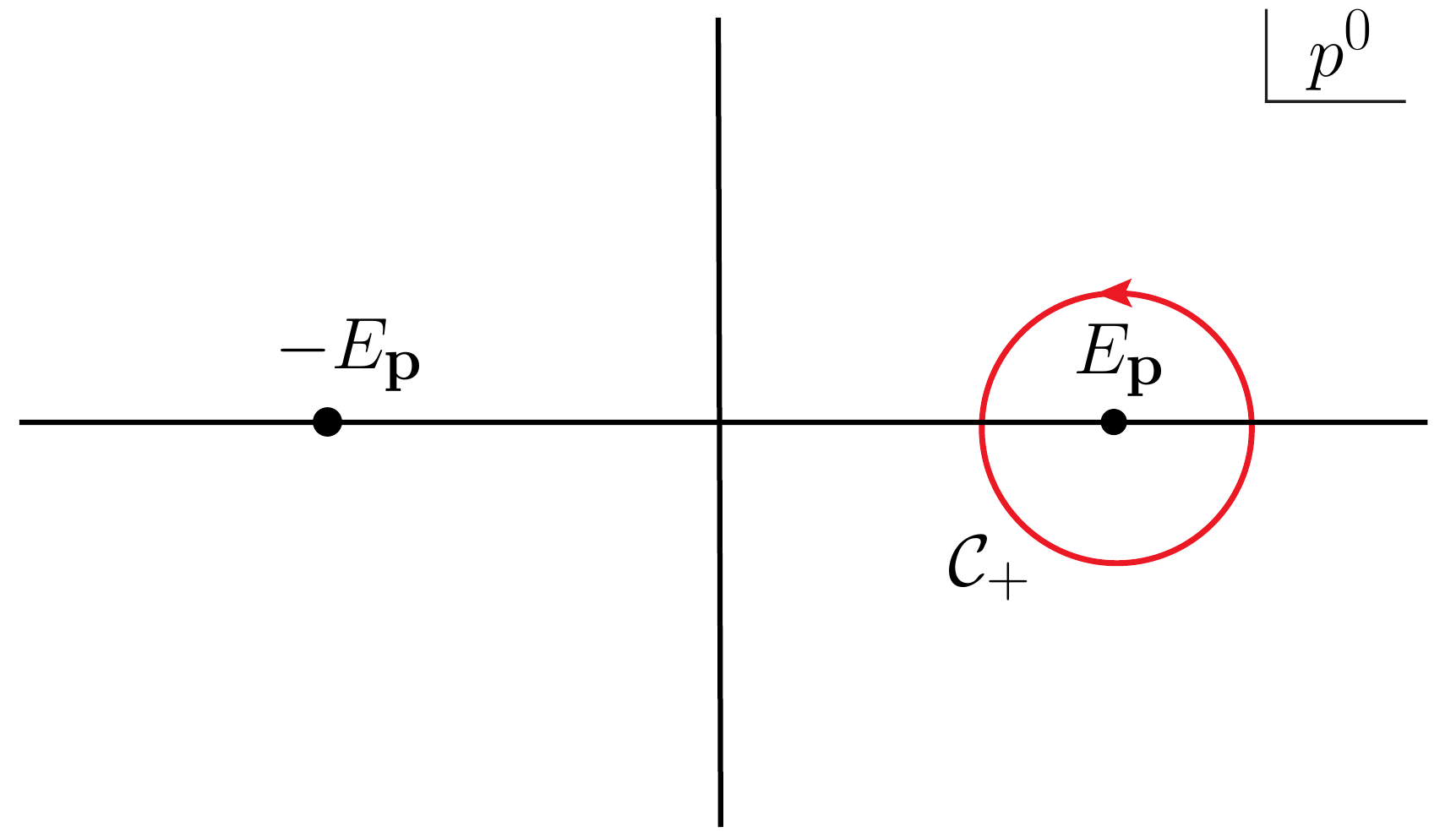}
\caption{The contour ${\cal C}_+$ wraps around the pole at $p^0 = E_{\bf p}$ in the complex $p^0$ plane, 
defining the Wightman propagator $S_+(x,y)$.}
\label{Fig 1}
\end{figure}
 The corresponding Green functions, including the Feynman propagator, are 
similarly defined with different contours around the poles.\footnote{Closed contours wrapping round either 
or both of the poles give solutions of the homogeneous Dirac equation (including the Wightman and anticommutator
(Pauli-Jordan) functions), whereas open contours along the real $p^0$ axis diverting above or below the poles
give Green functions, for example the retarded/advanced Green functions or the Feynman propagator.}

Similarly, we have
\begin{align}
S_{-\a\b}(x,y) ~&=~\langle 0|\bar\psi_\a(x)\,\,\psi_\b(y) |0\rangle \nonumber \\
&= \int \frac{d^3{\bf p}}{(2\pi)^3}\frac{1}{2E_{\bf p}} \,\left(\c^T.p - m\right)_{\a\b} e^{-ip.(x-y)} \ ,
\label{b9}
\end{align}
with the contributions coming from the positron creation and annihilation operators $b_{\bf p}^s, b_{\bf p}^{s\dagger}$
and the spinors $v_\a^s(p)$.  In the same way, the Wightman function for the charge conjugate fields is
\begin{align}
S_{+\a\b}^{\mathsf C}(x,y) ~&=~\langle 0|\psi_\a^{\mathsf C}(x)\,\,\bar\psi_\b^{\mathsf C}(y) |0\rangle \nonumber \\
&= \int \frac{d^3{\bf p}}{(2\pi)^3}\frac{1}{2E_{\bf p}} \,\left(\c.p + m\right)_{\a\b} e^{-ip.(x-y)} \ ,
\label{b10}
\end{align}
derived from the positron operators and the $u_\a^s(p)$ spinors.

One way to phrase the requirement of causality in QFT (usually called `microcausality' in this context)
is to demand that the VEV of the anti-commutator of two spinor fields should vanish when they are
spacelike separated. From the results above, still taking $p^0 = E_{\bf p}$, we find
\begin{align}
\langle0|\{\psi_\a(x),\,\bar\psi_\b(y)\} |0\rangle ~&=~\left(i\c.\partial_x + m \right) \,
\int \frac{d^3{\bf p}}{(2\pi)^3}\frac{1}{2E_{\bf p}} \,\left(e^{-ip.(x-y)} \,-\,e^{ip.(x-y)}\right) \nonumber \\
&{} \nonumber \\
&\rightarrow~ 0~~{\rm for}~(x-y)^2 < 0 \ .
\label{b11}
\end{align}
To see this, note that for spacelike $(x-y)$ {\it only}, we can transform $(x-y)\rightarrow \,-(x-y)$ 
by a Lorentz transformation and rotation, then since the integration measure is Lorentz invariant
it follows that the two terms cancel.\footnote{We have illustrated this for free fields for simplicity.
The argument could be generalised for interacting theories including self-energy contributions 
by the substitution $\c.p+m \rightarrow A(p^2) \c.p + B(p^2)$, etc.}

The key point is that this cancellation requires the existence of antiparticles with the {\it exact} 
mass and spin, and opposite charge, of the corresponding particles.
(Recall the second term arises from the positron operators $b_{\bf p}^s, b_{\bf p}^{s\dagger}$ whereas 
the first term involves the electron operators $a_{\bf p}^s, a_{\bf p}^{s\dagger}$.) 
Any differences, however small, of these properties would entail the violation of causality.

Notice, however, the key role of Lorentz invariance in this conclusion. We return to this in 
section \ref{sect 2.3} where we discuss the possibility of Lorentz breaking.

\subsection{Lorentz Symmetry and $\mathsf{CPT}$}\label{sect 2.2}

The preceding discussion has considered a QFT formulated on Minkowski spacetime, 
obeying the natural symmetries of Lorentz invariance $SO(1,3)$ and spacetime translations $T^4$.
This was sufficient to demonstrate that the existence of antiparticles with the exact mirror
properties of their particles is required to preserve causality. Notice that as yet there has been
no need to invoke $\mathsf{CPT}$ symmetry.

Lorentz symmetry played the key role. Spin $1/2$ fermions arise as spinor representations of the
Lorentz group, or more precisely its double cover $SL(2,C)$.  Left and right-handed Weyl spinors transform 
under the $(\tfrac{1}{2},0)$ and $(0,\tfrac{1}{2})$ representations of $SL(2,C)$, while the Dirac spinor
describing the electron/positron is the representation $(\tfrac{1}{2},0) \oplus (0,\tfrac{1}{2})$.
The fact that spin $1/2$ particles are described as Lorentz group representations will be
crucial when when we introduce gravity, especially in reference to $\mathsf{CPT}$.

In addition to the continuous symmetries, we may also discuss the discrete transformations of
charge conjugation ($\mathsf{C}$), parity ($\mathsf{P}$) and time reversal ($\mathsf{T}$) in the
Dirac theory. While these are independently symmetries of QED, in the full standard model each is 
known to be broken. However, the combination $\mathsf{CPT}$ has a special status and is conserved 
in any local, relativistic QFT. 

The actions of $\mathsf{C}$, $\mathsf{P}$ and $\mathsf{T}$ on the creation and annihilation operators 
are simple:
\begin{align}
&a_{\bf p}^s ~\xrightarrow{\mathsf{C}}~  b_{\bf p}^s \ , ~~~~~~~~ 
a_{\bf p}^s ~\xrightarrow{\mathsf{P}}~ a_{-\bf p}^s \ , ~~~~~~~~ 
a_{\bf p}^s ~\xrightarrow{\mathsf{T}}~ a_{-\bf p}^{-s} \ , \nonumber \\
&b_{\bf p}^s ~\xrightarrow{\mathsf{C}}~ a_{\bf p}^s \ , ~~~~~~~~ 
b_{\bf p}^s ~\xrightarrow{\mathsf{P}}~ - b_{-\bf p}^s \ , ~~~~~~~~ 
b_{\bf p}^s ~\xrightarrow{\mathsf{T}}~ b_{-\bf p}^{-s} \ . \nonumber \\
&{}
\label{b12}
\end{align}
Inserting these relations into the expansion (\ref{b2}) for the fields, and using appropriate identities 
for the $u_\a^s(p)$ and $v_\a^s(p)$ spinors \cite{Peskin:1995ev}, we find the corresponding, but less transparent,
relations (just writing the combined $\mathsf{PT}$ transformation for simplicity):
\begin{align}
&\psi(x)~\xrightarrow{\mathsf{C}}~\psi^{\mathsf C}(x) ~=~ C\bar\psi^T(x) \ , \nonumber \\
&\psi(x)~\xrightarrow{\mathsf{PT}}~\psi^{\mathsf{PT}}(x) ~=~   \PTzero \psi(-x) \ , 
\label{b13}
\end{align}
where $C$ is as above and $\PTzero  = - \c^0 \c^1 \c^3$, with similar expressions for $\bar\psi$.\footnote{In general,
there are phases associated with each of the transformations $\mathsf{C}$, $\mathsf{P}$ and $\mathsf{T}$,
whose product is constrained to be 1. These phases are all set to 1 here, but we need to retain the relative minus sign
in the parity transformations of fermions and antifermions in (\ref{b12}) to ensure the correct parity assignments for
fermion-antifermion bound states \cite{Peskin:1995ev}.}
Combining these, we find the key $\mathsf{CPT}$ transformations:
\begin{align}
&\psi^{\mathsf{CPT}}(x) ~=~ \PTzero \psi^{\mathsf{C}}(-x) ~=~  - \c^5 \psi^*(-x) \ , \nonumber \\
&\bar\psi^{\mathsf{CPT}}(x) ~=~ - \bar\psi^{\mathsf{C}}(-x) \PTzero  ~=~ \bar\psi^*(-x) \c^5 \ .
\label{b14}
\end{align}
Writing $\psi^{\mathsf{CPT}}$ explicitly in terms of the creation and annihilation operators,
\begin{equation}
\psi_\a^{\mathsf{CPT}}(x) = \int \frac{d^3{\bf p}}{(2\pi)^3}\frac{1}{\sqrt{2E_{\bf p}}}  
~ \sum_s\left(- b^{-s}_{\bf p} {u^s_\a(p)}^* e^{ip.x} \,+\,
a^{-s\dagger}_{\bf p} {v^s_\a(p)}^* e^{-ip.x} \right) \ . 
\label{b15}
\end{equation}
Note that $\mathsf{C}$ and $\mathsf{P}$ are realised by linear, unitary operators
whereas $\mathsf{T}$, and therefore $\mathsf{CPT}$, is an anti-linear, anti-unitary transformation.

We can now prove a fundamental result using $\mathsf{CPT}$ invariance to relate the physics of 
particles and antiparticles. If $\mathsf{CPT}$ is a symmetry of the QFT, and not spontaneously broken
by the vacuum state, then the Wightman propagator satisfies
\begin{equation}
\langle 0| \psi(x) ~ \bar\psi(y)|0\rangle ~=~
\langle 0| \psi^{\mathsf{CPT}}(x) ~ \bar\psi^{\mathsf {CPT}}(y)|0\rangle^* \ .
\label{b16}
\end{equation}
The complex conjugate on the r.h.s.~arises because of the anti-unitary property of $\mathsf{CPT}$.
Evaluating this, we find\footnote{Notice that the spin flip 
under $\mathsf{CPT}$ in the creation and annihilation operators is not apparent in the propagators
themselves because of the spin sum.}
\begin{align}
\langle 0| \psi^{\mathsf {CPT}}(x) ~ \bar\psi^{\mathsf {CPT}}(y)|0\rangle^*  
&~=\left( - \PTzero \,\langle 0| \psi^{\mathsf {C}}(-x) ~ \bar\psi^{\mathsf {C}}(-y)|0\rangle \,\PTzero\, \right)^* 
\nonumber \\
&~=\left( - \PTzero \int \frac{d^3{\bf p}}{(2\pi)^3}\frac{1}{2E_{\bf p}}\,
\left(A(p^2) \c.p + B(p^2) \right) \, \PTzero\, e^{ip.(x-y)} \right)^* 
\nonumber \\
&~= \left(\,\int \frac{d^3{\bf p}}{(2\pi)^3}\frac{1}{2E_{\bf p}}\,
\left(A(p^2) \c^*.p + B(p^2) \right) \, e^{ip.(x-y)}\, \right)^* \nonumber \\
&~=~ \langle 0| \psi^{\mathsf {C}}(x) ~ \bar\psi^{\mathsf {C}}(y)|0\rangle \ .
\label{b17}
\end{align}
We have allowed here for self-energies through $A(p^2)$, $B(p^2)$ so this result holds for an 
interacting theory, not simply for free fields. Crucially, the derivation assumes both Lorentz invariance 
{\it and} translation invariance (since we have assumed the propagator is a function of $(x-y)$ 
in the second step). Under these assumptions, which must be reconsidered in the presence of gravity
\cite{McDonald:2015iwt}, we have therefore shown that $\mathsf{CPT}$ invariance implies that 
the propagators of particles and antiparticles in Minkowski spacetime are identical, {\it i.e.}
\begin{equation}
\langle 0| \psi(x) ~ \bar\psi(y)|0\rangle ~=~ \langle 0| \psi^{\mathsf {C}}(x) ~ \bar\psi^{\mathsf {C}}(y)|0\rangle \ .
\label{b18}
\end{equation}

The general relation between Lorentz invariance and $\mathsf{CPT}$ symmetry is expressed in the 
famous $\mathsf{CPT}$ theorem \cite{Pauli:1940, Bell, Luders, Pauli:1955}.
The theorem may be proved rigorously in axiomatic field theory, but is fundamentally simple.
The essential statement is that in a {\it local}, {\it Lorentz invariant} QFT,
$\mathsf{CPT}$ is an exact symmetry. A simple proof is given in \cite{Weinberg:1995mt},
where it is shown, based on the transformations (\ref{b14}) together with their equivalents for scalar
and vector fields, that any local product of scalar, spinor and vector fields is necessarily invariant
under the combined $\mathsf{CPT}$ transformation. This simply means that any Lorentz invariant interaction
we may write down in an effective Lagrangian is $\mathsf{CPT}$ invariant.

To illustrate this, we can construct a table of the basic fermion bilinear operators
relevant to the analysis of low-energy antimatter experiments, together with their $\mathsf{C}$, $\mathsf{P}$,
$\mathsf{T}$ and $\mathsf{CPT}$ transformations \cite{Peskin:1995ev}.
 It is clear from Table \ref{TableCPT} that the Lorentz scalar operators are
$\mathsf{CPT}$ even, in accord with the $\mathsf{CPT}$ theorem. 

\begin{table}[h!]
\begin{center}
\begin{tabular} { c c c c c c c c c c c c c }
\\
\hline\hline\\
 \multicolumn{1}{c} {\raisebox{1.5ex} { } } &
  \multicolumn{1}{c} {\raisebox{1.5ex} {  }  }  &
   \multicolumn{1}{c} {\raisebox{1.5ex} {  }  }  &
 \multicolumn{1}{c} {\raisebox{1.5ex} {$\bar\psi \psi$}  } &
   \multicolumn{1}{c} {\raisebox{1.5ex} {  }  }  &
 \multicolumn{1}{c} {\raisebox{1.5ex} {$i\bar\psi \c^5 \psi$ } } &
   \multicolumn{1}{c} {\raisebox{1.5ex} {  }  } &
\multicolumn{1}{c} {\raisebox{1.5ex} {$\bar\psi \c^\m \psi$}  } &
  \multicolumn{1}{c} {\raisebox{1.5ex} {  }  }  &
  \multicolumn{1}{c} {\raisebox{1.5ex} {$\bar\psi\c^\m \c^5\psi$ }  } & 
    \multicolumn{1}{c} {\raisebox{1.5ex} {  }  }  &
\multicolumn{1}{c} {\raisebox{1.5ex} {$\bar\psi\s^{\m\n}\psi$ }  }  &
\\
\hline 
\\
$\mathsf{C}$	&&&   $+1$ && $+1$ && $-1$ && $+1$  && $-1$ 	\\
$\mathsf{P}$ &&&   $+1$  &&  $-1$ && $(-1)^\m$ && $-(-1)^\m$ && $(-1)^\m (-1)^\n$    \\
$\mathsf{T}$ &&&   $+1$  &&  $-1$ && $(-1)^\m$ && $(-1)^\m$ && $-(-1)^\m (-1)^\n$  \\
$\mathsf{CPT}$ &&&   $+1$  &&  $+1$ && $-1$ && $-1$ && $+1$  \\
\hline\hline
\end{tabular}
\end{center}
\caption{{$\mathsf{C}$, $\mathsf{P}$, $\mathsf{T}$ and $\mathsf{CPT}$ transformations for the
basic fermion bilinear operators. The notation is: $(-1)^\m = 1$ for $\m=0$ and $-1$ for $\m=1,2,3$.  }}
\label{TableCPT}
\end{table}

\subsection{Breaking Lorentz Invariance and $\mathsf{CPT}$}\label{sect 2.3}

So far we have reviewed how Lorentz invariance, $\mathsf{CPT}$ and translation invariance are fundamental
to the local, relativistic QFTs which successfully describe particle physics in Minkowski spacetime.
The fact that quantum fields are representations of the Lorentz group $SO(1,3)$ (or for spinor fields, $SL(2,C)$)
is essential in the first place to have particle states with definite masses and spins, since these are the eigenvalues 
of the two Casimir operators of the Lorentz group. The existence of antiparticles with exactly the same mass
and spin and opposite charge was then shown to be required to maintain causality. We also presented a direct
proof that $\mathsf{CPT}$ symmetry, together with Lorentz and translation invariance, implies the equality
of the propagators for particles and antiparticles. Finally, the $\mathsf{CPT}$ theorem ensures that $\mathsf{CPT}$
is an exact symmetry of any local, Lorentz invariant QFT.

The question then arises how this tightly-woven theoretical structure could be unravelled in the event that
experiments contradicted its clear predictions, {\it e.g.}~in violating charge neutrality in antihydrogen or 
finding a difference in the atomic spectroscopy of hydrogen and antihydrogen? 

The simplest way to describe a possible breaking of Lorentz invariance, while preserving the fundamental
structure of causal fields in representations of the Lorentz group, is to write a phenomenological Lagrangian 
including operators which are not Lorentz invariant. For QED, we therefore consider
\begin{align}
{\cal L}_{\rm LV} =\,& {\cal L}_{\rm QED} - \tfrac{1}{4}(k_F)_{\m\r\n\s}F^{\m\r}F^{\n\s}
+\tfrac{1}{2} (k_{AF})^\r \e_{\r\m\n\s} A^\m F^{\n\s}  \nonumber \\
& - a_\m \bar\psi \c^\m \psi - b_\m \bar\psi\c^5 \c^\m \psi -  \tfrac{1}{2} H_{\m\n} \bar\psi\s^{\m\n}\psi
+ c^{\m\n}i\bar\psi \c_\m D_\n\psi + d^{\m\n} i\bar\psi \c_5\c_\m D_\n \psi \ .
\label{b19}
\end{align}
This is just the restriction to QED\footnote{\label{fSME} Note that here we omit three further 
$\mathsf{CPT}$ odd operators, 
$i e^\m\bar\psi D_\m\psi $,  $f^\m\bar\psi \c^5 D_\m\psi$
and $\tfrac{i}{2} g^{\m\n\l}\bar\psi \s_{\m\n} D_\l\psi$, which could arise in QED
alone but which are not obtained as a restriction of the
SME \cite{Colladay:1998fq}.} of the full ``Standard Model Extension'' (SME) of Kosteleck\'y
and collaborators \cite{Colladay:1996iz, Colladay:1998fq}.
This has been the subject of an extensive programme of research over many years,
with stringent experimental bounds being established on many of the Lorentz-violating couplings 
\cite{Kostelecky:2008ts}. Note the profligacy of the parameter count in these theories once Lorentz symmetry
is lost -- the QED Lagrangian ${\cal L}_{\rm LV}$ alone has some 70 independent parameters.

The experimental consequences of some of these additional Lorentz-violating interactions 
for antimatter will be discussed below. First, we need some comments on the theoretical basis
of (\ref{b19}).

A key point is that only {\it local} operators have been included. Locality is one of the axioms
underlying the $\mathsf{CPT}$ theorem, so if locality were not valid this would break the link
between Lorentz invariance and $\mathsf{CPT}$. We briefly consider a model with non-local
operators later in this sub-section, but elsewhere in the paper we simply accept locality 
and the $\mathsf{CPT}$ theorem, in particular that a violation of $\mathsf{CPT}$ necessarily 
entails a violation of Lorentz invariance.

Next, note that it is implicit in (\ref{b19}), where the fields $\psi(x)$ are the usual spinor representations
of the Lorentz group and admit the decomposition (\ref{b2}) in terms of creation and annihilation
operators, that the charge and mass of particles and their antiparticles are identical. This is a key feature 
of the SME and allows it, with the provisos below, to have an interpretation as a causal theory.
We could imagine instead writing the expansion (\ref{b2}) with different masses for the particles
and antiparticles, so the momenta associated with the operators $a_{\bf p}^s$ and $b_{\bf p}^s$
satisfied energy-momentum relations with a different mass.  This was investigated by Greenberg
in \cite{Greenberg:2002uu}, with the conclusion, unsurprising in view of the discussion of causality
in section \ref{sect 2.1} and in what follows below, that such a `theory' would not be causal
and in a certain sense non-local. This justifies the restriction to the less destructive breaking of
Lorentz and $\mathsf{CPT}$ invariance parametrised in the SME Lagrangian (\ref{b19}).

Finally, we should distinguish between {\it spontaneous} and {\it explicit} breaking of Lorentz invariance.
One way to view the coefficients $a_\m$, $b_\m$, $\ldots$ in (\ref{b19}) is as the spontaneous 
Lorentz-violating VEVs of vector or tensor fields in some fundamental Lorentz invariant theory. 
${\cal L}_{\rm LV}$ would then represent an effective Lagrangian describing this theory at low energies
where the dynamics of these new fields can be neglected.

More generally, the philosophy of low-energy effective Lagrangians is to write an expansion in terms of 
operators of increasing dimension. Operators of dimension $> 4$ will have coefficients with 
dimensions of inverse powers of mass and will therefore be suppressed by the mass scale characteristic 
of some unknown dynamics at high energy. (A familiar example is the chiral Lagrangian for low-energy QCD.) 
This motivates us to restrict ${\cal L}_{\rm LV}$ only to the soft or renormalisable Lorentz-violating operators, 
with dimension $\le 4$, shown in (\ref{b19}). With this restriction, the theory is known as the 
{\it minimal} SME.  If operators of dimension $> 4$ are included, it is referred to as non-minimal.
We consider some examples of non-minimal operators in section \ref{subsubsection:LVCPTV}.

Alternatively, we can simply break Lorentz invariance {\it explicitly}, in which case the coefficients 
$a_\m$, $b_\m$, $\ldots$ are merely constants with no relation to underlying covariant fields. 
For low-energy phenomenology the distinction is not so important, but the issue of whether (\ref{b19}) is 
to be viewed as a complete theory in itself or simply as an effective Lagrangian with a Lorentz-invariant 
UV completion, is important for its theoretical interpretation, particularly as regards causality. 

As can be seen from Table \ref{TableCPT}, and in accord with the $\mathsf{CPT}$ theorem,
the Lorentz-violating operators in (\ref{b19}) divide into those that conserve $\mathsf{CPT}$ and those
that violate it, while of course all the operators in the original Lorentz-invariant ${\cal L}_{\rm QED}$ 
are $\mathsf{CPT}$ conserving. Of the complete set of Lorentz-violating operators, those with couplings
$a_\m, b_\m , k_{AF}$ are $\mathsf{CPT}$ odd, while those with 
$c^{\m\n}, d^{\m\n}, H^{\m\n}, k_F$ are $\mathsf{CPT}$ even.
Also note that $a_\m$ couples to a $\mathsf{C}$ odd operator, while the corresponding operator
with $b_\m$ is $\mathsf{C}$ even.

We can analyse ${\cal L}_{\rm LV}$ in the same way as presented above for the standard Lorentz-invariant
Dirac theory. To illustrate some of the theoretical issues, we show some results in the
especially simple case where only $a_\m$ is taken to be non-zero.

The modified Dirac equation is then 
\begin{equation}
\left(i\c.\partial - \c.a - m\right) \psi ~=~ 0 \ ,
\label{b20}
\end{equation}
while the expansion of the field $\psi(x)$ in creation and annihilation operators becomes\footnote{
Note that in this special case, we could equally well move the ${\bf a}$ dependence from the spinors
entirely into the exponent through a change of integration variable ${\bf p} \rightarrow {\bf p}-{\bf a}$,
though this will not be possible in the general Lorentz-violating theory.}
\begin{align}
\psi(x) = 
&~\int \frac{d^3{\bf p}}{(2\pi)^3}\, \sum_s \biggl(
\frac{1}{\sqrt{2E_{{\bf p}-{\bf a}}}} \,a^s_{{\bf p}-{\bf a}} \,
u^s\left(E_{{\bf p}-{\bf a}}, {\bf p} - {\bf a}\right) \, e^{-i p_u.x} 
\nonumber \\
&~~~~~~~~~~~~~~~~~~~~+\, \frac{1}{\sqrt{2E_{{\bf p}+{\bf a}}}} b^{s\dagger}_{{\bf p}+{\bf a}} 
v^s\left(E_{{\bf p}+{\bf a}}, {\bf p} + {\bf a}\right) 
e^{i p_v.x} \biggr) \ , 
\label{b21}
\end{align}
where $p_u = \left(E_{{\bf p}-{\bf a}} + a^0, {\bf p}\right)$ and 
$p_v = \left(E_{{\bf p}+{\bf a}} - a^0, {\bf p}\right)$, 
with  $E_{{\bf p}-{\bf a}} = \sqrt{({\bf p} - {\bf a})^2 + m^2}$.
The dispersion relations are:
\begin{equation}
\left(p_u - a\right)^2 - m^2 = 0 \ , ~~~~~~~
\left(p_v + a\right)^2 - m^2 = 0 \ ,
\label{b22}
\end{equation}
while the spinors $u$, $v$ are the same functions as usual, with shifted arguments.

Analogous expressions can be written for the $\mathsf{C}$ and $\mathsf{CPT}$ conjugates
$\psi^{\mathsf{C}}(x) = C\bar\psi^T(x)$ and $\psi^{\mathsf{CPT}}(x) = - \c^5 \psi^*(-x)$,
making the transformations (\ref{b12}) for the creation/annihilation operators together
with the substitution $a_\m \rightarrow - a_\m$ since this is $\mathsf{C}$ odd 
(see Table \ref{TableCPT}).

The Wightman propagators are easily found:
\begin{align}
S_+(x,y) &= \langle 0|\psi(x) \,\bar\psi(y)|0\rangle    \nonumber \\ 
&=\int \frac{d^3{\bf p}}{(2\pi)^3}\frac{1}{2E_{{\bf p}-{\bf a}}}\,
\left( \c.\left(E_{{\bf p}-{\bf a}}, {\bf p} - {\bf a}\right) + m\right) \,e^{-i p_u.(x-y)} \nonumber \\
&= -i \int_{{\cal C}_+^u} \frac{d^4 p}{(2\pi)^4} \, \frac{\c.(p-a) + m}{(p-a)^2 - m^2}\, e^{-ip.(x-y)} \ , 
\label{b23}
\end{align}
where ${\cal C}_+^u$ circles the pole at $p^0 = E_{{\bf p}-{\bf a}} + a^0$ (Figure \ref{Fig 2}).
\begin{figure}[h!]
\centering
\includegraphics[scale=0.6]{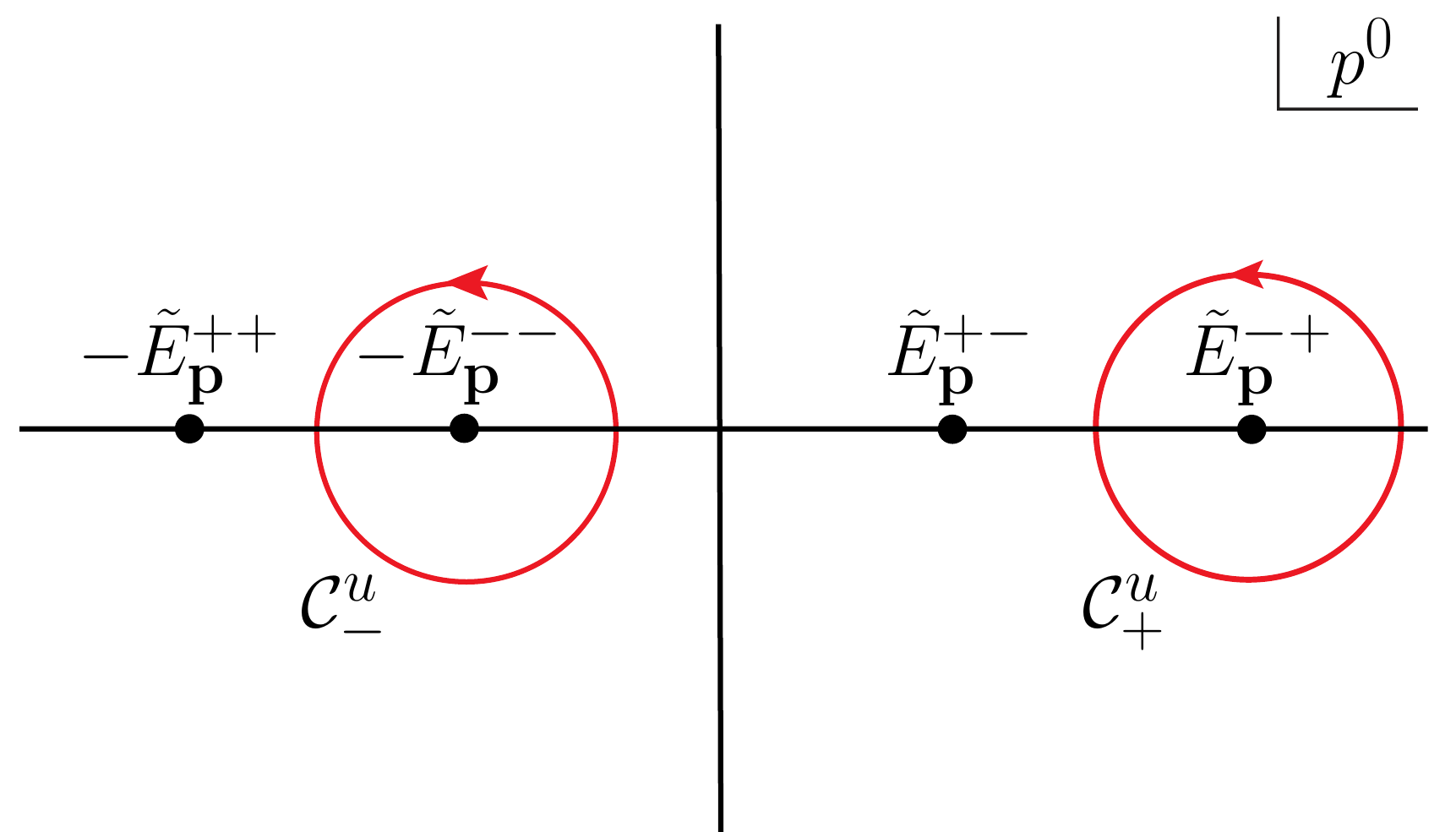}
\caption{The contours ${\cal C}_+^u$ and ${\cal C}_-^u$ in the complex $p^0$ plane 
relevant for calculating the Wightman and anticommutator functions in the Lorentz-violating theory 
with $a_\m$ non-zero.  The abbreviated notation for the poles is
$\tilde{E}_{\mathbf{p}}^{-+} = E_{\mathbf{p} - \mathbf{a}} + a^0$ and
$\tilde{E}_{\mathbf{p}}^{--} = E_{\mathbf{p} - \mathbf{a}} - a^0$, {\it etc.} }
\label{Fig 2}
\end{figure}

The equivalent result for $\langle 0| \psi^{\mathsf{C}}(x) \,\bar\psi^{\mathsf{C}}(y)|0\rangle$ 
follows with the substitution $a_\m \rightarrow - a_\m$, the poles being shifted accordingly.
Unsurprisingly, in this Lorentz-violating theory
the Wightman propagators for the fields $\psi$ and $\psi^{\mathsf{C}}$ are different -- electrons and positrons 
have different dispersion relations, given in (\ref{b22}), and propagate differently.

The derivation in (\ref{b17}) for the Wightman propagator for $\psi^{\mathsf{CPT}}$ remains unchanged, so we have
\begin{align}
\langle0|\psi^{\mathsf{C}}(x)\, \bar\psi^{\mathsf{C}}(y)|0\rangle &=  
\langle0|\psi^{\mathsf{CPT}}(x)\, \bar\psi^{\mathsf{CPT}}(y)|0\rangle^* \nonumber \\
&\ne \langle0|\psi(x)\, \bar\psi(y)|0\rangle \ ,
\label{b24}
\end{align}
as expected in a theory with $\mathsf{CPT}$ violation.

This immediately raises the issue of causality. Given the r\^ole of Lorentz invariance and the exact equivalence
of the properties of particles and antiparticles in establishing microcausality in section \ref{sect 2.1},
we have to question whether the introduction of the Lorentz-violating terms in ${\cal L}_{\rm LV}$ 
necessarily violates microcausality.

For the simple case with only $a_\m$ non-zero, this is readily answered.
After some calculation we can explicitly write the anticommutator function as
\begin{equation}
\langle0|\{\psi(x),\bar\psi(y)\}|0\rangle 
= \left(i \c.\partial_x - \c.a + m\right) 
\int\frac{d^3{\bf p}}{(2\pi)^3} \left( \frac{1}{2 E_{{\bf p}-{\bf a}}} e^{-i p_u.(x-y)} 
- \frac{1}{2 E_{{\bf p}+{\bf a}}} e^{i p_v.(x-y)} \right)  \ ,
\label{b25}
\end{equation}
which follows from integrating around both the contours ${\cal C}_+^u$ and ${\cal C}_-^u$ in Figure \ref{Fig 2}.
It is critical here that these contours circle the poles which are both shifted to the right by $+ a_0$.
After some changes of variable, it then follows that
\begin{equation}
\langle0|\{\psi(x),\bar\psi(y)\}|0\rangle = e^{-i a.(x-y)}\,
\left(i\c.\partial_x + m \right) \int \frac{d^3{\bf p}}{(2\pi)^3}\frac{1}{2E_{\bf p}} \,\left(e^{-ip.(x-y)} \,-
\,e^{ip.(x-y)}\right) \Big|_{p^0 = E_{{\bf p}}} \ ,
\label{b26}
\end{equation}
where, as a result of the specification of the contours above, the $a_\m$ dependence factorises.
This is the key step. The remaining integral is then exactly as appeared in the Lorentz-invariant theory, 
and vanishes for spacelike $(x-y)$.

Perhaps counter-intuitively, we therefore find that microcausality continues to hold even in the theory
with Lorentz and $\mathsf{CPT}$ violation induced by a non-vanishing $a_\m$ coupling, 
despite the apparently different propagation characteristics of electrons and positrons.
In fact, there is a reason for this \cite{Colladay:1996iz}. The couplings $a^\mu$ in ${\cal L}_{\rm LV}$ can be removed 
by a field redefinition if we write $\psi(x) = \exp(-i\a.x) \xi(x)$ and rewrite the Lagrangian in terms of
the new field $\xi(x)$.  Physical electrons and positrons are then defined in terms of creation and annihilation
operators in $\xi(x)$.

In general, the dispersion relations following from ${\cal L}_{\rm LV}$ are {\it quartic} in the momenta, 
which doubles the number of poles in the complex $p^0$ plane in the calculation of the Wightman
propagators and other Green functions. Nevertheless, a similar argument to that shown above for
the simpler case of $a_\m$ demonstrates that the Lorentz-violating theory with the couplings
$b_\m$ or $H_{\m\n}$ also satisfies microcausality \cite{Colladay:1996iz, Kostelecky:2000mm}.

The situation is more subtle when the couplings $c^{\m\n}$, $d^{\m\n}$ to operators with derivatives are 
present. In this case, the poles may not lie on the real $p^0$ axis, and it is easy to identify special cases 
({\it e.g.} $c^{00} < 0$) where microcausality is violated and superluminal phase velocities arise for
electrons/positrons with momentum exceeding some threshold \cite{Colladay:1996iz, Kostelecky:2000mm}. 
This is also evident for certain of the photon couplings $(k_F)_{\r\s\m\n}$ and $(k_{AF})^\r$. 
This means that the theory (\ref{b19}) as it stands is {\it not} causal. 
However, if ${\cal L}_{\rm LV}$ is regarded only as a low-energy effective Lagrangian, then it remains possible 
that it may admit a causal UV completion. As emphasised in \cite{Shore:2003zc, Shore:2007um},
and the series of papers \cite{Hollowood:2007ku, Hollowood:2008kq, Hollowood:2009qz, Hollowood:2011yh} 
in a gravitational context, it is the UV limit of a QFT which determines whether or not
it is causal. In particular, in a theory with non-trivial dispersion relations, causality requires the 
high-momentum limit of the phase velocity to be less than $c$ ({\it not} the group velocity, which is 
irrelevant for causality in general). There exist perfectly causal QFTs which nevertheless have a low-energy 
effective Lagrangian exhibiting superluminal propagation and apparent microcausality violation.

It follows that, except for the restricted case with only $a_\m, b_\m, H_{\m\n}$ non-zero,
causality requires that ${\cal L}_{\rm LV}$ is to be regarded as an effective Lagrangian only.
Imposing causality of the fundamental theory of nature does not then preclude Lorentz and $\mathsf{CPT}$ 
violations from arising in low-energy experiments. Indeed, an experimental measurement indicating a non-vanishing
value for one of the Lorentz-violating couplings in ${\cal L}_{\rm LV}$ may provide an important clue as to the 
nature of the fundamental theory at high, perhaps Planckian, energies.

\vskip0.4cm

\noindent{\it Non-local quantum field theories}:
\vskip0.2cm

Locality is one of the axioms of the $\mathsf{CPT}$ theorem and it is interesting to consider
how our analysis, in particular the result that Lorentz violation is a necessary condition
for $\mathsf{CPT}$ violation, is impacted if we allow for non-locality.
String theory may initially be thought to exploit this loophole, but even here the theory at low energies,
well below the string or Planck scales, is still described by a local effective Lagrangian satisfying the
$\mathsf{CPT}$ theorem. A form of non-locality is intrinsic to quantum mechanics through entanglement,
but again this is not of a form that impacts on the $\mathsf{CPT}$ theorem. Other more exotic mechanisms
for $\mathsf{CPT}$ violation, {\it e.g.} involving non-trivial spacetime topology 
\cite{Klinkhamer:1999zh, Klinkhamer:2017hms}, are also not relevant here. 

It is, however, interesting to investigate the consequences of adding explicit non-local operators
directly into the QED Lagrangian. In particular, an ingenious proposal has been put forward in
\cite{Chaichian:2011fc, Dolgov:2012cm, Chaichian:2012hy} (for a review, see \cite{Fujikawa:2016her}) of a non-local, 
$\mathsf{CPT}$ violating but Lorentz invariant extension of QED. 
At energies well below the non-locality scale, this theory exhibits different
effective masses for the electron and positron, with the obvious consequences for the
hydrogen and antihydrogen spectra. However, due to the non-locality, the theory is expected to be 
non-unitary \cite{Chaichian:2012hy, Chaichian:2012ga, Fujikawa:2004rt}, while the realisation of causality 
and microcausality is also a potential issue. It therefore has probably to be viewed as only an effective theory 
from the point of view of analysing antimatter experiments at the energy scales of atomic physics, although 
it is not clear how it could be embedded in some more fundamental theory either at still higher
energies above the non-locality scale or in higher dimensions.

The model proposes adding the following non-local operator to the QED Lagrangian 
\cite{Chaichian:2012hy}:
\begin{equation}
S_{nloc} \,=\, - i\m \int d^4 x \int d^4 y\, \bar\psi(x) \left[\theta(x^0 - y^0) -\theta(y^0 - x^0) \right] 
\, \D_\ell(x-y) \, \Phi(x,y)\, \psi(y) \ ,
\label{bbb1}
\end{equation}
where
\begin{equation}
\D_\ell(x-y) \,=\, \d\left((x-y)^2 - \ell^2\right) \,-\, \d\left((x-y)^2\right) \ ,
\label{bbb2}
\end{equation}
is the non-local kernel.\footnote{The extra $\ell$-independent delta function in (\ref{bbb1}) is included
for technical reasons to cancel an infra-red divergence which would otherwise appear in the evaluation
of the form-factor $f(p)$ below. Note also that the $\ell$-dependent delta function links spacetime
points which are timelike separated.} The non-locality length scale is $\ell$ and we denote the corresponding 
high energy scale as $\L = 1/\ell$. The coefficient $\m$ is an arbitrary parameter with mass dimension 3.
Here,
\begin{equation}
\Phi(x,y) \,=\, \exp \left[i e \int_y^x dz^\m\, A_\m(z)\right]
\label{bbb3}
\end{equation}
is the non-integrable (path-dependent) phase factor \cite{Schwinger:1951nm, Wu:1975es} necessary
to preserve gauge invariance in the point-split bilinear fermion operator.

The new term (\ref{bbb1}) is hermitian (for $\m$ real) and manifestly Lorentz invariant.
Using the results in section \ref{sect 2.2} for the action of $\mathsf{C}$, $\mathsf{P}$ and
$\mathsf{T}$ (remembering that $\mathsf{T}$ is anti-unitary), we readily see that it has
$\mathsf{C}= -1$, $\mathsf{P}=1$ and $\mathsf{T} = 1$. It follows that it is odd under each of
$\mathsf{C}$, $\mathsf{CP}$ and $\mathsf{CPT}$. 

Neglecting the phase factor for the moment, we first show how the non-local operator (\ref{bbb1})
gives rise to a different effective mass for electrons and positrons. 
First, it follows directly from (\ref{bbb1}) that the modified Dirac equation is
\begin{equation}
\left( i \c.\partial - m\right) \psi(x) \,-\, i \m \int d^4 y \, 
\left[\theta(x^0 - y^0) - \theta(y^0 - x^0)\right]
\, \D_\ell(x-y) \psi(y) \,=\, 0 \ .
\label{bbb4}
\end{equation}
From section \ref{sect 2.2}, we see that the equivalent equation for the charge conjugate spinor
$\psi^{\mathsf{C}} = C \bar\psi^T(x)$ is 
\begin{equation}
\left( i \c.\partial - m\right) \psi^{\mathsf{C}}(x) \,+\, i \m \int d^4 y \, 
\left[\theta(x^0 - y^0) - \theta(y^0 - x^0)\right]
\, \D_\ell(x-y) \psi^{\mathsf{C}}(y) \,=\, 0 \ .
\label{bbb5}
\end{equation}
Note the change in sign of the contribution from the new non-local term, reflecting the fact that
it is $\mathsf{C}$ odd.

Now, writing $\psi(x) = e^{-i p.x} U(p)$, we find
\begin{equation}
\bigl(\c.p - m - f(p)\bigr) U(p) \,=\, 0 \ ,
\label{bbb6}
\end{equation}
where we define the Lorentz invariant factor $f(p) = i \left(f_+(p) - f_-(p)\right)$ with
\begin{equation}
f_{\pm}(p) \,=\, \m \int d^4 z \, e^{\pm i p.z} \theta(z^0) \D_{\ell}(z) \ .
\label{bbb7}
\end{equation}
For spacelike $p^2 < 0$, it follows (by evaluating in the frame $p^\m = (0, {\bf p})$) that
$f(p)$ vanishes. For timelike $p^2 > 0$, we choose the frame $p^\m = (p^0, {\bf 0})$ for which
\begin{equation}
f(p) \,=\, - 4\pi \m \int_0^\infty dz \left( \frac{z^2}{\sqrt{z^2 + \ell^2}} 
\sin\left({p^0\sqrt{z^2 + \ell^2}}\right)
\,-\, \frac{z^2}{\sqrt{z^2}} \sin\left({p^0\sqrt{z^2}}\right) \right)\ ,
\label{bbb8}
\end{equation}
which is clearly odd under $p^0 \rightarrow -p^0$. The second term arises from the inclusion of the
$\ell$-independent delta function in $\D_{\ell}(x-y)$.
After some manipulations \cite{Fujikawa:2014qra}, we can express this as a smooth, finite function
in the form
\begin{align}
f(p)\,&=\, 4\pi\, \frac{\m}{\L^2} \left[\theta(p^0)  - \theta(-p^0)\right]\theta(p^2) \nonumber \\
&\times \left[\frac{\pi}{4} \,+\, \int_0^1 du\, \left(u - \frac{1}{2u}\right) \sin\left(\frac{|p^0|u}{\L} \right)
\,+\, \int_1^\infty du\, \frac{1}{2u\left(\sqrt{u^2 -1} + u\right)^2}\, \sin\left(\frac{|p^0|u}{\L} \right) \right] 
\nonumber \\
&=\pi^2 \frac{\m}{\L^2} \,\left[\theta(p^0)  - \theta(-p^0)\right]\theta(p^2)\,
\left[ 1 \,+\, O\left(\frac{|p^0|}{\L}\right) \right] \ .
\label{bbb88}
\end{align}

In general, this modifies the electron dispersion relation in a non-trivial way. However, for 
energies $p^0 \simeq m$ well below the non-locality scale $\L$, it simply gives the electron an
effective mass
\begin{equation}
m_- \,\simeq\, m \,+\, \pi^2 \frac{\m}{\L^2} \ ,
\label{bbb9}
\end{equation}
while from (\ref{bbb5}), the correction to the positron mass has the opposite sign,
\begin{equation}
m_+ \,\simeq\, m \,-\, \pi^2 \frac{\m}{\L^2} \ .
\label{bbb10}
\end{equation}
The electron and positron therefore have a $\mathsf{CPT}$ violating mass splitting 
$\D m \simeq 2\pi^2 \m/\L^2$. Of course, the size of this splitting is completely arbitrary,
as even when we fix the non-locality scale, $\m$ remains a free parameter in the model.

Now consider the effect of the non-integrable phase factor $\Phi(x,y)$ in (\ref{bbb3}), which we recall
is necessary to preserve gauge invariance. A full treatment is technically challenging \cite{Chaichian:2012hy}.
However, if we simply expand the exponential and just consider the linear term, we see that
it induces a non-local $\bar{\psi} \psi A$ coupling of the photon to an electron-positron pair. 
This can be evaluated as,
\begin{equation}
\int d^4 x \, e^{-ik.x}\,\langle e^-(p)\,e^+(p')|J^\m(x)|0\rangle ~=~
(2\pi)^4 \d^4(k-p-p')\, e \, \bar{u}(p) \left[ \c^\m + F^\m(p,p')\right] v(p') \ ,
\label{bbb11}
\end{equation}
where $J^\m(x)$ is the electromagnetic current and $F^\m(p,p')$ is a form-factor,
\begin{equation}
F^\m(p,p') \,=\, - \frac{\partial}{\partial p^\prime_\mu} \int_0^1 d\eta\,f\bigl((\eta-1)p + \eta p'\bigr) \ .
\label{bbb12}
\end{equation}
We therefore find that the model necessarily implies a $\mathsf{C}$, $\mathsf{CP}$ and
$\mathsf{CPT}$ violating form-factor modification of the elementary pointlike electron-positron-photon
interaction in QED. 

Overall then, this non-local model has radical consequences even for the low-energy physics
of electrons and positrons well below the non-locality scale. These are tightly constrained by existing 
precision measurements in QED. Theoretically, it shows that the introduction of non-locality can circumvent
the $\mathsf{CPT}$ theorem and allow $\mathsf{CPT}$ violating particle-antiparticle mass differences
and couplings without violating Lorentz invariance. However, as mentioned above, the model as it stands
is unlikely to be a unitary, causal QFT and it remains quite unclear how it could be embedded in a 
fundamental theory satisfying these basic principles.

\subsection{General Relativity and Equivalence Principles}\label{sect 2.4}

So far, we have not considered gravity. However, the precision now being attained by laboratory experiments 
involving antimatter means that a full understanding of their implications for fundamental physics requires us 
to consider gravitational effects.

The established, and phenomenally successful, theory of gravity in classical physics is general relativity (GR). 
It follows that we should use GR from the outset as the framework to discuss gravitational effects in antimatter
experiments, not resort immediately to ad-hoc descriptions based on the superseded concepts of Newtonian 
theory, {\it e.g.}~that gravity couples to mass rather than, as in GR, the energy-momentum tensor.

The essential insight of GR is to replace the notion of an independent gravitational force by the
introduction of curved spacetime as the arena on which the laws of nature are formulated.
Here, spacetime is taken to be a pseudo-Riemannian manifold,\footnote{The terminology ``pseudo'' 
simply indicates the Minkowski signature of the metric $(-+++)$ rather than the Euclidean signature. 
In what follows, we just refer to the Minkowski signature manifolds as Riemannian for simplicity.} {\it i.e.}~it 
admits a metric of the standard quadratic form,
\begin{equation}
ds^2 = g_{\m\n} dx^\m dx^\n \ .
\label{b27}
\end{equation}
The fundamental theorem of Riemannian geometry then asserts that there is a unique, torsion-free, 
metric-preserving connection (the Levi-Civita connection) enabling parallel transport on the manifold.
This connection has the property that around any given point on the spacetime manifold
we can choose a local orthonormal frame (Riemann normal coordinates) in which {\it locally} the metric tensor reduces
to the Minkowski metric and the Christoffel symbols (which depend on the first derivatives of the metric)
vanish. In this sense, we can characterise the essential feature of the spacetime of GR as ``locally flat''.
The key symmetry principle, therefore, is that while in the presence of gravity we must abandon
global Lorentz invariance and translation invariance, we maintain {\it local} Lorentz invariance.

In physics terms, this implements geometrically the requirement that at each point in spacetime, there exists
a {\it local inertial frame} (LIF).  ``Local'' in this context means relative to the scale at which tidal gravitational
effects depending on the spacetime curvature (which depends on the second derivatives of the metric)
become important.
The existence of LIFs is the central principle on which GR is based, and dictates the choice of Riemannian
manifolds as the geometric description of spacetime.  

To couple matter to gravity we therefore have to formulate QFT on a curved spacetime. We outline this here
for a Dirac field. The immediate problem, as discussed earlier, is that the spinor field is a representation of the
$SL(2,C)$ covering group of the Lorentz group. But Lorentz invariance is no longer a global symmetry of 
spacetime in GR as it is in special relativity. However, because of the local Lorentz symmetry, we can still 
describe fermions by defining spinor fields with respect to the local orthonormal frame at each point in the curved
spacetime. This is done using the vierbein formalism. We define the vierbein $e_a{}^\m(x)$ such that
\begin{equation}
\eta_{ab} = g_{\m\n}(x) e_a{}^\m(x) e_b{}^\n(x)  \ ,
\label{b28}
\end{equation}
where $\eta_{ab} = {\rm diag}(-1,1,1,1)$ is the Minkowski metric. The coupled gravity--Dirac action in GR 
then becomes (ignoring the cosmological constant, which will not play a r\^ole here)
\begin{equation}
S = \int d^4 x \sqrt{-g} \left(\frac{R}{16\pi G} + \bar\psi\left(i \c^a e_a{}^\m D_\m - m\right) \psi \right) \ .
\label{b29}
\end{equation}
The covariant derivative on spinors is
\begin{equation}
D_\m \psi = \left(\partial_\m - \tfrac{i}{4} \omega_\m{}^{bc}\s_{bc}\right) \psi \ ,
\label{b30}
\end{equation}
where $\s^{ab} = \tfrac{i}{2} \left[\c^a,\c^b\right]$ and the spin connection is defined as
$\omega_\m{}^b{}_c = e_\n{}^b \left(\partial_\m e_c{}^\n + \Gamma^\n_{\m\r} e_c{}^\r \right)$.

The first key point here is that it is only the existence of local orthonormal frames, itself a
property of Riemannian spacetime,
that allows us to describe spin 1/2 particles and Dirac fields in the usual way at all.

Next, note that in this conventional GR action, the gravity-matter coupling is through the connection {\it only},
not the curvature. Since the Christoffel symbols $\Gamma^\n_{\m\r} \sim 0$ in a {\it local} inertial frame
({\it i.e.}~in Riemann normal coordinates), it follows that in a LIF $D_\m \psi \sim \partial_\m \psi$ and the 
Dirac action takes its special relativistic form. This is how conventional GR implies the 
Strong Equivalence Principle, {\it viz.}~that the laws of physics take their special relativistic form in the
local inertial frame at any given point in spacetime.

We can also note immediately the implications for $\mathsf{CPT}$ symmetry.
Since in GR we only have local Lorentz invariance, it follows that the discrete symmetries $\mathsf{P}$,
$\mathsf{T}$ and therefore $\mathsf{CPT}$ are only defined as transformations in the local Minkowski space 
at each spacetime point. (See {\it e.g.}~ref. \cite{McDonald:2014yfg} for further elaboration.)
In particular, they have nothing to do with the extended nature of the curved spacetime. 
$\mathsf{P}$ or $\mathsf{T}$ invariance of the action (\ref{b29}) does not involve any sort of space
or time reflection symmetry of the curved spacetime manifold.

The dynamical Einstein equations are derived by varying the action (\ref{b29}) with respect to the metric $g_{\m\n}$,
giving
\begin{equation}
G_{\m\n} \equiv R_{\m\n} - \tfrac{1}{2} R g_{\m\n} = 8 \pi G T_{\m\n} \ ,
\label{b31}
\end{equation}
where $T_{\m\n} = \tfrac{i}{2} \bar\psi \c^a e_a{}^{\{\m} D^{\n\}}\psi$ is the covariantly-conserved energy-momentum
tensor for the Dirac field and $R_{\m\n} (R)$ is the Ricci curvature tensor (scalar).
Note that the Einstein tensor $G_{\m\n}$ is automatically conserved by
virtue of the Bianchi identity, matching the r.h.s.~of (\ref{b31}).

This illustrates another key point. In GR, the gravitational field couples to the energy-momentum tensor, {\it not} the 
mass. It is described by a tensor field, the metric $g_{\m\n}(x)$, which in quantum theory corresponds to a 
massless spin 2 particle (the graviton) propagating with the speed of light, as has been beautifully 
confirmed by the recent discovery of gravitational waves \cite{Abbott:2016blz}.
The interaction is always attractive.

This is in marked contrast to the old Newtonian view, based on the force equation $F = G m_1 m_2/r^2$. 
Apart from invoking action at a distance, this encourages the misleading idea of mass as a 
`gravitational charge' by comparing with the analogous equation in electrostatics. However, in electromagnetism
the force is mediated by a spin 1 particle, the photon, and so the force between charges may be either
attractive or repulsive. This is not the case with gravity. Moreover, the frequently invoked distinction between
`gravitational' and `inertial' mass is not relevant once we have abandoned the obsolete Newtonian force equation
and adopted GR as the theory of gravity. There is only one mass
in the GR action (\ref{b29}), {\it viz.}~the Lorentz invariant particle mass in the LIF in which the spinor field is defined.

We now collect some elementary consequences of GR that we will need later in analysing the experiments.
First, it follows from the conservation of the energy-momentum tensor that free particles follow geodesics
in curved spacetime. That is, 
\begin{equation}
\frac{d^2 x^\m}{d\l^2} + \Gamma^\m_{\r\s} \frac{dx^\r}{d\l} \frac{dx^\s}{d\l} = 0 \ ,
\label{b32}
\end{equation}
where $x^\m(\l)$ describes the particle trajectory parametrised by the affine parameter $\l$. 
Clearly, in the LIF where $\Gamma^\m_{\r\s} \sim 0$, this path becomes a straight line. 

The geodesic equation is simply derived from the covariant conservation of the energy-momentum
tensor, $\nabla_\m T^{\m\n} = 0$, using the explicit form of the 
energy-momentum tensor for a free particle (see {\it e.g.}~ref. \cite{Hobson:2006se}).
An alternative derivation, which we shall use later in discussing possible WEPff violations, 
is to find the equation of motion by extremising the action for a point particle in curved
spacetime, {\it viz.}
\begin{equation}
S =-m \int ds = -m \int d\l\,\sqrt{g_{\m\n} \frac{dx^\m}{d\l} \frac{dx^\n}{d\l}} \ .
\label{b32a}
\end{equation}
A short calculation then shows,
\begin{equation}
\d S= m \int d\l\, \left(\frac{ds}{d\l}\right)^{-1}\d x_\m \biggl[\, \frac{d^2 x^\m}{d \l^2}  \,+\, 
\frac{1}{2} g^{\m\a}
\left(g_{\a\r,\s} + g_{\s\a,\r} - g_{\r\s,\a}\right) \frac{dx^\r}{d\l} \frac{dx^\s}{d\l} \biggr] \ .
\label{b32b}
\end{equation}
Identifying the second term as the Christoffel symbol, the geodesic equation follows as
$\d S/\d x^\m = 0$.

Note that the mass $m$ occurs here simply as an overall factor in $S$ and does not appear 
in the geodesic equation. Put otherwise, the {\it same} mass parameter multiplies both the 
``acceleration'' term and the ``gravity'' term in the equation of motion. As we now see, this is 
how the essentially Newtonian formulation of the equivalence principle as the identity
$m_i = m_g$ of the ``inertial'' and ``gravitational'' masses is realised.

To show this, we reproduce the equation of motion for a particle falling in the Earth's gravitational field.
The gravitational field in the exterior region of a spherically symmetric mass $M$ is described by
the Schwarzschild metric,
\begin{equation}
ds^2 = - \left(1 - \frac{2GM}{r}\right) dt^2 + \left(1 - \frac{2GM}{r}\right)^{-1} dr^2 + r^2 d\theta^2
+ r^2 \sin^2(\theta) d\phi^2 \ .
\label{b33}
\end{equation} 
In the weak field limit, we may write the metric in the form $g_{\m\n} = \eta_{\m\n} + h_{\m\n}$ and work
to first order in $h_{\m\n}$. Now, with the standard simplifications for a 
slow-moving (non-relativistic)  particle, and since $\Gamma^\m_{00} = -\tfrac{1}{2} g^{\m j}g_{00,j}$ 
for a static metric, the geodesic equation (\ref{b32}) quickly reduces to\footnote{Note that
eq. (\ref{b34}) refers to the space and time coordinates as they appear in the Schwarzschild metric
-- as explained below, these have to be translated into the physical measurements made by an
individual observer in order to confront with experiment. However, this translation only applies
a correction of $O(GM/r)$ and may be neglected since the r.h.s.~is already of that order.}
\begin{equation}
\frac{d^2 x^r}{dt^2} = \tfrac{1}{2} \partial_r h_{00} \ , 
\label{b34}
\end{equation}
for a radially falling particle. This matches the Newtonian equation with potential 
$U(r) = -\tfrac{1}{2} h_{00} = - GM/r$.  (Note we are using $c=1$ units throughout.)
At the Earth's surface, $U = - GM/R  \simeq - 7 \times 10^{-10}$.
Again, notice that the mass of the falling particle (being equal on both sides of (\ref{b34}))
cancels out from the equation of motion in accordance with the foundations of GR. 
This experimental prediction of GR therefore realises the universality of free-fall (WEPff).

An important issue in GR is to relate the coordinates used to describe the spacetime metric
(and in which calculations are most readily performed) with the physical measurements of 
space and time made by individual observers. (See, {\it e.g.}~\cite{Hobson:2006se} for a particularly
clear account.) This gives rise to {\it gravitational time dilation} and will be a key factor 
in interpreting the results of spectroscopic frequency measurements in curved spacetime.

In theoretical terms, we regard each observer (whether freely-falling or not) 
as equipped with a local orthonormal frame
$\hat e_a$ ($a = 0,1,2,3)$ whose components with respect to the coordinate basis
define a vierbein $\hat e_a{}^\m$ as described above.
The timelike frame vector is chosen to lie along the wordline of the observer:~specifically,
$\hat e_0{}^\m = u^\m$ where $u^\m = dx^\m/d\s$  is the observer's 4-velocity 
and $\s$ is the proper time along the observer's worldline (so $d\s^2 = - g_{\m\n} dx^\m dx^\n$ 
and the 4-velocity is normalised by $g_{\m\n}u^\m u^\n = -1$). 
A spacetime coordinate interval $dx^\m$ is then measured by this observer as the {\it projection
onto this local orthonormal frame}.  That is, 
\begin{equation}
d\hat x^a = \eta^{ab} \hat e_b{}^\m g_{\m\n} dx^\n \equiv  \hat e^a{}_\m dx^\m \ ,
\label{b35aa}
\end{equation}
where $\hat e^a{}_\m$ is the inverse vierbein field.

It follows immediately that if a frequency source, {\it e.g.}~an atomic spectral
transition, is moving through spacetime (so the coordinate interval to be measured is $dx^\m$), then a
{\it comoving} observer will measure the corresponding time interval as $d\hat t = d\s$,
the {\it proper time along the worldline} of the moving source. 
Explicitly, this fundamental observation follows from the formalism above as
\begin{equation}
d\hat t = \eta^{00} \,\hat{e}_0{}^\m \,g_{\m\n} \,dx^\n = - u^\m\,g_{\m\n}\,dx^\n = d\s \ .
\label{b36a}
\end{equation}

Now consider the measurement of the time interval between two events at the same point P, 
separated by a coordinate interval $dt$. In the rest frame, the time measurement is 
simply the corresponding proper time interval $d\t = \sqrt{-g_{00}}\, dt$, defined from the 
metric as $ds^2 = -d\t^2$. In a moving frame, with coordinate velocity $\textsl{v}^i = dx^i/dt$,
the measured time between the two events is, for a diagonal metric,
\begin{equation}
d\hat t = \eta^{00}\, \hat e_0{}^0\, g_{00}(P)\, dt = - u^0\, g_{00}(P) \,dt \equiv  \frac{d\t}{d\s} \,d\t \ .
\label{b35}
\end{equation}
To evaluate this time dilation factor, note that
\begin{equation}
u^0 = \frac{dt}{d\s}  = \left(-g_{00} - g_{ij} {\textsl{v}}^i {\textsl{v}}^j\right)^{-1/2} \ ,
\label{b37}
\end{equation}
and so, 
\begin{equation}
\frac{d\t}{d\s} = \left(1 + g_{ij}{\textsl{v}}^i {\textsl{v}}^j/g_{00}\right)^{-1/2} \ .
\label{b37a}
\end{equation}
We recognise that {\it locally} this factor is just the expression in curved spacetime coordinates
of the special relativistic time dilation factor $\c(v^2)$ in the orthonormal frame at P,
as follows from the vierbein definition $d\hat{x}^a = \hat{e}^a{}_\m dx^\m$ above.\footnote{As
a further demonstration of consistency, we should find the same time dilation factor for the
relativistically equivalent situation where we instead consider two spacetime events 
{\it on the worldline of the moving observer}, separated by coordinate interval $dt$ and 
therefore $dx^i = \textsl{v}^i dt$. In the comoving frame, the measured time is the proper time
interval along the worldline, $d\s$. In the stationary frame, the formalism above gives
the measured time interval as 
\begin{equation*}
d\tilde{t} = - \tilde{u}^\m\, g_{\m\n}(P) \,dx^\n \ ,
\end{equation*}
where $\tilde{u}^\m$ is the 4-velocity of the stationary observer, that is $\tilde{u}^0 = dt/d\t$
and $\tilde{u}^i = 0$. Since the metric is assumed diagonal, this reduces to
\begin{equation*}
d\tilde{t} = - \tilde{u}^0\, g_{00}(P)\, dt = d\t = \frac{d\t}{d\s}\, d\s \ .
\end{equation*}
So, as required by the relativity principle, we indeed recover the same time dilation factor as above. }

To illustrate this further, consider a frequency measurement at a given space position 
$r= r_O$ in the Schwarzschild metric. An observer (O) 
stationary at $r=r_O$ will therefore measure the time $d\hat t_O$ corresponding to 
the coordinate time interval (inverse frequency) $dt$ as
\begin{align}
d\hat t_O &= - u_O^0 \, g_{00}(r_O) \,dt \,=\, d\t \nonumber \\
&= \sqrt{-g_{00}}\, dt  \simeq \left(1- \frac{GM}{r_O}\right)\, dt \ ,
\label{b36}
\end{align}
In the last equality we have quoted the result only to first order in the local gravitational potential 
$U_O = - GM/r_O$, as we do from now on.  So as noted above, the observer fixed with respect to the 
frequency source measures a time interval $d\hat t_O = d\t$, the proper time.

This fixed observer is not, however, freely-falling and we may also want to consider measurements made 
in the LIFs corresponding to such observers, whose worldlines satisfy the geodesic equation.
Consider therefore an observer (A) freely-falling radially along a trajectory with boundary condition 
$dr/dt = 0$ at $r=\infty$. Solving the geodesic equation, we find this observer has normalised 4-velocity
$u_A^\m = \Bigl(\left(1-2GM/r_O\right)^{-1}, $ $-\left(2GM/r_O\right)^{1/2},0,0\Bigr)$ at $r=r_O$.
So observer A measures the time interval considered above as
\begin{equation}
d\hat t_A = - u_A^0\, g_{00}(r_O) \, dt = dt \ .
\label{b38}
\end{equation}

Alternatively, consider the freely-falling observer (B) in a circular orbit with constant radius $r_O$.
This observer has 4-velocity
$u_B^\m = \left(dt/d\s\right)\, \left(1,0,0,\omega\right)$, where the angular frequency 
$\omega = d\phi/dt$ and the orbital velocity is $v = \omega r = \left(GM/r\right)^{1/2}$. 
It follows from (\ref{b37}) that $dt/d\s = \left(1 - 3GM/r_O\right)^{-1/2}$.
So observer B measures the time interval as 
\begin{equation}
d\hat t_B = - u_B^0 \,g_{00}(r_O)\, dt \simeq \left(1 - \frac{1}{2}\frac{GM}{r_O}\right)\, dt \ .
\label{b39}
\end{equation}

Now, since the coordinate time $dt$ is not a measured quantity, what is important here is 
only the {\it ratios} of measurements amongst the different observers. Every time measurement
is only a ratio with respect to another clock, which is also affected by the spacetime curvature.
So here, we find to leading order in $GM/r_O$,
\begin{equation}
d\hat t_A = \left(1 + \frac{GM}{r_O}\right)\, d\hat t_O \ , ~~~~~~
d\hat t_B = \left(1 + \frac{1}{2}\frac{GM}{r_O}\right)\, d\hat t_O \ . 
\label{b40}
\end{equation}

In both cases, the freely-falling observers, who are moving relative to the fixed location of
the frequency source, measure a {\it greater} time interval than the fixed observer. Comparing 
with (\ref{b37a}), we can check directly that these results may be interpreted as the local special 
relativistic time dilation factor $\gamma(v^2)$, where $v^2 = g_{ij}\frac{dx^i}{d\t} \frac{dx^j}{d\t}$
is the appropriate squared physical velocity measured with the fixed observer's time.  
It is perhaps worth emphasising that the different freely-falling observers do measure 
different times. 

Note also that since the ratio of these time measurements $d{\hat t}_O$, $d{\hat t}_A$ and 
$d{\hat t}_B$ is simply determined by the ratio of the relevant 4-velocity components
$u_O^0$, $u_A^0$ and $u_B^0$, with the metric factor $g_{00}(r_O)$ cancelling, 
these results illustrate an important feature of GR, {\it viz.}~that
purely {\it local} measurements are not dependent on the absolute
value of the gravitational potential.\footnote{The local flatness of Riemannian spacetime
means that at each point we can construct Riemann normal coordinates in which the metric
has the form
\begin{equation*}
g_{\m\n}(x) = \eta_{\m\n} + \frac{1}{3} R_{\m\r\n\s}x^\r x^\s + O(x^3) \ ,
\end{equation*}
since $\Gamma^\l_{\m\n} \sim 0$ in these coordinates. The curvature tensor involves the
second derivatives of the metric (and hence the potential $U(r) = -GM/r$). It follows 
that physical measurements in a local laboratory of size $\ell$ only depend on the curvature
at order $O(|U| \ell^2/L^2)$ where $L$ is the curvature length scale, which is hugely suppressed
relative to the potential itself  (see {\it e.g.}~\cite{Visser:2018omi} for a recent comment).}

These examples involve observers at the same point in the gravitational potential.
Considering observers at different heights gives rise to the {\it gravitational redshift} effect.
To derive this, consider a light wave emitted from a source (E) at $r=r_E$ radially upwards to
the observer (O) at $r_O = r_E + h$. It is straightforward to see from the null geodesic equation 
that the coordinate time $dt$ between successive wave maxima is the same at the receiver (O)
as at the emitter (E).\footnote{For a photon following a null geodesic, its trajectory is characterised by
$g_{00} dt^2 + g_{rr}dr^2 = 0$, so the coordinate time to travel from E to O is simply
\begin{equation*}
t_O - t_E = \int_{r_E}^{r_O} dr\,\left(1 - \frac{2GM}{r}\right)^{-1} \ ,
\end{equation*}
where, since the metric is static, the integral on the r.h.s.~is a function of the space coordinates only.
The time of flight for successive maxima is therefore the same, so the coordinate wave period $dt$
is the same at the receiver and emitter.}
Using the results above, we see that the period of the wave measured by an observer E fixed at the 
location of the emitter is therefore $d\hat t_E = \left(1-GM/r_E\right)\, dt$, while measured at the
receiver O it is $d\hat t_O = \left(1-GM/r_O\right)\, dt$.
The ratio of observed frequencies is therefore
\begin{align}
\frac{\n_O}{\n_E} = \frac{d\hat t_E}{d\hat t_O} 
&= \left(1-\frac{GM}{r_E}\right)\,\left(1 - \frac{GM}{r_O}\right)^{-1} \nonumber \\ &
\simeq 1 - \frac{GMh}{r_E^2} \ ,
\label{b41}
\end{align}
for $h \ll r_E$. The observer at height $h$ above the emitter therefore measures the light {\it redshifted}
relative to the measurement at the emitter, by a factor $\D \n/\n \simeq GMh/r_E^2$.
Notice that for these fixed observers at different spacetime points, the measured effect 
depends only on the {\it difference} of the gravitational potentials, not their absolute values, 
and is therefore suppressed by an additional factor $h/r_E$.
This was first observed experimentally in the famous Pound-Rebka experiment \cite{Pound}
using gamma rays and exploiting the M\"ossbauer effect. 

Another interesting gravitational redshift test may be made using atom matter-wave interferometry
\cite{Muller:2010zzb, Hohensee:2011wt}. To see the principle involved, consider two identical
frequency sources (ideal clocks) P and Q at positions $r_P$ and $r_Q$ with velocities $\textsl{v}_P$
and $\textsl{v}_Q$ respectively. A simple calculation combining (\ref{b41}) with (\ref{b35}) and
(\ref{b37a}) shows that the difference in frequencies associated with P and Q as measured in the
laboratory frame is given by
\begin{equation}
\frac{\D \n_{P-Q}}\n \,\equiv\, \frac{\n_P - \n_Q}{\n} \,=\, U(r_P) - U(r_Q) - 
\frac{1}{2}\left(\textsl{v}_P^2 - \textsl{v}_Q^2\right) \ ,
\label{b41a}
\end{equation}
to leading order in the weak field, low-velocity approximation. At this order we can simply take
the denominator factor $\n$ to be the common flat-spacetime limit of the source frequencies.

Now, if P and Q start from a common position at $t=0$ and follow different trajectories before being brought
back to a common point at time $T$, they will acquire a phase difference given by 
\begin{align}
\D \phi \,&=\, \int dt \,\D \w_{P-Q}  \nonumber \\
&=\, \w \int_0^T dt\,\left[g\left(r_P - r_Q\right) 
- \frac{1}{2} \left(\textsl{v}_P^2 - \textsl{v}_Q^2 \right) \right] \ ,
\label{b41b}
\end{align}
where $\w$ is the angular frequency and $g = GM/R^2$ is the Earth's gravitational acceleration.

Notice that this result can be derived\footnote{Explicitly,
\begin{align*}
\int d\s_P &= \int dt\,\left[- \left(g_{00}(P) + g_{ij}(P) \textsl{v}_P^i \textsl{v}_P^j\right)\right]^{1/2} \\
&= \int dt\, \bigl(-g_{00}(P)\bigr)^{1/2} \, \left(1 + g_{ij}(P) \textsl{v}_P^i \textsl{v}_P^j / g_{00}(P)
\right)^{1/2} \\ 
&\simeq \int dt \left(1 - \frac{GM}{r_P} - \frac{1}{2} \textsl{v}_P^2 \right) \ ,
\end{align*}   
to first order in the small quantities $GM/r_P$ and $\textsl{v}_P^2$. The result (\ref{b41b}) follows 
immediately as
\begin{align*}
\D \phi &= \w \int \left(d\s_P - d\s_Q\right) \\
&= \w \int dt \left[ U(r_P) - U(r_Q)  - \frac{1}{2} \left( \textsl{v}_P^2 - \textsl{v}_Q^2\right) \right] \ .
\end{align*}     }
very elegantly as
\begin{equation}
\D \phi \,=\, \w \int \left(d\s_P - d\s_Q\right) \ .
\label{b41c}
\end{equation}
That is, the phase difference measures the difference in the proper times along the distinct trajectories
of P and Q. 

In the atom interferometry experiment \cite{Muller:2010zzb, Hohensee:2011wt}, a laser cooled atom 
projected vertically is subjected to three pulses from a pair of crossed laser beams. The first puts the atom 
into a superposition of quantum states with different momenta, so they follow different trajectories.
The second pulse reverses their spatial separation and brings them back to a common point at which the 
third laser pulse records the phase difference of the superposed states. The matter waves oscillate
with the Compton frequency $\w_C = mc^2/\hbar$.

The experiment can be arranged so that the gravitational redshift contribution to the phase difference 
(\ref{b41b}) can be isolated, leaving
\begin{equation}
\D \phi_{redshift} \,=\, \w_C \int_0^T dt\,g \, \D r(t) \ ,
\label{b41d}
\end{equation}
where $\D r(t)$ is the vertical separation of the trajectories and we continue to set $c=1$.

This experiment allows a very high precision test of WEPc in the laboratory. Further details and prospects 
for repeating it with antihydrogen in order to test WEPc directly in a neutral, pure antimatter system
will be discussed in section \ref{sect 3.3}.

This concludes our brief summary of the fundamental principles of GR, together with
some of the basic theory of time meaurements which we will need later in describing the effect 
of curved spacetime on atomic spectroscopy measurements in the Earth's gravitational field 
and in orbits around the Earth or Sun. In the next section, we discuss some implications of
any possible violation of the predictions of GR, in particular any anomalous gravitational 
measurements which would distinguish between matter and antimatter.

\subsection{Breaking General Relativity and the Equivalence Principles}\label{sect 2.5}

As we have seen, GR provides an elegant and compelling view of gravity whose predictions,
however radical and counter-intuitive, have passed all experimental tests for more than a century. 
Nevertheless, as a purely classical theory, we know that eventually GR must be completed into 
a full quantum theory of gravity, and we may expect that even at the level of a low-energy effective
Lagrangian small deviations from the simplest formulation of GR could arise. 
While such modifications of the dynamics of GR are relatively easy to implement, as we shall see
what is not so straightforward is to envisage changes to the theory that distinguish matter and
antimatter in a way that could be observable in the current generation
of antimatter experiments. An interesting discussion of the fundamental issues involved is given
in \cite{Blas:2018qdl}.    

A conservative approach is to start by retaining the core principles of GR -- the description of gravity 
as a Riemannian manifold and local Lorentz invariance -- but modifying the dynamics described 
by the action (\ref{b29}).

The action (\ref{b29}) was determined using the criterion that the matter couplings to gravity are through
the connection only, not the curvature. This means that in a LIF, the equations of motion take their 
special relativistic form, thereby implementing the Strong Equivalence Principle.  
The simplest extension of GR is therefore to violate the SEP by including explicit curvature terms
in the action. These transform covariantly so we maintain local Lorentz invariance, but the 
{\it dynamics} is now modified to be sensitive to the curvature at each spacetime point.

For example, for the free Dirac theory we can generalise the action to:
\begin{align}
S &= \int d^4 x \sqrt{-g} \biggl(\frac{R}{16\pi G} + \bar\psi\left(i
    \c^\m D_\m - m\right) \psi \nonumber \\
&~~~~~~+e\,R\,\bar\psi \psi + f\,\bar\psi D^2 \psi  + a\,\partial_\m R\, \bar\psi \c^\m \psi
+ b\,R\,\bar\psi i\c^\m \overleftrightarrow{D}_\m \psi \nonumber \\
&~~~~~~+ c\,R_{\m\n}\,\bar\psi i \c^\m \overleftrightarrow{D}^\n \psi 
+ d\,D_\n\bar\psi i  \c^\m \overleftrightarrow{D_\m}  D^\n\psi  + \ldots \biggr) 
\ ,
\label{b42}
\end{align}
with the obvious notation $\c^\m = \c^a e_a{}^\m$.
Here, in the spirit of regarding (\ref{b42}) as a low-energy effective Lagrangian for some more 
fundamental UV complete theory, we have included all the operators of dimension 4, 5 and 6 which are
quadratic in the spinor field and, for ease of illustration only, conserve parity. 
The coefficients $a, \ldots,f$ carry the appropriate inverse dimensions of mass, potentially set by the scale of
the UV completion.

A notable feature of this SEP-violating effective Lagrangian for fermions is that it does not involve the
Riemann curvature tensor $R_{\m\r\n\s}$ itself, only the contracted forms --  the Ricci tensor $R_{\m\n}$ and
scalar $R$. A general discussion of such SEP-violating actions is given in \cite{Shore:2004sh,McDonald:2014yfg}.

This is not true of the extension to QED, since there are explicit SEP-violating couplings of the photon field strength 
to the Riemann curvature:
\begin{align}
S &= \int d^4 x \sqrt{-g} \biggl(-\tfrac{1}{4}F_{\m\n}F^{\m\n} + \tilde a R F_{\m\n} F^{\m\n} 
+ \tilde b R_{\m\n} F^{\m\r} F^\n{}_\r  \nonumber \\
&~~~~~~+ \tilde c\, R_{\m\r\n\s} F^{\m\n} F^{\r\s} 
+ \tilde d\, D_\m F^{\m\r} D_\n F^{\n}{}_\r + \ldots  \biggr)\ .
\label{b43}
\end{align}

In the same spirit, we can also modify the purely gravitational dynamics by adding terms of quadratic or
higher order in the curvature, {\it e.g.}
\begin{equation}
S = \int d^4 x \sqrt{-g} \biggl(\hat a\, R^2 + \hat b\, R_{\m\n} R^{\m\n} + \hat {c}\, R_{\m\r\n\s} R^{\m\r\n\s} 
+ \ldots \biggr) \ .
\label{b44}
\end{equation}
There is a large literature on modified (higher-derivative) gravity theories of this type, 
motivated in part by superstring theory.

The extra terms in (\ref{b42}) and (\ref{b43}) modify the energy-momentum tensor and the 
equations of motion for the fields, which no longer necessarily follow geodesics. 
These couplings may be different for different particles, and may in principle even distinguish matter 
and antimatter. So in general, the universality of free-fall (WEPff) is lost in this GR extension.

On the other hand, it is not so clear how this modified dynamics would affect the gravitational time
dilation results described in the last section. These depend purely on the nature of the metric
and spacetime, and since all matter is still coupled to the same metric the universality of clocks
(WEPc) seems at first sight to be maintained. However, we now have to consider the possibility of 
these new gravitational interactions dynamically modifying the spectrum of anti-atoms, for example, 
inducing physical differences between time/frequency measurements with different atomic/anti-atomic
clocks.  

The SEP-violating QED action is in some respects analogous to the Lorentz-violating action considered in 
section \ref{sect 2.3} and the phenomenological consequences may in some cases be similar.
There are, however, many important differences, notably that the r\^ole of the Lorentz-violating couplings
is played here by the covariant curvature tensors. This particularly affects the $\mathsf{C}$, $\mathsf{P}$
and $\mathsf{T}$ character of the operators.  These can be read off from Table \ref{TableCPT}, 
remembering that the spinor quantities are all referred to the local Minkowski frame using the 
implicit vierbein.  A careful analysis is given in \cite{McDonald:2014yfg}. Importantly, all these
operators are $\mathsf{CPT}$ invariant.

A particularly interesting example is the term
\begin{equation}
{\cal L}_a = a\,\partial_\m R\, \bar\psi \c^\m \psi \ .
\label{b45}
\end{equation}
This is $\mathsf{C}$ and $\mathsf{T}$ odd, $\mathsf{P}$ even, and therefore $\mathsf{CP}$ odd, 
$\mathsf{CPT}$ even.
The fact that, uniquely amongst those in (\ref{b42}), this operator is $\mathsf{CP}$ odd, allows it to modify
the propagation (dispersion relations) for fermions and antifermions {\it differently}.\footnote{The divergence of
the Ricci scalar $\partial_\m R$ acts here as a background vector field coupling to the current $J^\m = \bar\psi \c^\m \psi$,
so the coupling depends on the corresponding charge and is equal and opposite for particles and antiparticles.
We return to this idea in section \ref{sect 2.6} when we consider extra background fields, such as a ``gravivector'',
which could have gravitational strength interactions distinguishing matter and antimatter.} We can therefore 
conclude that in this SEP-violating GR extension, there is an induced matter-antimatter asymmetry in a 
background gravitational field with a time-varying Ricci scalar \cite{McDonald:2014yfg, McDonald:2015iwt}.

This is a radical consequence of what is a natural and relatively mild modification of GR, in which the
core geometrical structure is maintained while the dynamics is extended to include direct curvature
couplings. That said, the theoretical consistency of the theory based on (\ref{b42}), (\ref{b43}) needs 
to be analysed critically. Just as in the Lorentz-violating QED extension, if the action is viewed as 
a fundamental theory in itself it is at risk of violating causality -- indeed (\ref{b43}) implies 
birefringent superluminal propagation of photons. Once more, we are therefore led to the point of view
of regarding (\ref{b42}), (\ref{b43}) as a low-energy effective Lagrangian whose intrinsic inconsistencies may be 
modified by a suitable UV completion \cite{Shore:2003zc, Shore:2007um}.

Now, it is a remarkable fact that even if we start from the usual QED action (\ref{b29}) in curved spacetime, 
loop corrections to the electron and photon propagators generate precisely the operators in the effective
action (\ref{b42}), (\ref{b43}) and (\ref{b44}), with the exception of the $\mathsf{CP}$-violating 
operator.\footnote{The
scale of these quantum effects is $O(\a \l_c^2/L^2)$, where $\a$ is the fine structure constant, 
$\l_c = 1/m$ is the Compton wavelength of the virtual particles arising in the self-energy or vacuum
polarisation Feynman diagrams, and $L$ is the curvature scale. Of course, these quantum effects only
become significant in an epoch of extremely high curvatures when the quantum scale becomes 
comparable with the curvature scale. In the BSM model of 
\cite{McDonald:2015ooa, McDonald:2015iwt, McDonald:2016ehm},
this is set by the heavy neutrino mass. Remarkably, in the high curvature and temperature regime
of the early universe, this mechanism may generate a sufficient difference in the dispersion relations
of the light neutrinos and antineutrinos to give rise to the current observed baryon-to-photon ratio 
of $O(10^{-10})$.} This means that conventional QFT in curved spacetime
{\it automatically} violates the SEP at low energies due to radiative quantum corrections,
while maintaining causality and unitarity 
\cite{Hollowood:2007ku, Hollowood:2008kq, Hollowood:2009qz, Hollowood:2011yh}.
Furthermore, in a BSM theory with $\mathsf{CP}$-violating couplings, even the operator (\ref{b45}) 
is generated. This provides the theoretical basis for a recently proposed mechanism -- radiatively-induced
gravitational leptogenesis -- for generating the observed matter-antimatter asymmetry of the universe
\cite{McDonald:2015ooa, McDonald:2015iwt, McDonald:2016ehm}.

We therefore see that in principle it is perfectly possible for nature to be described by a SEP-violating
low-energy effective Lagrangian, exhibiting WEPff violation and perhaps also independently WEPc violation,
while maintaining the fundamental principles of local QFT and GR. 

Unfortunately, there is an obvious problem as far as terrestrial antimatter experiments are concerned,
{\it viz.}~that the Schwarzschild spacetime describing the gravitational field on the Earth's surface 
is Ricci flat ($R_{\m\n} = 0$, $R=0$), although the Riemann tensor itself is non-vanishing. 
The modified Dirac action (\ref{b42}) is therefore insensitive to the Earth's gravitational field,
although the photon action (\ref{b43}) and light propagation is affected. The only way out would be 
for an extended pure gravity action to generate a modified solution with non-vanishing Ricci tensor
which could couple to fermions. 

Nevertheless, this is only one approach to extending GR and violating the equivalence principles.
More radically, we could consider adding new fields and interactions, discussed briefly in the following 
section, or even modify the metric structure of spacetime and consider alternatives to the simple 
Riemannian picture.

In the spirit of the latter approach, we could consider bimetric, or multi-metric, theories where 
different fields couple to different metrics imposed on the background Riemannian manifold. 
An interesting gedanken experiment is then to reconsider the gravitational redshift scenario of
the previous section, but where the emitter and receiver couple to different metrics.
A simple case to illustrate the idea would be if the receiver, but not the emitter, was described by
an effective Lagrangian of the form (\ref{b42}) with the coefficient $c \neq 0$. This would correspond 
to an {\it effective metric}:
\begin{equation}
g_{\m\n}^{\rm eff} = g_{\m\n} + 2 c R_{\m\n} \ .
\label{b46}
\end{equation}
This would give an additional curvature-dependent contribution to the
redshift formula (\ref{b41}), violating WEPc.
Unfortunately, this particular operator is $\mathsf{CP}$ conserving and we do not have a model which would 
differentiate matter from antimatter in this simple way. 

While such curvature-dependent modifications to an effective metric are relatively easy to contemplate
within the worldview of GR, this is not true of the totally phenomenological approach used especially in some
of the early literature on equivalence principle violation in antimatter systems (see {\it e.g.}~\cite{Hughes:1990ay}).
These imagine an effective metric with $g_{00}$ component of the form
\begin{equation}
(g_{00})_{\rm eff} =  1 - \a_g \frac{2GM}{r} \ ,
\label{b47}
\end{equation}
where $\a_g$ is a parameter to be constrained by experiment, independently for matter and antimatter.
Clearly this directly enters the redshift formula.
It is not clear, however, that this is a useful parametrisation. In the first place, it is very hard to imagine 
a causal extension of GR in which $\a_g -1$ is non-zero and different for particles and antiparticles.
Moreover, this would imply that local experiments would be sensitive 
not just to the local curvature but to the absolute value of the gravitational potential $U(r)=-GM/r$ itself. 
Yet this becomes greater for ever more distant astronomical structures, ranging from the Sun to the
Galaxy or even the Virgo Cluster. Numerical values are discussed later in \ref{sect 3.3.2}, but for now
we simply note that
the loss of universality implied by different non-vanishing $\a_g$ parameters for different fields
would therefore not only violate the equivalence principle but would involve a rather bizarre non-locality 
of matter-gravity interactions observed on Earth. This parametrisation is, however, still quite widely used to 
quantify measurements in experiments testing the equivalence principle.  We return to this issue in 
section \ref{sect 3.3}.

\vskip1cm

\subsection{`Fifth' Forces, $B-L$ and Supergravity}\label{sect 2.6}

This brings us to the possibility that any anomalous results that may be seen in future antimatter experiments 
are due not to a modification of the fundamental principles of our existing theories but due to new interactions,
so-called `fifth forces', beyond the standard model (BSM).

Since our focus is on phenomena which would distinguish matter and antimatter, we are led to consider new
interactions that are mediated by the exchange of a spin 1, or vector, boson. Whereas spin 2 bosons (such as the 
graviton, corresponding to a tensor field) and spin 0 bosons (scalars, such as the Higgs) couple equally to
particles and antiparticles, spin 1 bosons couple to a vector current with a strength proportional to the 
corresponding charge. Just as in QED with photon exchange, this charge is opposite for particles and
their antiparticles.

It is then immediately apparent that while they would distinguish matter and antimatter, such new vector
interactions would necessarily affect the dynamics of purely matter systems. They are therefore already
highly constrained by comparisons of ordinary matter interactions with conventional theories, irrespective
of future tests with antimatter.

Here, we consider especially two theoretically well-motivated classes of theory involving new vector interactions.
The first is a BSM theory with gauged $B-L$ symmetry, which contains an 
additional $Z'$ boson compared to the standard model, as well as right-handed neutrinos.
The second is ${\cal N}\ge 2$ supergravity, in which the graviton lies in a supersymmetry multiplet with a 
spin 1 `{\it gravivector}' boson.

\subsubsection{Gauged $B-L$ theories}

A particularly compelling class of BSM theories which could produce long-range forces distinguishing 
matter and antimatter are those where the $B-L$ symmetry of the standard model is gauged.
The full gauge group is extended to $SU(3)_C \times SU(2)_L \times U(1)_Y \times U(1)_{B-L}$, with the 
introduction of a new abelian gauge field and corresponding spin 1 boson $Z'$. We denote the new gauge
coupling by $g'$ and the corresponding `fine structure constant' $\alpha' = g^{\prime 2} / 4\pi$.

To understand the theoretical motivation, we need to briefly review some key ideas involving anomalies
in QFT. An anomaly arises when a symmetry which is exact in the classical theory is broken in the full
quantum theory. There are many ways to understand the origin of anomalies, which fundamentally involve 
the deep mathematical structure of gauge theories, but in the simplest perturbative picture they arise from
the behaviour of 3-point Feynman diagrams where the vertices are currents 
$J_\m^a = \bar\psi \gamma_\m T^a \psi$, for some symmetry generators $T^a$, and left or right-handed
fermions $\psi_{L,R}$ go round the triangle; see Figure \ref{Fig 3}.
\begin{figure}[h!]
\centering
\includegraphics[scale=0.6]{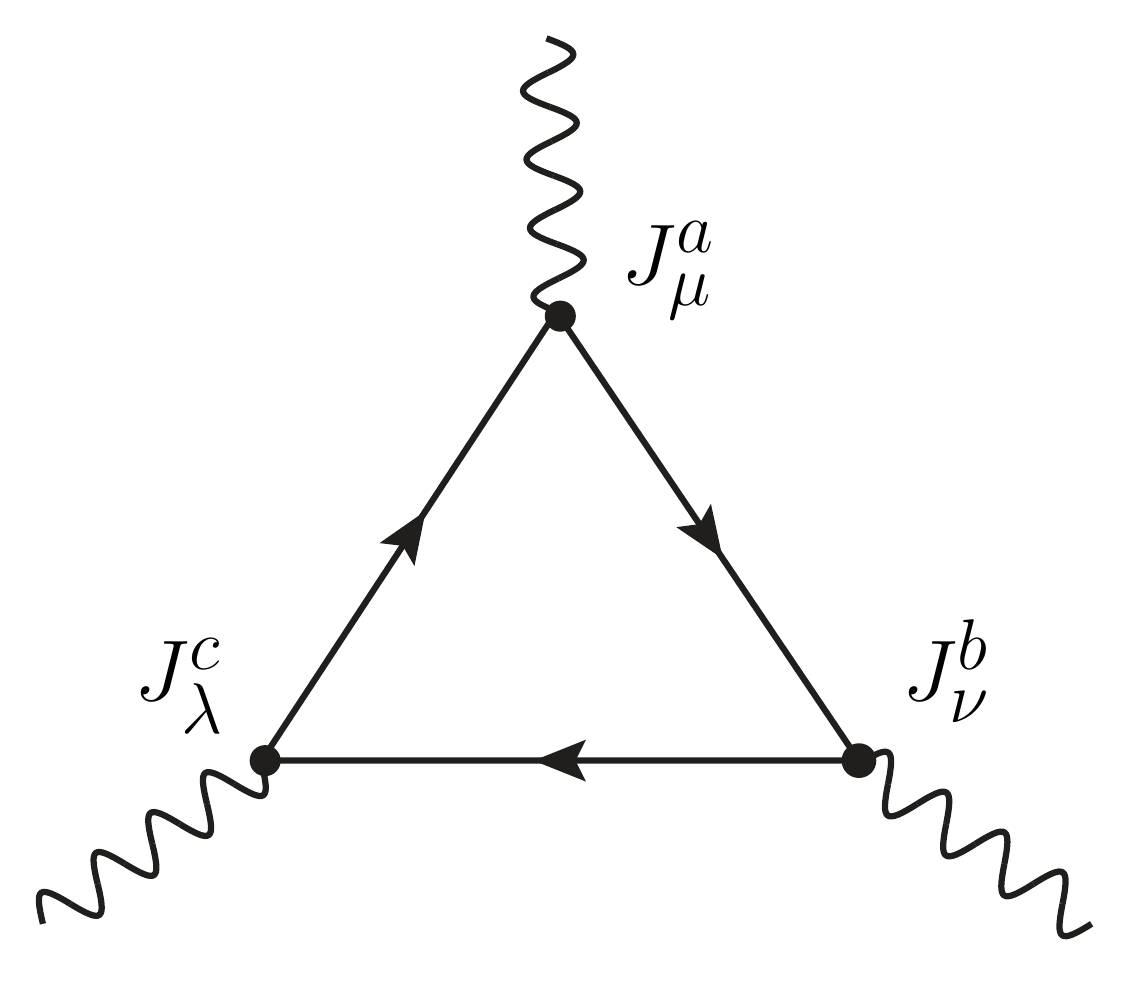}
\caption{Feynman diagram for a 3-current Green function $\langle 0|J_\m^a ~J_\n^b ~ J_\l^c |0\rangle$
with $L$ and $R$-handed fermions in the loop. Gauge fields $A_\m^a, A_\n^b$ and  $A_\l^c$ couple to the 
currents if the corresponding symmetries are gauged. This diagram may contribute an anomaly which breaks 
conservation of the currents.}
\label{Fig 3}
\end{figure}
An otherwise conserved current $J_\m^a$ will be anomalous if the coefficient ${\cal A}$ in the 
triangle diagram is non-zero, where 
\begin{equation}
{\cal A} = \sum_{L\,{\rm reps}} T^a \{ T^b,T^c\} - \sum_{R\,{\rm reps}} T^a \{ T^b,T^c\}  \ ,
\label{b48}
\end{equation}
and the sums are over the representations of the $L$ and $R$-handed fermions in the corresponding
symmetry groups.\footnote{This notation is over-simplified for clarity -- the generators 
$T^a$, $T^b$, $T^c$ in (\ref{b48}) may refer to different symmetry groups.}

If these anomalies affect only {\it global} currents, such as the axial current in QED where the anomaly
determines the physical decay $\pi^0 \rightarrow \gamma \gamma$, the quantum theory remains consistent.
However, if the anomalies affect {\it local} currents, {\it i.e.}~if a gauge field is coupled to an anomalous current,
then {\it unitarity} will be broken in the full QFT. Any realistic particle physics theory must therefore only
involve gauge fields which couple to anomaly-free conserved currents. Clearly, this places important 
constraints on the fermion content of the theory.

In the standard model, both baryon number $B$ and lepton number $L$ are conserved at the classical level
but are anomalous in the full quantum theory. However, the anomaly ${\cal A}$ cancels when one vertex is the 
combined $B-L$ current while the others refer to the gauge symmetries of the standard model.
In fact, $B-L$ is the only anomaly-free conserved global symmetry in the standard model.

This provides the motivation to extend the standard model gauge group to include $U(1)_{B-L}$. 
However, unitarity now demands that all triangle diagrams with one or more $U(1)_{B-L}$ vertices must have a
vanishing anomaly coefficient. In particular, for the $U(1)_{B-L}^3$ triangle, the anomaly is
\begin{align}
{\cal A} &= \sum_{L\,{\rm reps}}   Q_{B-L}^3    -   \sum_{R\,{\rm reps}} Q_{B-L}^3  \cr
&= \left( 3.\,2\left(\frac{1}{3}\right) + 2\left(-1\right) \right) 
- \left(3\left(\frac{1}{3}\right) + 3 \left(\frac{1}{3}\right) + \left(-1\right) \right) \cr
&= -1 \ ,
\label{b49}
\end{align}
where we have shown the $B-L$ charges and $SU(3)_C$,  $SU(2)_L$ multiplets for one generation of quarks 
and leptons. To obtain an anomaly-free theory with ${\cal A}=0$ we must therefore add a new R-handed 
neutrino for each flavour generation, so the quark and lepton representations are precisely balanced.

To summarise, the fundamental principle of {\it unitarity} requires that the standard model extension with
$U(1)_{B-L}$ gauged requires three right-handed neutrinos as well as the new gauge boson $Z'$. 
Of course, this is a very welcome bonus as it permits neutrinos to acquire non-vanishing masses as
required by experiment (in contrast to the standard model itself where only L-handed neutrinos and
R-handed antineutrinos exist and are therefore required to be massless).

We now focus on the gauge sector and the new $Z'$ boson. There are two distinct realisations of the theory,
one with unbroken $U(1)_{B-L}$ and the other where $U(1)_{B-L}$ is spontaneously broken through the 
interaction with a new Higgs field.

\vskip0.4cm
\noindent (i) {\it Unbroken $U(1)_{B-L}$}:
\vskip0.2cm
In this model \cite{Heeck:2014zfa}, 
the L and R-handed neutrinos combine into three flavours of massive Dirac neutrinos with masses
determined, as for the other leptons, by arbitrary Yukawa couplings to the $SU(2)_L$ doublet Higgs field $\phi$, 
that is
\begin{equation}
{\cal L}_D = y_D^{ij} \left( \overline{\ell_L^i}\,\tilde{\phi}\, \nu_R^j \,+\, {\rm h.c.} \, \right) \ ,
\label{b50}
\end{equation}
in standard notation, for generations $i = 1,2,3$.
Since the Higgs has zero $B-L$ charge, its VEV does not break the $U(1)_{B-L}$ gauge symmetry
and at first sight it appears the $Z'$ must remain massless.

However, for abelian gauge theories (only), there is an alternative to the Higgs mechanism which can give a 
mass to the $Z'$. This is the Stueckelberg mechanism. It involves the introduction of a scalar field analogous 
to a conventional Higgs field but where its modulus (corresponding to the physical Higgs boson) is
non-dynamical while its phase (corresponding to the Goldstone boson) provides the extra degree of freedom
required to give a massive spin 1 boson. Essentially, it is a gauged $U(1)$ non-linear sigma model,
in contrast to the Higgs theory which is a gauged linear sigma model.

The details, and the proof of renormalisability and unitarity -- which work only for a $U(1)$ gauge theory -- need
not concern us here (see ref.~\cite{Ruegg:2003ps}). The essential point is simply that there exists a 
theoretically consistent model in which the $U(1)_{B-L}$ gauge symmetry remains unbroken while the 
$Z'$ boson acquires a mass, which is unconstrained by theory. In this model, the light neutrinos
are of Dirac type.

\vskip0.4cm
\noindent (ii) {\it Spontaneously broken $U(1)_{B-L}$}:
\vskip0.2cm
In this case (see for example \cite{Basso:2008iv}), 
we introduce a new complex $SU(2)_L$ singlet Higgs field $\Phi$ with $B-L$ charge $Q_{B-L}= -2$.
If this acquires a VEV, the $Z'$ boson acquires a mass in the usual way and there remain two physical Higgs
bosons in the spectrum. The $\Phi$ field can couple to the R-handed neutrino fields, giving them a
Majorana mass proportional to its VEV in addition to the Dirac mass (\ref{b50}).
The interaction term in the Lagrangian is 
\begin{equation}
{\cal L}_M = y_M^{ij} \left( \overline{\left(\nu_R^i\right)^c}  \,  \nu_R^j\, \Phi \,+\, {\rm h.c.}\, \right) \ .
\label{b51}
\end{equation}
This interaction allows explicitly $B-L$ violating processes involving the R-handed neutrinos,
such as neutrinoless double beta decay.
The spectrum comprises two massive Majorana neutrinos for each fermion generation.

\vskip0.2cm
What this discussion illustrates is that it is highly non-trivial to add a new `fifth force' interaction to the 
standard model, even one as seemingly innocuous as a coupling to baryon or lepton number. 
There are fundamental principles, in this case unitarity (via anomaly freedom), which severely
constrain the particle content and interactions on purely theoretical grounds.

The phenomenology of these two variants of gauged $U(1)_{B-L}$ theory therefore differs in key aspects
but both have been extensively studied, especially in the parameter range of most interest to particle 
physicists exploring neutrino phenomenology and the implications of a potential new $Z'$ boson in the 
mass range of interest at the LHC \cite{Basso:2008iv}. 

Here, we are more interested in two mass windows for the $Z'$ which could impact on low-energy
antimatter experiments. The first is $m_{Z'} \sim 1$ keV, which implies a short-range interaction with range 
$\l = 1/m_{Z'} \sim 10^{-10} \,{\rm m}$, sufficient to influence the internal structure of atomic systems.

The second is $m_{Z'} \lesssim 10^{-14}$ eV, corresponding to a new long-range interaction with range
$\l \gtrsim 10^7$ m. This (including the massless limit $m_{Z'} = 0$) would imply a macroscopic interaction
between the Earth, which has a huge $B-L$ charge, and a single atom or molecule with non-vanishing 
$B-L$ at the surface. This force is in principle measurable in free-fall experiments of the type which may
be realised for suitable antimatter systems at the CERN AD.

Existing constraints on the strength of a new $U(1)_{B-L}$ interaction over the entire range of $Z'$ mass from 
zero to GeV/TeV values are summarised in \cite{Heeck:2014zfa}. For $m{_Z'} \sim 1$ keV, limits arise 
from low-energy electron neutrino scattering, particularly elastic scattering of solar neutrinos, 
which (for the unbroken theory) gives $\alpha' \lesssim 10^{-13}$,
and from stellar astrophysics, which gives a more stringent bound 
$\alpha' \lesssim 10^{-31}$ from an analysis of energy-loss mechanisms involving the $Z'$ and new neutrinos.
See Figure 3 of \cite{Heeck:2014zfa} for a clear compilation of the experimental bounds.

For a very light $Z'$, existing equivalence principle experiments on ordinary matter, especially 
highly precise torsion balance experiments, constrain the  $U(1)_{B-L}$ gauge coupling to the range 
$\alpha' \lesssim 10^{-49}$. The extremely small value arises because of the
large $B-L$ charge of the Earth, equal to the number of neutrons, which is of order 
$Q_{B-L}^{\rm Earth} \sim 10^{51}$. Such long-range forces are generally parameterised in the form of a modified
gravitational potential derived from single-boson Born exchange,
\begin{align}
V(r) ~&=~ - \frac{G_\infty m_1 m_2}{r} + \frac{\alpha' Q^1_{B-L} Q^2_{B-L}}{r} e^{- m_{Z'}r} \cr
&=~ - \frac{G_\infty m_1 m_2}{r} \, \left( 1 - \tilde{\alpha} \, e^{-r/\l} \right) \ ,
\label{b52}
\end{align}
where $\tilde\alpha = \alpha' Q^1_{B-L}  Q^2_{B-L} /G_{\infty} m_1 m_2$.
Note that we have introduced the `fundamental' Newton constant $G_\infty$ here, since an infinite range 
fifth force would change the effective constant $G_N$ measured in experiment. The same ambiguity
would affect the definition of the Planck mass from $G = 1/ M_{pl}^2$.
Also note crucially that the $Z'$ exchange force is {\it attractive} when the $B-L$ charges of the 
interacting bodies are {\it opposite}, and repulsive when they are the same, in exact analogy
with the electric force between charged particles.

\subsubsection{${\cal N}\ge 2$ supergravity}

Another class of theories which naturally involve extra vector boson interactions are supergravities
with ${\cal N}\ge 2$. Supersymmetry is an extension of the Poincar\'e symmetry of flat spacetime 
to include new generators ${\cal Q}_\a^a$, where $\a$ is a spinor index and $a= 1,\ldots {\cal N}$
counts the number of supersymmetries, which satisfy {\it anticommutation} relations. 
This extended spacetime symmetry is known as a graded Lie algebra. If supersymmetry is promoted 
to be a {\it local}, rather than global, symmetry, the resulting quantum field theory is supergravity.

The impact of supersymmetry on the spectrum of a QFT is to relate fields, or particles, with spins differing by 
$1/2$. So the simplest ${\cal N}=1$ supergravity has one spin 2 graviton and one spin $3/2$ Majorana gravitino.
These supermultiplets become progressively longer as the number of supersymmetries is increased.
${\cal N}=2$ supergravity has one graviton, two gravitinos, and one spin 1 {\it gravivector} (all massless,
and balancing $2 + 2 = 4$ bosonic and $2\times 2 = 4$ fermionic degrees of freedom).
Ultimately we arrive at ${\cal N}=8$ supergravity, with a spectrum comprising
1 spin 2, 8 spin 3/2, 28 spin 1, 56 spin 1/2 and 70 spin 0 massles particles (128 bosonic and 
128 fermionic degrees of freedom). Theories with ${\cal N}> 8$ necessarily involve fields with 
spins greater than 2 and are not evidently unitary.

For our purposes, we are interested in the potential physical effects of a gravivector on low-energy
antimatter experiments. The simplest theory to consider is ${\cal N}=2$ supergravity in which
(setting aside the gravitinos for the moment) the gravitational force due to the exchange of the 
massless graviton is complemented by a new `fifth force' mediated by gravivector exchange.
Matter fields describing the standard model fermions must be introduced into the theory 
separately as ${\cal N}=2$ supermultiplets with mass $m$, comprising two Majorana fermions $\chi^i$
and two complex scalar fields.

The construction of supergravity Lagrangians is highly technical and the details need not concern us 
here. The key point is that the supersymmetry algebra relates the matter coupling of the gravivector 
to that of the graviton.\footnote{Briefly, the global ${\cal N}=2$ supersymmetry algebra acting on the massive matter 
supermultiplets contains a central charge, which is matched to a gauge transformation on the
gravivector field induced by the local ${\cal N}=2$ supersymmetry transformation on the 
graviton supermultiplet. This identification of the gravivector as the gauge boson for the ${\cal N}=2$
central charge fixes its coupling to the matter supermultiplet.
The full Lagrangian for ${\cal N}=2$ supergravity coupled to a massive matter
supermultiplet was first derived explicitly in \cite{Zachos:1978iw,Zachos:1979uh}.}
Letting $\k^2 = 4\pi G$, the coupling of the gravivector $A_\m$ to the massive
Dirac fermion $\psi = \tfrac{1}{\sqrt{2}}\left(\chi^1 + i \chi^2\right)$ in the matter supermultiplet
is of the familiar form,
\begin{equation}
{\cal L}_{\rm int} = i g'\, \bar\psi \gamma^\m \psi \, A_\m \ ,
\label{b53}
\end{equation}
with $g' = \k m$, that is $\alpha' = m^2/M_{pl}^2$.

From this point, the analysis mirrors that presented above for the $U(1)_{B-L}$ gauge theory, with the
gravivector boson playing the r\^ole of the $Z'$. The gravivector is massless in the theory with unbroken 
supersymmetry, but may be given a mass $m_V$ (along with the gravitinos) by breaking supersymmetry
through a super-Higgs mechanism. By evaluating the 1-graviton and 1-gravivector Born exchange diagrams,
we deduce the following effective potential between matter/antimatter fermions,
\begin{equation}
V(r) \,=\, - \frac{G m_1 m_2}{r}\, +\,\frac {g_1 g_2}{4\pi \,r} \, e^{-m_V r} \ ,
\label{b54}
\end{equation}
with $g = \pm \k m$ for a fermion (antifermion) respectively. 

Rewriting (\ref{b54}) and substituting for the couplings $g_1$, $g_2$, we find the remarkable 
result \cite{Zachos:1979uh, Scherk:1979aj, Scherk:1980gq},
\begin{equation}
V(r) \,=\, - \frac{G m_1 m_2}{r} \, \left( 1 \, \mp\, e^{-m_V r}\, \right) \ ,
\label{b55}
\end{equation}
where the $-$ sign gives the potential between two fermions or two antifermions,
with the $+$ sign for a fermion-antifermion interaction.
With a massless gravivector, there is therefore an {\it exact cancellation} between the usual gravitational 
force and the new gravivector-induced interaction between pairs of elementary fermions or pairs of antifermions. 
In this sense, the gravivector force is `antigravity'. Notice, however, that this is exactly the opposite
of the popular use of `antigravity', which speculates (without foundation) about a repulsive 
gravitational interaction between matter and antimatter. For a fermion-antifermion pair, the 
gravivector force is attractive and doubles the strength of the usual gravitational attraction.

A similar effect occurs in higher supergravities. For example, in ${\cal N}=8$ supergravity there are 
contributions from the graviton, gravivector and spin 0 graviscalar fields. The spin 2 and spin 0 exchange 
diagrams are always attractive, while the spin 1 exchange may be attractive or repulsive.
For unbroken ${\cal N}=8$ supergravity, the coupling is $g = 2 \k m$  and the contributions again cancel
exactly for a fermion-fermion or antifermion-antifermion interaction \cite{Scherk:1980gq}.

For long-range interactions, the gravivector is therefore constrained by the same experimental tests 
of the weak equivalence principle (WEPff) on ordinary matter that constrain the $Z'$ interaction 
in $U(1)_{B-L}$ gauge theory. However, in supergravity, as we have seen, the coupling 
$\alpha' = m^2/ M_{pl}^2$ is {\it fixed}, so the only free parameter is the gravivector mass 
$m_V$.\footnote{Using a light quark mass, the gravivector coupling is of order 
$\alpha' =m_q^2/M_{pl}^2 \sim 10^{-42}$.
Comparing this with the above limit $\alpha' \lesssim 10^{-49}$ on the coupling of a light $Z'$ boson 
with range greater than the Earth's radius, shows immediately that in this theory a massless or light
gravivector of this range is already ruled out by existing equivalence principle tests on matter.}

An additional complication in applying the potential (\ref{b55}) to matter comprised of nucleons
is that the fundamental interaction is between the gravivector and the elementary quarks in the
bound-state nucleon, whereas the graviton couples to the full energy-momentum tensor including
the gluons (the gravivector-gluon coupling vanishes). 
Since the quark masses constitute approximately 1 percent of the nucleon mass,
there is scope for the gravivector and graviton interactions with nucleons to differ by a 
factor of O$(10^{-2})$.  A comprehensive account of the experimental limits on gravivector (and graviscalar)
interactions in ${\cal N}=2$ and ${\cal N}=8$ supergravity from
equivalence principle tests on matter, including these considerations,
is given in \cite{Bellucci:1996wz}. This concludes that the range of the gravivector interaction
is bounded by $\l \lesssim 1$ m, with the gravivector mass $m_V \gtrsim 10^{-6}$ eV.

The conclusion is that while the strength of the gravivector interaction is fixed by supersymmetry 
to be of the same order as the gravitational force, its range is already experimentally constrained
{\it in these theories} to be too short to have any appreciable effect on single-atom free-fall experiments 
dependent on the mass of the Earth, such as the WEPff tests proposed in antimatter experiments.

\subsubsection{$S$, $V$, $T$ background fields}

Although these supergravities are not themselves realistic BSM theories encompassing the standard model,
they do suggest we investigate the experimental consequences of potential gravivector and graviscalar
interactions in more generality. In this section, we therefore adopt a purely phenomenological approach
and consider the implications of theories which may possess additional vector and scalar background fields 
with arbitrary couplings and ranges in addition to the standard tensor gravitational field.\footnote{There 
are a great many proposals for modified gravity theories, from the original Brans-Dicke scalar-tensor theory
\cite{Brans:1961Sx} to the generalised Horndeski models \cite{Horndeski:1974wa}
and further extensions involving vector fields such as the TeVeS \cite{Bekenstein:2004ne} and
scalar-tensor-vector \cite{Moffat:2005si} models. For a recent review, 
see {\it e.g.} \cite{Clifton:2011jh}. As explained above, since our interest centres on antimatter experiments,
we are concerned here with theories with new vector fields coupling directly to a fermion current.}

The general static effective potential between two particles (labelled 1 and 2) is then a straightforward 
extension of (\ref{b54}), {\it viz.}
\begin{equation}
V(r) = - \left( \, \frac{G m_1 m_2}{r} - \frac{g_1^V g_2^V}{4\pi r} e^{-r/\l_V} 
+ \frac{g_1^S g_2^S}{4 \pi r} e^{-r/\l_S} \, \right) \ ,
\label{b56}
\end{equation}
where we have included a single vector and scalar field with couplings and ranges $g^V, \l_V$ and
$g^S, \l_S$ respectively. We could readily add further independent vectors and scalars to $V(r)$ if desired.

If the two particles are in motion with 4-velocities $u_1$ and $u_2$, with corresponding relativistic
factors $\c_1$ and $\c_2$, then this potential is modified:
\begin{equation}
V(r) = - \frac{1}{\c_1 \c_2}\, \biggl( \,\frac{Gm_1 m_2}{r} \left( 2 \left(u_1. u_2\right)^2 - 1\right) 
 \,-\, \frac{g_1^V g_2^V}{4 \pi r} \,u_1. u_2 \,e^{-r/\l_V} \,+\, \frac{g_1^S g_2^S}{4 \pi r}\, e^{-r/\l_S} \, \biggr) \ .
\label{b57}
\end{equation}
More simply, if we consider one particle at rest, so that $u_1.u_2 = \c$ (where $\c$ refers to the 
relative velocity), we can write simply
\begin{equation}
V(r) = - \, \biggl( \,\frac{1}{\c} \left( 2 \c^2 - 1\right) \frac{Gm_1 m_2}{r}  
\,-\, \frac{g_1^V g_2^V}{4 \pi r}  \,e^{-r/\l_V} 
\,+\, \frac{1}{\c} \frac{g_1^S g_2^S}{4 \pi r}\, e^{-r/\l_S} \, \biggr) \ .
\label{b58}
\end{equation}
These forms would also accommodate the $U(1)_{B-L}$ interaction discussed above.

Consider now massless gravivectors and graviscalars with infinite-range potentials and couplings of 
gravitational strength.  In the scenarios envisaged in the extended supergravity theories above
\cite{Scherk:1979aj, Scherk:1980gq}, the couplings of the tensor, vector and scalar interactions
would balance leaving a net zero force between pairs of matter particles.

More realistically \cite{Macrae:1984zf} (see also the extended development of these ideas by
Goldman, Hughes and Nieto \cite{Goldman:1986hk, Nieto:1987, Nieto:1991xq}),
we could consider a scenario where the gravivector and graviscalar couplings are equal and these
interactions cancel. This would occur generically if the vector and scalar arose from dimensional reduction
of a five-dimensional theory, as occurs in some supergravity and BSM models.
In this case, the gravitational force between matter particles would obey the usual Einstein tensor
gravity, whereas that between matter and antimatter would experience a stronger {\it attraction} due to the 
addition of the gravivector and graviscalar forces. This possibility has been extensively promoted
to motivate independent free-fall experiments on antimatter.

To assess this scenario, we first have to recognise that we cannot engineer an exact cancellation
{of the gravivector and graviscalar interactions for both the static and moving cases (\ref{b56}) and (\ref{b58})
due to the different velocity-dependences of the potential for vectors and scalars. 
Even in this scenario, therefore, the size of the gravivector and graviscalar interactions would still
be constrained to some extent by existing equivalence principle tests on matter.
This would limit the possible deviation from WEPff in antihydrogen free-fall experiments.

As a quick estimate, equivalence principle tests using lunar laser ranging 
of the Earth-Moon system in the Sun's gravitational field establish the validity of WEPff to $1:10^{13}$
\cite{Williams:2012nc}. Since the relative velocity of the Earth and Moon is approx.~1kms$^{-1}$, corresponding
to $\c - 1 \sim 10^{-10}$, this would constrain the size of the vector and scalar couplings in
 (\ref{b56}), (\ref{b58}) to give $\D g/g \lesssim 10^{-3}$, where $\D g$ is the difference in the gravitational
acceleration of matter $(g)$ and antimatter.

Secondly, there are several possibilities for the scalar couplings, not all of which match those allowed for a 
vector \cite{Adelberger:1991ke}. 
For example, while the absence of anomalies only allows a vector coupling to the combination
$B-L$, a scalar could couple to $B$ or $L$ independently. Alternatively, a graviscalar could couple to the
full trace of the energy-momentum tensor $T^\m{}_\m$ rather than simply the mass as shown 
in the supergravity models above. More generally, a fundamental scalar would couple to 
$\mathsf{C}$ (and $\mathsf{CPT}$) even operators whereas the vector coupling, as in (\ref{b53}), is 
$\mathsf{C}$ (and $\mathsf{CPT}$) odd.

A further serious difficulty with this scenario is that even if a cancellation could be achieved at the level
of fundamental interactions, the most stringent EP tests involve bound states, ranging from nucleons
(as bound states of quarks and gluons) and atoms to macroscopic bodies. So while a gravivector may couple
directly to the masses of the constituent elementary particles, the corresponding tensor graviton
and graviscalar would also couple to the full energy-momentum of the bound state.
Moreover, through loop corrections to the energy states ({\it e.g.}~the Lamb shift \cite{Alves:2009jx})
or through the parton distribution functions describing the nucleon structure, these bound states
are already sensitive to the existence of antimatter. These corrections have been analysed 
in some detail in \cite{Adelberger:1991ke, Wagner:2012ui, Alves:2009jx},
where constraints as low as $\D g/g \lesssim 10^{-9}$ are quoted. For example, by considering the 
motion of electrons in the atom, an improved bound from the differing velocity dependence of 
scalar and vector interactions of $\D g /g \lesssim 10^{-7}$ is quoted in \cite{Alves:2009jx}.

While there is clearly a significant level of model-dependence in all these analyses, the overall picture
is clear that the high precision of existing EP tests places severe constraints on scenarios which attempt 
to hide new gravitational strength vector and scalar forces so that they manifest themselves only
in dedicated antimatter experiments. This perhaps emphasises the importance of aiming at high
precision in gravity experiments on antimatter as well as in spectroscopy.

The impact of new `fifth force' vector and scalar interactions would not be limited to WEPff tests. 
For example, a long-range $U(1)_{B-L}$ vector interaction would produce an analogue of the Stark effect 
on the spectrum of hydrogen, and oppositely on antihydrogen.
Moreover, WEPc tests would in principle also be sensitive to new long-range vector forces, which 
would modify the Schwarzschild metric around the Earth to Reissner-Nordstr\"om type.
Spectroscopic measurements analogous to those measuring a gravitational redshift would
then exhibit frequency shifts proportional to the coefficient $g_{00}$ of this new metric.
These effects will be considered in the following sections.

\subsection{Matter-antimatter Asymmetry and $\mathsf{CPT}$}\label{sect 2.7}

One of the most important outstanding issues in cosmology is to understand the 
origin of matter-antimatter asymmetry, {\it i.e.}~why the observable universe
is composed overwhelmingly of matter with only a negligible antimatter component.
The issue is quantified by the observed value of the key cosmological parameter
$\eta = n_B/n_\gamma$, the baryon-to-photon ratio. Here, $n_B$ and $n_\gamma$ 
(essentially the entropy) are the number densities of baryons and photons in 
the present universe. The observed value \cite{Aghanim:2018eyx,Tanabashi:2018oca}
is $\eta \simeq 6.1 \times 10^{-10}$.

Explaining the matter-antimatter asymmetry is frequently cited as a motivation for
experimental searches for $\mathsf{CPT}$ violation, specifically with antihydrogen.
Here, we give a brief appraisal of the possible relevance of $\mathsf{CPT}$
violation in explaining the observed asymmetry.

The theoretical requirements for a mechanism to generate matter-antimatter 
asymmetry were set out by Sakharov \cite{Sakharov:1967dj}
in the well-known conditions for baryogenesis (or leptogenesis\footnote{A lepton
asymmetry generated at high temperatures in the early universe can be 
converted to the present-day baryon asymmetry through non-perturbative
``sphaleron'' \cite{Klinkhamer:1984di} interactions at the electroweak scale. 
While violating both $B$ and $L$ separately, these processes still conserve 
$B-L$ symmetry.  Many explanations 
of the observed baryon-to-photon ratio therefore focus on leptogenesis as the 
fundamental mechanism.}), {\it viz.}~the theory must exhibit:
\begin{enumerate}
\item $B$ (or $L$) violation;
\item  $\mathsf{C}$ and $\mathsf{CP}$ violation;
\item processes out of thermal equilibrium.
\end{enumerate}

At zero temperature, the standard model is $B$ and $L$ conserving (up to negligible 
instanton effects), but for temperatures at or above the electroweak scale, which
arise in the early universe, non-perturbative sphaleron-induced processes
allow substantial $B$ and $L$ violation \cite{Kuzmin:1985mm}. This is the origin of the
``electroweak baryogenesis'' mechanism, in which baryogenesis would take place 
during a first-order electroweak phase transition. However, the nature of the
electroweak phase transition (which is a smooth crossover in the standard model) 
and the magnitude of $\mathsf{CP}$ violation in the CKM matrix mean that this 
mechanism cannot reproduce the observed value of the baryon-to-photon ratio
within the standard model.

In beyond-the-standard-model (BSM) theories, on the other hand, $B$ or $L$ violation 
is readily achieved, {\it e.g.}~through the non-equilibrium decay of heavy gauge bosons 
or Higgs fields in grand unified theories (GUTs) or the decay of additional 
R-handed neutrinos. Both of these are theoretivcally well-motivated, models with 
sterile neutrinos with heavy Majorana masses providing a natural `see-saw' mechanism for generating 
the light neutrino masses. BSM modifications can also render the electroweak phase transition 
first-order, realising the electroweak baryogenesis mechanism. There are many related 
BSM models of baryogenesis or leptogenesis which, for suitably chosen values
of the parameters characterising the theory, can give rise to the observed 
matter-antimatter asymmetry within a standard cosmological framework. 
(See \cite{Buchmuller:2005eh,Shaposhnikov:2009zzb,Garbrecht:2018mrp}
for a selection of reviews.)
Indeed, already in 2009, the review \cite{Shaposhnikov:2009zzb} quoted more than 40
viable proposals for generating the matter-antimatter asymmetry.
 
What then is the relevance of $\mathsf{CPT}$ violation for this discussion?
The Sakharov conditions were formulated under the {\it assumption} of exact
$\mathsf{CPT}$ symmetry. Allowing for $\mathsf{CPT}$ violation means that
it is possible to establish a matter-antimatter asymmetry {\it in thermal equilibrium}
\cite{Cohen:1987vi}.
That is, the third condition can be replaced by
\begin{itemize}
\item [3.] $\mathsf{CPT}$ violation.
\end{itemize}
Note, however, that the first two conditions are still required -- even with 
$\mathsf{CPT}$ violation, we still need a BSM theory exhibiting $B$ or $L$
violating interactions as well as $\mathsf{C}$ and $\mathsf{CP}$ violation. 

To illustrate what such a model of leptogenesis could look like, consider the
following term for the light neutrinos (restricting initially to one flavour for simplicity)
in the minimal Lorentz and $\mathsf{CPT}$ violating SME discussed in 
section \ref{sect 2.2}:
\begin{equation}
{\cal L}_a = a_\mu^{(3)}\, \overline{\nu_L} \gamma^\mu \nu_L \ .
\label{bnew27}
\end{equation}
The dimension-three operator $\overline{\nu_L} \gamma^\mu \nu_L$ is $\mathsf{C}$
violating, $\mathsf{CP}$ odd, and $\mathsf{CPT}$ odd.
It is the conserved current for (neutrino) lepton number, so the space integral of its
time component, $\int d^4 x \,\overline{\nu_L} \gamma^0 \nu_L$, is the corresponding
charge, {\it i.e.}~lepton number. The coupling $a_0^{(3)}$ therefore acts as a 
chemical potential. This can also be seen directly from the discussion in section
\ref{sect 2.3}, where we saw how an interaction of the form ${\cal L}_a$ would
modify the dispersion relation differently for neutrinos and antineutrinos. 
At non-zero temperature, this modifies the corresponding particle distributions
in exactly the way characteristic of a chemical potential in statistical mechanics.

Now, in the high temperature environment of the early Universe, and providing
there exist lepton number violating reactions to maintain thermal equilibrium, 
this effective chemical potential $\mu = a_0^{(3)}$ implies different equilibrium 
number densities for neutrinos and antineutrinos\footnote{Explicitly, the net 
lepton number is found from the statistical distributions (neglecting neutrino masses 
and curvature effects) as 
\begin{align*}
n_\ell = n_\nu - n_{\bar{\nu}} &= g_\nu \int \frac{d^3 {\bf p}}{(2\pi)^3}\,
\left[\frac{1}{e^{(E-\mu)/T} + 1} - \frac{1}{e^{(E+\mu)/T} + 1} \right] 
\nonumber \\
&= \frac{g_\nu}{2\pi^2} \int_0^\infty dE\,E^2 \, 
\left[\frac{1}{e^{(E-\mu)/T} + 1} - \frac{1}{e^{(E+\mu)/T} + 1} \right] 
\nonumber \\
&\simeq \frac{g_\nu}{2\pi^2}\, \frac{2\mu}{T}\, T^3 \, \int_0^\infty dx\, x^2 
\frac{e^x}{(e^x +1)^2}   ~~~~
= ~~\frac{1}{3} \mu T^2 \ .
\end{align*}
The photon number density is $n_\gamma = \frac{2\zeta(3)}{\pi^2} T^3$.
For comparison, the entropy density, which is alternatively used to normalise the
baryon asymmetry, is $s = \frac{2\pi^2}{45} g_{*s} T^3$, where $g_{*s}$ 
is the effective number of light degrees of freedom at the scale $T$. },
resulting in a net lepton
number density $n_{\ell} \sim \mu T^2$. As the universe cools, the rate of these 
interactions falls until at some decoupling temperature $T_D$ they fall below the 
rate of expansion, given by the Hubble parameter \cite{Kolb:1990vq}. 
In most scenarios, we expect 
$T_D$ in the range $100\,{\rm GeV} \lesssim T_D \lesssim 10^{12}\,{\rm GeV}$,
its value determined by the BSM dynamics.
At this point, thermal equilibrium 
is no longer maintained and the lepton number density freezes out at the value
$n_{\ell} \sim \mu T_D^2$. Given that the photon density varies with temperature
as $n_\gamma \sim T^3$, the resulting lepton-to-photon ratio is then frozen at
the value $\eta_{\ell} \equiv n_{\ell}/n_\gamma \sim a_0^{(3)}/T_D$.
Provided that $T_D$ is above the electroweak scale, this lepton number
asymmetry may be subsequently converted to a baryon number asymmetry
by sphaleron interactions,
yielding a final baryon-to-photon ratio $\eta$ diluted by a factor of around 
$10^2$  relative to $\eta_{\ell}$.

To convert this into a realistic model, we need to remember that, as discussed
in section \ref{sect 2.3}, in the case of a single fermion flavour we can make 
a field redefinition in the kinetic term in the Lagrangian to remove the 
interaction ${\cal L}_a$. We therefore need to extend ${\cal L}_a$ to three
flavours of neutrinos, {\it viz.} ${\cal L}_a \rightarrow (a_\mu^{(3)})_{ij}
\overline{\nu_L}^{\,i} \gamma^\mu \nu_L^j$, together with flavour mixing.

The mechanism described above is essentially that already studied in detail as
``spontaneous leptogenesis'', where the role of the coupling $a_\mu$
is played by a time-dependent VEV of a scalar field, with the equivalence
$a_\mu \leftrightarrow \partial_\mu \phi$, or ``gravitational leptogenesis'',
where $a_\mu \leftrightarrow \partial_\mu R$ and the coupling is replaced by the
time derivative of the Ricci scalar in an expanding universe (as described briefly
in section \ref{sect 2.5}).

The same mechanism could also be employed with flavour-mixed quarks, 
generating a baryon asymmetry directly. In this case, we could envisage a 
lower decoupling temperature, but an absolute requirement is that the baryon
asymmetry must be established before the onset of big-bang nucleosynthesis
at $T = 10\,{\rm MeV}$, the successful theory of which provides a stringent 
constraint on the value of $\eta$.

At this point, we can check whether this mechanism could be realistic, given
the experimental constraints on the relevant minimal SME parameters.  
Of course, this makes the assumption that these couplings remain constant
over the evolution of the Universe.  This is non-trivial, since if they are regarded 
as VEVs of some time-dependent fields, their values would evolve and could
be markedly higher at the time of lepto(baryo)genesis than their current values.
From the SME data tables \cite{Kostelecky:2008ts}
(Table D26), bounds on the relevant neutrino coefficients of
$(a_0^{(3)})_{ij} \lesssim 10^{-20}\,{\rm GeV}$ are quoted. With $T_D > 100\,{\rm GeV}$,
the resulting prediction for $\eta_{\ell} \sim a_0^{(3)}/T_D$ would be many orders
of magnitude too small.  Constraints on $(a_0^{(3)})_{ij}$ in the quark sector would
similarly exclude the direct baryogenesis scenario under these assumptions.

However, in \cite{Bertolami:1996cq}, where this type of $\mathsf{CPT}$ violating model was
first considered, the potential of higher-dimension operators in the 
SME effective Lagrangian to produce the observed value of the asymmetry
was discussed. For example, we could consider interactions of the form
${\cal L}_a^{(5)} = -a_{\mu\rho\sigma}^{(5)} \bar{\psi} \gamma^\mu \partial^\rho
\partial^\sigma \psi$, where here we can simply take $\psi$ to be the electron
field. This would be the leading-order electron coupling of this type, since
the dimension 3 coupling $a_\mu^{(3)}$ can be removed by a field redefinition and is
not physical.  (This interaction is described in the context of $\overline{\rm H}$
spectroscopy in section \ref{subsubsection:LVCPTV} below.)
Because of the extra derivatives in the operator, the same analysis would
predict a corresponding lepton asymmetry of order $|a_{0\rho\sigma}^{(5)}|\, T_D$.

Now, in the spirit of effective actions, where higher order couplings should
be suppressed by powers of the fundamental UV scale $M$,  this may be expected
to give a smaller contribution than $a_0^{(3)} /T_D$ by a power of $T_D^2/M^2$.
However, if we set this aside, we could simply ask what the current experimental 
bounds on the non-minimal couplings would allow. In order to extract 
phenomenology from this operator, we can either make a non-relativistic
expansion, which leads amongst others to the coupling $a_{200}^{\rm NR}$
which enters the transition frequency for the $1S$\,-\,$2S$ transition in H and 
$\overline{\rm H}$ (see section \ref{subsubsection:LVCPTV}), or an ultra-relativistic 
expansion \cite{Kostelecky:2013rta} in which the relevant parameter is
$\mathring{a}^{{\rm UR}(5)}$.  The ultra-relativistic regime is appropriate for an analysis of
the astrophysical effects of modified dispersion relations, which would allow reactions
such as $\gamma\rightarrow e^+ e^-$ which are ruled out by the observation of multi-TeV gamma rays 
\cite{Konopka:2002tt, Kostelecky:2013rta}. 
The couplings $a_{200}^{\rm NR}$ and $\mathring{a}^{{\rm UR}(5)}$ are closely
related, but not identical, and very different bounds for the electron are quoted in the
the SME data tables \cite{Kostelecky:2008ts} (Table D7). Neither corresponds
exactly to the combination of $a_{\mu\rho\sigma}^{(5)}$ couplings determining the lepton 
asymmetry in the model above. Nevertheless, it is clear that the measurement of a non-vanishing
$a_{200}^{\rm NR}$ coefficient in the well-understood setting of $1S$\,-\,$2S$ $\overline{\rm H}$
spectroscopy would provide a strong impetus to models of leptogenesis
and baryogenesis based on $\mathsf{CPT}$ violation.

Finally, note that here we have only considered a relatively simple model for illustration. Other possibilities
combining $\mathsf{CPT}$ violation, also involving other SME couplings, and more exotic 
BSM physics can readily be constructed and have been actively studied 
(see {\it e.g.}~\cite{Mavromatos:2018map} for examples and a review) and may be added 
to the extensive list of proposals for a theory of baryogenesis.

To summarise this discussion, we have seen several general points to keep in mind
while considering the relevance of $\mathsf{CPT}$ violation to leptogenesis and
baryogenesis:
\begin{itemize}
\item [(i)] There are already a large number of realistic proposed mechanisms
of baryogenesis {\it without} $\mathsf{CPT}$ violation, but they all need some 
BSM physics.
\item [(ii)] With $\mathsf{CPT}$ violation, a baryon or lepton asymmetry 
may be generated {\it in thermal equilibrium}, in contrast to $\mathsf{CPT}$
conserving theories, via a mechanism similar to spontaneous or
gravitational baryo(lepto)genesis. However, these models still require new 
BSM physics with as yet undiscovered particles.
\item [(iii)] Existing constraints on SME parameters appear to rule out models
restricted to the minimal SME. Non-minimal SME couplings are in general
less constrained, however, and the different dependence on the decoupling temperature
in models involving them offers the potential to achieve the required asymmetry, provided the couplings
are not too suppressed by the high-energy scale underlying the SME effective Lagrangian.
\item[(iv)] In the minimal SME, the parameters which are probed in $\overline{\rm H}$
spectroscopy (notably the space components $b_3^{e,p}$)
are different from those ({\it e.g.}~the time components $a_0^{\nu,q}$) which
enter the simplest leptogenesis or baryogenesis models. 
At higher order, the non-minimal couplings accessible to spectroscopy are
very closely related, though not identical, to those involved in the simplest models
of baryo(lepto)genesis. Any measurement of $\mathsf{CPT}$ violation in 
$\overline{\rm H}$ spectroscopy would therefore have a significant impact on
the development of theories seeking to explain the matter-antimatter asymmetry 
of the universe.
\end{itemize}

{\section{Antihydrogen}\label{sect 3}

So far, only the ALPHA collaboration has performed experiments to measure some of the properties 
of antihydrogen relevant for testing the fundamental physics described above. 
Accordingly, we will use their data as exemplars in our discussion. 
As summarised in section \ref{sect 1}, the ALPHA results to date
comprise:  a limit on the charge neutrality of the anti-atom \cite{Amole14ch,Ahmadi:2016ozp}; 
demonstration of a method to investigate the gravitational interaction of antihydrogen \cite{Amole13} 
(a study accompanied by detailed investigations of the trajectories of antihydrogen atoms held in a magnetic 
minimum neutral atom trap \cite{Zhmoginov,Zhong}); detection and spectroscopy of antihydrogen
ground state hyperfine transitions \cite{Amole12,Ahmadi17}, the two-photon Doppler-free $1S$\,-\,$2S$
transition \cite{ALPHA1S2S,Ahmadi:2018eca}, and the observation of the Lyman-$\alpha$ ($1S$\,-\,$2P$) transition 
\cite{ALPHA1S2P} and a determination of the Lamb shift \cite{Ahmadi:2020ael} in the anti-atom.

These investigations fall into two broad categories. As fundamental physics tests, the spectroscopy 
measurements and charge limits are primarily sensitive to potential Lorentz and $\mathsf{CPT}$ violations, though
in principle could also be affected by a new $U(1)_{B-L}$ force. 
The gravitational studies are tests of WEPff, including `fifth forces'.  Clearly these are free-fall experiments, 
but it should be kept in mind that even the charge investigations also involve assumptions regarding the form 
and evolution of the trajectories of the trapped anti-atom which could be influenced by a WEPff violation.

\subsection{Charge Neutrality of Antihydrogen}\label{sect 3.1}

One of the most direct tests of fundamental principles with antihydrogen, and one of the first 
to be confirmed experimentally \cite{Ahmadi:2016ozp}, is its electric charge neutrality.
Evidently, the net electric charge of the anti-atom is expected to be zero, interpreted as the 
sum of the charges of the antiproton ($-1$) and positron ($+1$). The question we address here is 
what fundamental principles are tested by an experimental measurement of a putative antihydrogen
charge, here denoted as $Q$.

\subsubsection{Antihydrogen charge measurement}

ALPHA has performed two studies that have been used to set a limit on $Q$ via an analysis of the 
behaviour of the anti-atom in the presence of applied electric ($E$)
and magnetic ($B$) fields, 
were it to have a charge. The first 
\cite{Amole14ch} consisted of a retrospective analysis of an experiment in which trapped 
antihydrogen atoms were released from the neutral atom trap via magnetic field reduction, 
in the presence of electric fields (biased in different trials either axially to the left or the right 
of the magnetic trap centre, and denoted as $E_L$ and $E_R$ respectively) whose function was to aid 
in distinguishing antihydrogen annihilations from those of any antiprotons remaining in the trap region. 
A search was made for a possible electric field-induced shift in the measured axial antihydrogen annihilation 
distributions, $\langle z \rangle_{\Delta}$, caused by a non-zero $Q \propto \langle z \rangle_{\Delta}/(E_R - E_L)$. 
From a measured mean axial deflection of $4.1 \pm 3.4$\,mm from the centre of the atom trap a limit was found 
as $Q = (-1.3 \pm 1.1 \pm 0.4) \times 10^{-8}e$ including statistics and systematics, with a 1$\sigma$ 
confidence.

In their second experiment \cite{Ahmadi:2016ozp} ALPHA used a stochastic heating method in which the 
application of random time-varying electric fields would result in stochastic energy kicks to a charged 
anti-atom such that it would eventually leave its shallow trapping well. In particular, an antihydrogen atom
with charge $Q$ would gain (from $N$ kicks of voltage change $\Delta\phi$) a kinetic energy of 
$|Q|e\Delta\phi\sqrt{N}$ such that it will leave a well of depth 
$E_{well}$ if $|Q| \gtrsim E_{well}/e\Delta\phi\sqrt{N}$. The measured parameter was the survival probability 
of the anti-atoms in the trap, when compared to null trials when no stochastic field was applied, 
and the result was that the charge was bounded as $|Q| < 0.71$ ppb ($1\sigma$): a 20-fold improvement 
on the previous study \cite{Amole14ch}.

In both of these experiments simulations of antihydrogen trajectories were used extensively  in the 
derivation of the final results for $Q$, and in studies of the pertinent uncertainties - particularly those 
due to systematic effects. Specifically, it was assumed that the antihydrogen motion is such that its 
magnetic moment adiabatically follows the (spatially varying) magnetic field of the neutral trap such that 
its motion can be investigated using classical mechanics. The combined equation of motion for the 
centre of mass position
$\mathbf{r}$ of antihydrogen, with an inertial mass $m_i$ and charge $Q$ (in units of the 
fundamental charge $e$) and so-called gravitational mass $m_g$ (see
section \ref{sect 3.3.1}), in spatially and temporally 
varying electric and magnetic fields ($\mathbf{E}(\mathbf{r},t)$ and  $\mathbf{B}(\mathbf{r},t)$ respectively) is,
\begin{equation}
m_i\frac{{\rm d}^2\mathbf{r}}{{\rm d}t^2} = \boldsymbol{\nabla}(\boldsymbol{\mu} \cdot 
\mathbf{B}(\mathbf{r},t)) + Qe[\mathbf{E}(\mathbf{r},t) + \frac{\mathbf{dr}}{dt}
\times\mathbf{B}(\mathbf{r},t)] - m_g g{\mathbf{\hat{y}}},
\label{d1}
\end{equation}
where $\boldsymbol{\mu}$ is the magnetic moment and
$g$ is the local acceleration due to gravity which is assumed to act in the $y$-direction.
 Here the familiar, essentially Newtonian, notion that a possible WEP anomaly can expressed via a 
``gravitational'' mass $m_g$ distinct from the inertial mass has been applied, with 
ALPHA defining the ratio $F=m_g/m_i$ and setting limits on this parameter from an analysis of 
their experiment, as will be described below.

Ahmadi {\em et al.} \cite{Ahmadi:2016ozp} have also described how this value for $Q$ can be 
used in conjunction with the limit on the so-called antiproton charge anomaly \cite{Hori11} (defined in
obvious notation as $(|q_{\overline{p}}| - e)/e$) to reduce the corresponding quantity for the positron 
by a factor of 25 to 1 ppb.~($1\sigma$). The assumption in the analysis is that the value of $m_i$ used 
is that of hydrogen.

Such charge  anomaly limits are usually quoted as tests of $\mathsf{CPT}$ 
(see {\it e.g.} \cite{Tanabashi:2018oca}), 
since, as described in section \ref{sect 2.1}, charge differences equal to zero between particles and antiparticles  
(and hence $Q=0$) are strict requirements of the $\mathsf{CPT}$ theorem. We comment critically on this below.
We also draw attention here to the work of Hughes and Deutch \cite{HughesDeutch} who used measurements 
of cyclotron frequencies and spectroscopic data to derive limits on the electric charges of positrons 
and antiprotons. The direct limit set on $Q$ described here could also be used to augment and update 
the jigsaw of data they present.

\subsubsection{Theoretical principles}

Of course, Lorentz and $\mathsf{CPT}$ symmetry does imply that the charges of $\overline {\rm H}$ and H should 
be the same. However, charge neutrality of antihydrogen is not really a test of $\mathsf{CPT}$ in the sense 
that even in the Lorentz and $\mathsf{CPT}$ violating effective theories described in section \ref{sect 2}, 
the charges of the proton and antiproton, and the electron and positron, are still required 
to be equal and opposite. In fact, this is a much deeper property of relativistic QFT which 
ultimately is necesssary for {\it causality}.

As explained in section \ref{sect 2.1}, the existence of antiparticles with precisely the same mass 
and opposite charge is a requirement of causality, necessary to ensure the vanishing of the
relevant correlators for spacelike separation (microcausality). It is therefore almost impossible 
to see how any difference in magnitude of the charge of the electron and positron could be 
compatible with our present understanding of QFT.

Even if we momentarily set these structural issues aside and continue to entertain the idea that
the electron and positron charges could be different, further difficulties proliferate.
Given the existence of the coupling of the photon to an $e^+ \, e^-$ pair, either the photon
would be charged or, in contradiction to a huge body of experimental evidence, electric
charge conservation would be violated. In the former case, an accelerating electron would
lose charge by synchrotron radiation, leading to further paradoxes.  Indeed, experimental limits
on the charge of the photon are extremely strong, the PDG quoting a bound of $10^{-35} e$
\cite{Tanabashi:2018oca}.

A second issue, which would apply equally to hydrogen, is then the equality of the magnitude 
of the charges of the antiproton and positron  (or equivalently, the proton and electron).
In the standard model, this is again ensured by a deep principle, in this case {\it unitarity}.
As discussed in section \ref{sect 2.6}, the absence of anomalies in gauged currents imposes 
a set of constraints on the charges of the fermions which appear as internal lines in the 
3-current triangle diagrams. This imposes a precise balance between the quark and lepton
charges. If this is broken, we would lose conservation of a gauged current, which in turn 
would lead to a loss of unitarity in the QFT. 

We see, therefore, that the experimental measurement of charge neutrality of antihydrogen
should be viewed as a test of {\it causality} and {\it unitarity} in the standard model QFT.
While $\mathsf{CPT}$ invariance does imply the equality of the charge of hydrogen and antihydrogen, 
charge neutrality would still be required even in a $\mathsf{CPT}$-violating QFT.

The only remaining loophole would seem to be our assumption that the measured charge 
of the bound states is given precisely by the sum of the charges of their constituents, either
the antiproton and positron for the antihydrogen atom or the valence quarks 
$\bar{u} \bar{u} d$ for the antiproton. This is referred to as the ``assumption of charge superposition''
in \cite{Ahmadi:2016ozp}. It could conceivably be possible to imagine some
sort of charge screening mechanism which could invalidate this for a particular experimental 
meaurement. However, such a screening effect would have had to avoid detection everywhere
else in the particle physics of hadrons, or indeed in hydrogen, and seems extremely unlikely 
to be a factor in interpreting the antihydrogen charge neutrality experiment.
Fortunately, these experiments have indeed validated the expected null result to high precision.

\subsection{Antihydrogen $1S$\,-\,$2S$, $1S$\,-\,$2P$ and $1S$ Hyperfine 
Spectroscopy}\label{sect 3.2}

The advent of precision antihydrogen spectroscopy has opened a new window to test 
fundamental principles such as Lorentz and $\mathsf{CPT}$ invariance, to extend experimental
tests of GR to antimatter systems, and to search for new long-range forces.

The gold standard spectroscopic measurement in this field is the two-photon $1S$-$2S$ 
transition in hydrogen, for which the transition frequency is known to a few parts in $10^{15}$
\cite{Matveev:2013orb}, many orders of magnitude more precise than state of the art 
theoretical calculations in QED. A direct comparison of the antihydrogen and hydrogen 
spectra therefore provides a more precise test of QED and $\mathsf{CPT}$ invariance than follows
from theory and hydrogen spectroscopy alone, further motivating these antimatter experiments.

\subsubsection{Antihydrogen spectroscopy} \label{sect 3.2.1}

Here, we give a brief overview of the experimental arrangement for spectroscopy in the ALPHA apparatus. 
Details, including a level diagram (Figure \ref{Fig 4}) with state labels, of the transitions involved in 
spectroscopy are given in section~\ref{subsubsection:LVCPTV} below. 

Historically, only microwave radiation could be coupled into the ALPHA apparatus, giving access to 
spin-flip transitions between ground-state hyperfine levels by positron spin resonance (PSR)~\cite{Amole12}. 
These transitions result in a reversal of the magnetic moment of the antihydrogen atom which will then no longer 
be trapped, leading to an annihilation signal registered by the silicon vertex detector. In addition to being 
the first ever interrogation of the internal structure of an anti-atom, this experiment is also of great 
significance because it shows that a spectroscopic signal can be produced from only one trapped anti-atom. 
The experiment also gave rise to the appearance and disappearance
detection protocols described in section \ref{sect 1}, which have been used in various 
combinations in all subsequent spectroscopic measurements. The ALPHA-2 upgrade includes a waveguide 
which enables efficient delivery of microwave radiation directly into the trap structure, and thereby higher 
transition rates. The microwave radiation is also used to drive the electron cyclotron resonance in trapped 
plasmas to determine the magnitude of the central magnetic field (approx. \SI{1}{\tesla}). The field can be 
determined with a precision of 1 ppm and an accuracy of 4 ppm. When the microwaves are tuned to the 
relevant PSR frequency (just below \SI{30}{\giga\hertz} at the B-field used in experiments), any unwanted 
population of hyperfine level $c$ states can be efficiently removed, allowing spectroscopy 
with spin polarised samples. 

Ground-state hyperfine spectroscopy is performed by first trapping ground-state antihydrogen atoms. 
During antihydrogen synthesis there is no control of the internal state of the anti-atom, and the trapped 
population typically contains equal amounts of anti-atoms in the weak-field seeking $1S_c$ and $1S_d$ states. 
Spectroscopy proceeds by injecting microwave radiation and stepping the frequency of the microwaves in small 
intervals over a frequency range (starting below resonance) that includes the onset of the $1S_c$\,-\,$1S_b$ transition 
which is effectively set by the magnitude of the central magnetic trap B-field. Once the frequency has been 
stepped over a range that covers all main spectral features, the frequency is then adjusted up by the hyperfine 
splitting (approx \SI{1.42}{\giga\hertz}) and the stepping continues to cover the  $1S_d$\,-\,$1S_a$ transition in a 
similar fashion.  Antihydrogen is detected in appearance mode during the frequency sweep, producing a 
spectrum which shows a rapid onset of resonance for both transitions. The onset frequency difference is 
used to determine the hyperfine splitting, leading to an uncertainty of four parts in \num{e4}. A refined 
measurement, making full use of the electron cyclotron resonance technique to determine the B-field, is possible.

The $1S$\,-\,$2S$ transition has long been the gold-standard in laser spectroscopy of atomic hydrogen due to 
the long lifetime (about \SI{1/8}{\second}) of the $2S$ state, and the cancellation of the Doppler effect 
(to first order) when a transition from the $1S$ state is induced by two counter-propagating photons. 
The flip-side is that, in addition to a narrow bandwidth, substantial laser power is required to drive the 
dipole-forbidden transition from the $1S$  ground state with experimentally relevant rates. While 
narrow band sources can nowadays straightforwardly be created from diode lasers, it is technically 
challenging to achieve sufficient power at the \SI{243}{\nano\meter} (UV) wavelength of the two-photon 
transition. In the ALPHA experiment the power from commercial laser sources is resonantly enhanced 
in an optical cavity surrounding the trapping region in a near-axial orientation. The enhancement cavity 
which is operated in the cryogenic region near \SI{4}{\kelvin} and in ultra-high vacuum, yields about 
\SIrange{1}{2}{\watt} of circulating laser power with a \SI{200}{\micro\meter} waist which is sufficient 
to cause photoionisation from the $2S$ state with a third photon. The ionised antihydrogen atom is 
no longer confined and thus, a signal on the silicon vertex detector is produced when the laser is on resonance. 
The laser system is referenced to atomic time via a GPS-scheduled quartz oscillator, with additional referencing 
from a locally operated Cs atomic clock, to provide the frequency accuracy of \num{8e-13}. The transition was 
observed~\cite{ALPHA1S2S} and later characterised with an uncertainty of 2 parts in a trillion~\cite{Ahmadi:2018eca}. 

While the $1S$\,-\,$2P$ transition in hydrogen and antihydrogen is arguably the simplest transition in any atomic 
system, the \SI{121.6}{\nano\meter} wavelength (the famous Lyman-alpha line) poses a formidable challenge 
for laser spectroscopy since there are no simple laser systems in this region of the spectrum. The short 
(about \SI{1.6}{\nano\second}) lifetime of the $2P$ state leads to a broad natural linewidth even before 
any further inhomogeneous broadening such as the Doppler effect. The Lyman-alpha line is nevertheless 
of great significance {\it e.g.} in astronomy and cosmology. For precision experiments with hydrogen and 
antihydrogen the line is of relevance because scattering red-detuned photons on the transition leads 
to Doppler cooling which in turn reduces line broadening. In order to achieve both a reliable and 
sufficiently intense source for excitation of the line, ALPHA has constructed a pulsed laser system based 
upon frequency tripling in Kr/Ar gas. The fundamental wavelength is produced from a pulsed solid-state 
(Ti:sapphire) laser which is seeded from a narrow-band diode laser. The current system produces up to 
\SI{0.8}{\nano\joule} inside the trapping region per \SI{12}{\nano\second} pulse at   \SI{121.6}{\nano\meter}, 
with a \SI{65}{\mega\hertz} linewidth. Due to positron spin-mixing (see subsection \ref{subsubsection:LVCPTV}), 
the excited atom may decay quickly to an untrapped state and a signal is produced on the silicon vertex detector. 
An added advantage of the pulsed laser system is that the appearance signal can be analysed in coincidence 
with the arrival of the pulse, leading to a measurement of the time-of-flight from the trap to the electrode wall, 
from which the velocity distribution of the trapped atoms can be reconstructed. The observation of the 
Lyman-alpha transition~\cite{ALPHA1S2P}, which was performed with 500 accumulated antihydrogen atoms, 
paves the way for antihydrogen fine structure spectroscopy and laser cooling, while more recently the
ALPHA collaboration has presented a detailed investigation of the fine structure of the $1S$\,-\,$2P$
transitions at a precision of 16 parts per billion, allowing a first determination of the Lamb shift for
antihydrogen \cite{Ahmadi:2020ael}.

\subsubsection{Lorentz and $\mathsf{CPT}$ violation}
\label{subsubsection:LVCPTV}

In this section, we analyse the possibility of observing Lorentz or $\mathsf{CPT}$ violation in 
antihydrogen spectroscopy, focusing on the $1S$\,-\,$2S$,  $1S$\,-\,$2P$ and the $1S$ hyperfine transitions.
We follow the systematic approach of parametrising potential Lorentz/$\mathsf{CPT}$ violating
effects in terms of the couplings in the SME effective action.
Our discussion here is not new\footnote{In this spirit, we give here only a limited set of
references to enable the reader to follow the quoted results. For a more complete survey
of the extensive SME literature and its applications to antimatter and spectroscopy, see
for example the review and compendium of limits on the couplings in \cite{Kostelecky:2008ts}.}, 
the aim being simply to provide an accessible presentation
of the key principles involved. 

The starting point is to approximate the full SME effective action by the corresponding
non-relativistic Hamiltonian appropriate for low-energy atomic physics. This has been carried
out by several authors \cite{Bluhm:1998rk, Kostelecky:1999mr, Kostelecky:1999zh, 
Altschul:2009bx, Yoder:2012ks},
including not just the coefficients of the renormalisable SME operators shown in (\ref{b19})
but also those from the full expansion of higher derivative operators in the extended
effective action \cite{Kostelecky:2013rta,Kostelecky:2015nma,Kostelecky:2018fmc}.

The non-relativistic Hamiltonian is obtained from the SME Lagrangian (\ref{b19}) in the
standard way as an expansion in powers of $p_i/m_e$, where $p_i$ is the electron/positron 
momentum, the novelty being simply that the familiar Dirac and QED action is extended
to include the additional Lorentz and $\mathsf{CPT}$-violating couplings. 
The leading terms in this expansion contribute at $O(1)$ in the fine structure constant
to the atomic energy levels, while those of $O(p_i p_j/m_e^2)$ give corrections of $O(\a^2)$.
As we shall see, the SME corrections to the $1S$-$2S$ transition frequency measured by ALPHA
only occur at $O(\a^2)$.

Following \cite{Kostelecky:1999mr, Kostelecky:1999zh, Yoder:2012ks}, we can write the relevant 
terms in the SME modified Hamiltonian appropriate for describing the antihydrogen atom as:
\begin{equation}
H_{\rm SME} = \sum_{\omega = e,p}\,\left[ A^\omega +2 B_k^\omega S^k 
+ \left( E_{ij}^\omega +2 F_{ijk}^\omega S^k\right) \frac{p_i p_j}{m_e^2} + \ldots  \right] \ ,
\label{d2}
\end{equation}
where $S^k = \tfrac{1}{2}\s^k$ is the spin operator.  In terms of the SME Lagrangian couplings,
\begin{align}
A^e &= -a_0^e - m_e \,c_{00}^e + \ldots  \nonumber \\
B_k^e &= - b_k^e - m_e \,d_{k0}^e - \tfrac{1}{2} \e_{kij} H_{ij}^e  \nonumber \\
E_{ij}^e &= - m_e \left( c_{ij}^e + \tfrac{1}{2} c_{00}^e \,\d_{ij} \right)  \nonumber \\
F_{ijk}^e &= -\tilde{d}_i^e \,\d_{jk} + \tfrac{1}{2} \left(\d_{ij} b_k^e - \d_{ik}b_j^e\right) + \ldots 
\label{d3}
\end{align}
with $\tilde{d}_i^e = m_e \left(d_{0i}^e + \tfrac{1}{2} d_{i0}^e\right) - \tfrac{1}{4} \e_{ijk}H_{jk}^e$.
Analogous expressions hold for the antiproton, but with an overall extra factor of $\e = m_e^2/m_p^2$ 
multiplying $E_{ij}^p$ and $F_{ijk}^p$ to compensate for the use of the electron mass in the momentum
factor in (\ref{d2}). As in (\ref{b19}), we omit the SME coefficients 
$e_\m$, $f_\m$ and $g_{\m\n\l}$ essentially for simplicity of
presentation.\footnote{To restore these couplings, make the following substitutions:
\begin{align*}
A ~&\rta~ A  \,-\, m\, e_0   \nonumber \\
B_k ~&\rta~ B_k \,-\, \tfrac{1}{2}m\,\e_{kmn}\,g_{mn0}   \nonumber \\
F_{ijk} ~&\rta~ F_{ijk} \,-\, m\,\e_{ikm}(g_{m0j} \,+\, \tfrac{1}{2}g_{mj0})\ . 
\end{align*}
This means that in the transition energies, the $b_3$-dependence derived from $B_k$
always occurs in the combination $b_3 + m g_{120}$, while the corresponding terms 
derived from $F_{ijk}$ come in the combination $b_3 - m(g_{102} -g_{201} + g_{120})$. 
(See also footnote \ref{fSME}.)
\label{fHAM}}.
Compared with the corresponding Hamiltonian for hydrogen with electrons and protons, 
we have simply made the substitutions
$a_\m \rta - a_\m$, $b_\m \rta b_\m$, $c_{\m\n} \rta c_{\m\n}$, $d_{\m\n} \rta - d_{\m\n}$,
$H_{\m\n} \rta -H_{\m\n}$, $g_{\m\n\l}\rta g_{\m\n\l}$ dictated by the
$\mathsf{C}$ conjugates of the corresponding operators.

So far, we have considered only the operators of dimension $\le 4$ in the SME Lagrangian,
corresponding to a renormalisable theory (the {\it minimal} SME). 
However, as discussed in section \ref{sect 2.3}, we should rather view the SME as an 
effective field theory describing the low-energy dynamics of some UV-complete fundamental QFT. 
The effective theory also includes (non-minimal) operators of
dimension $> 4$, with the corresponding dimensional couplings being suppressed by
powers of $1/\L$, where $\L$ is the scale of the fundamental theory, and so provides a 
systematic expansion for corrections to the leading-order low-energy physics. 

A fully comprehensive and systematic account of these higher-dimensional operators 
and their contribution to hydrogen and antihydrogen spectroscopy is given in 
\cite{Kostelecky:2013rta, Kostelecky:2015nma}.
To illustrate this, we focus on one particular dimension 5 operator which will play a r\^ole in
describing the $1S$-$2S$ antihydrogen transitions, {\it viz.}
\begin{equation}
L_{\rm SME}^{(5)} = \int d^4 x ~ a_{\m\r\s}^{(5)} \, \bar\psi \c^\m D^\r D^\s \psi \ , 
\label{d4}
\end{equation}
and consider its relevant contribution to the non-relativistic Hamiltonian for antihydrogen, 
\begin{equation}
H_{\rm SME}^{(5)} = - \sum_{\omega = e,p} \, \left(\, a^{(5)\,\omega}\, m_\omega^2
\,+\, a_{ij}^{(5)\,\omega}\, p_i p_j  \,\right).
\label{d5}
\end{equation}
The coefficients denoted here by $a^{(5)}$ and $a_{ij}^{(5)}$ are related to the original couplings
$a_{\m\r\s}^{(5)}$ in (\ref{d4}) by\footnote{To see this, note that the new coupling in (\ref{d4}),
together with the original $a^{(3)}_\m$ coupling in the minimal SME (\ref{b19}),
corresponds to a substitution $p_\m \rightarrow p_\m - \hat{a}_\m$ in the momentum-space Dirac
equation, where $\hat{a}_\m = a^{(3)}_\m + a^{(5)}_{\m\r\s}\,p^\r p^\s$.
Extending the usual energy-momentum
relation for the modified Dirac equation, and reversing the sign of the $a$-couplings as appropriate
for antimatter, we readily find that to first order in the SME couplings,
the Hamiltonian acquires an extra contribution \cite{Kostelecky:2013rta}
\begin{equation*}
H_{\rm SME}^{(a)} ~=~ - \frac{1}{E_0} \, \hat{a}_\m \,p^\m \ , 
\end{equation*}
with $E_0 = \sqrt{{\bf p}^2 + m^2}$. Expanding in the non-relativistic approximation
$E_0 = m + {\bf p}^2/2m + \ldots$ and retaining only even powers of $p^i$ (which contribute to the
matrix elements for spectroscopy), we then find
\begin{equation*}
H_{\rm SME}^{(a)} ~=~ -\Bigl(a^{(3)}_0 \,+\, a^{(5)}_{000}\, m^2\Bigr) \,-\, 
\Bigl(a^{(5)}_{000}\, |{\bf p}|^2 \,+\, 3\, a^{(5)}_{0ij}\, p^i p^j \Bigr) \,+\, \ldots \ ,
\end{equation*}
where we have used the total symmetry of the $a^{(5)}_{\m\n\l}$ couplings \cite{Kostelecky:2013rta}.
The results in (\ref{d55}) follow immediately.} 
\begin{align}
a^{(5)} ~&=~ a^{(5)}_{000}   \nonumber \\
a^{(5)}_{ij} ~&=~ a^{(5)}_{000}\,\d_{ij}  \,+\, 3\, a^{(5)}_{0ij} \ .
\label{d55}
\end{align}
The terms in (\ref{d5}) can then be added to the
Hamiltonian (\ref{d2}) where they correspond to the substitutions $A \rta A - a^{(5)}\,m^2$ and 
$E_{ij} \rta E_{ij} - a_{ij}^{(5)}\, m^2$.  Comparing with the $c_{00}$ and $c_{ij}$ terms, 
we see that in the effective field theory expansion this term would be expected to be suppressed
by $O(m_e/\L)$. In the usual interpretation of the SME as the low-energy effective theory 
corresponding to a fundamental theory at the string or Planck scales, this is clearly a 
tiny factor. Nevertheless, we will carry it forward in the analysis of the antihydrogen
transitions below.

To include the contributions from the Lorentz and $\mathsf{CPT}$ violating operators in the 
photon sector in (\ref{b19}), we need a slightly different approach.  First note that these terms 
in the SME Lagrangian modify the photon propagator through two-point interactions 
$-2i(k_F)_{\m\r\n\s} q^\r q^\s$ and $2 (k_{AF})^\r \e_{\m\r\n\s}q^\s$ respectively (where $q^\m$ 
is the momentum of the photon propagator).   This modifies the photon mediated interaction 
between the antiproton and the positron, changing the energy levels of the atom. 
This can be calculated in the framework of non-relativistic QED \cite{Caswell:1985ui}, 
the result quoted in \cite{Yoder:2012ks} being a change in the energy of a given state $|\psi_n\rangle$ 
given by
\begin{equation}
E_{SME}^\gamma = \frac{\a^2}{n^2}\,\langle \psi_n|
m_e (k_F)_{0i0j} \left(\hat{p}_i \hat{p}_j - \d_{ij}\right)   + (k_{AF})_i L_i |\psi_n\rangle \ ,
\label{d5a}
\end{equation}
where $L_i$ is the positron orbital angular momentum and $\hat{p}_i$ is its 
unit 3-momentum vector. However, this only includes the {\it spin-independent} contributions
to the effective positron-antiproton potential and a complete analysis of the bound state
involves many new vertices in the NRQCD framework \cite{Labelle:1996en}. In particular,
including the photon coupling to the so-called Fermi vertex gives a further $k_{AF}$ dependent
contribution of (compare \cite{Gomes:2016wqj}),
\begin{equation}
E_{SME,\,spin}^\gamma = \frac{\a^2}{n^2}\,\langle \psi_n| (k_{AF})_i \,\left( S_i + \hat{p}_i \hat{p}_j S_j\right)
|\psi_n\rangle \ .
\label{d5b}
\end{equation}
Evidently, these terms can be absorbed into extra $k_{AF}$ dependent pieces in $B_k^e$ and $F_{ijk}^e$
in the Hamiltonian (\ref{d2}). Their contribution to the antihydrogen transitions considered below
can therefore easily be included, but we do not display them explicitly here since, with hindsight, $k_{AF}$
is known to be bounded far more stringently from astrophysics than the known bounds 
from atomic physics on the matter terms in (\ref{d2}), especially $b_k^e$.

\vskip0.5cm
\noindent $1S$ {\it hyperfine transitions}:
\vskip0.2cm

To calculate the contributions of these Lorentz and $\mathsf{CPT}$ violating couplings to the 
antihydrogen spectrum and the specific transitions measured by ALPHA, we need
first to describe the appropriate hyperfine states, remembering that in ALPHA the
anti-atoms are confined in a magnetic trap. Our notation is
${\bf F} = {\bf I} + {\bf J}$, where ${\bf I}$ is the antiproton spin and 
${\bf J} = {\bf L} + {\bf S}$, where ${\bf L}$ and ${\bf S}$ are the positron orbital and spin
angular momentum respectively. States are labelled by the corresponding quantum numbers
as $|n\, \ell\, j \, f \, m_f\rangle$.  Restricting initially to $\ell = 0$ states (for which $j = 1/2$),
we will use both this `hyperfine' basis $|f \, m_f\rangle$ and the `spin' basis $|m_I \, m_s\rangle$
as convenient. Following \cite{Ahmadi17}, we also denote the antiproton spin $m_I = \pm 1/2$
by $\Uparrow$, $\Downarrow $ and the positron spin $m_s = \pm 1/2$ 
by $\uparrow$, $\downarrow $.

The hyperfine interaction coupling the antiproton and positron spins is determined by the
Hamiltonian $H_{\rm hyp} =  \mathcal{E}_{HF}\, {\bf I}\,.\, {\bf S}$. An elementary calculation shows that the 
degeneracy of the $1S$ states is split, with $\D E = - (3/4) \mathcal{E}_{HF}$ for the singlet state 
$|f \, m_f\rangle = |0 \, 0\rangle$ while $\D E = (1/4) \mathcal{E}_{HF}$ for the triplet $|1 \, m_f\rangle$
with $m_f = \pm 1, 0$. $\mathcal{E}_{HF}$ is therefore the hyperfine splitting of the $f=1$ and $f=0$ states 
at zero magnetic field.  Explicitly, 
\begin{equation}
\mathcal{E}_{HF} = \frac{2}{3}\, \mu_0\, g_e \,\mu_B \,g_p \,\mu_N \,|\psi_{n00}(0)|^2 \ ,
\label{d5c}
\end{equation}
where $\m_B = e/2m_e$ is the Bohr magneton, $\m_N = e/2m_p$, and the $g$-factors
are approximately $g_e = 2.002$ and $g_p = 5.585$.  $\psi_{n00}(0)$ is the $nS$ wave function 
at the origin, and $|\psi_{n00}(0)|^2 = 1/(\pi n^3 a_0^3)$ where $a_0 = 1/(\a m_e)$ is the Bohr radius.

In a constant magnetic field ${\bf B}$, the Hamiltonian including the Zeeman coupling becomes
\begin{equation}
H_{\rm hyp} = \mathcal{E}_{HF} {\bf I}\,.\,{\bf S} - g_e \m_B\, {\bf S}\,.\,{\bf B}
+ g_p \m_N \, {\bf I}\,.\,{\bf B} \ .
\label{d5aa}
\end{equation}
The energy eigenstates at non-zero ${\bf B} = B {\bf e}_z$  are linear combinations of
the $m_f = 0$ states, found by  diagonalising the Hamiltonian (\ref{d5aa}).

In the spin basis, they are
\begin{align}
|d\rangle &= |\Downarrow \,\, \downarrow\rangle   \nonumber \\
|c\rangle &= \cos\theta_n |\Uparrow \,\, \downarrow \rangle + 
\sin\theta_n |\Downarrow \,\, \uparrow\rangle   \nonumber \\
|b\rangle &= |\Uparrow \,\, \uparrow \rangle   \nonumber  \\
|a\rangle &= -\sin\theta_n |\Uparrow \,\, \downarrow\rangle  +  
\cos\theta_n |\Downarrow \,\, \uparrow \rangle  ~~\ ,
\label{d6}
\end{align}
where the mixing angle $\theta_n(B)$ is given by
$\tan 2\theta_n = \mathcal{B}_0 /n^3 B,$ with 
$\mathcal{B}_0 = \mathcal{E}_{HF}/g_e \m_B (1+\e_p) \,\simeq 50.7\, {\rm mT}$\,\cite{Bluhm:1998rk}.  
Here, we have defined $\e_p = g_p m_e/g_e m_p$.
The corresponding form in the hyperfine basis is given in the footnote.\footnote{For convenience, 
we also give the explicit
form of these states in the hyperfine basis:
\begin{align*}
|d\rangle &= |1 \,\, -1\rangle   \nonumber \\
|c\rangle &= \tfrac{1}{\sqrt{2}} \left( \left(\cos\theta_n + \sin\theta_n\right) |1 \,\, 0\rangle 
- \left(\cos\theta_n - \sin\theta_n\right) |0 \,\, 0\rangle \right)  \nonumber \\
|b\rangle &= |1 \,\, 1\rangle   \nonumber \\
|a\rangle &= \tfrac{1}{\sqrt{2}} \left( \left(\cos\theta_n - \sin\theta_n\right) |1 \,\, 0\rangle 
+ \left(\cos\theta_n +\sin\theta_n\right) |0 \,\, 0\rangle \right) \ . \nonumber 
\end{align*}
\label{footnoteHYP} }

Clearly, for zero field we have $\cos\theta_n = 1/\sqrt{2}$ and the $|c\rangle$ and $|a\rangle$
states are just the hyperfine $|1\,\, 0\rangle$ and $|0 \,\, 0\rangle$ states respectively.
In ALPHA, the magnetic field is of order $B \sim 1$T which is in the high field regime
where $\cos\theta_n \simeq 1$, so the states are very well approximated by 
$|c\rangle \simeq |\Uparrow \,\, \downarrow\rangle$ and
$|a\rangle \simeq |\Downarrow \,\, \uparrow\rangle$.

\begin{figure}[h!]
\centering\includegraphics[scale=0.53]{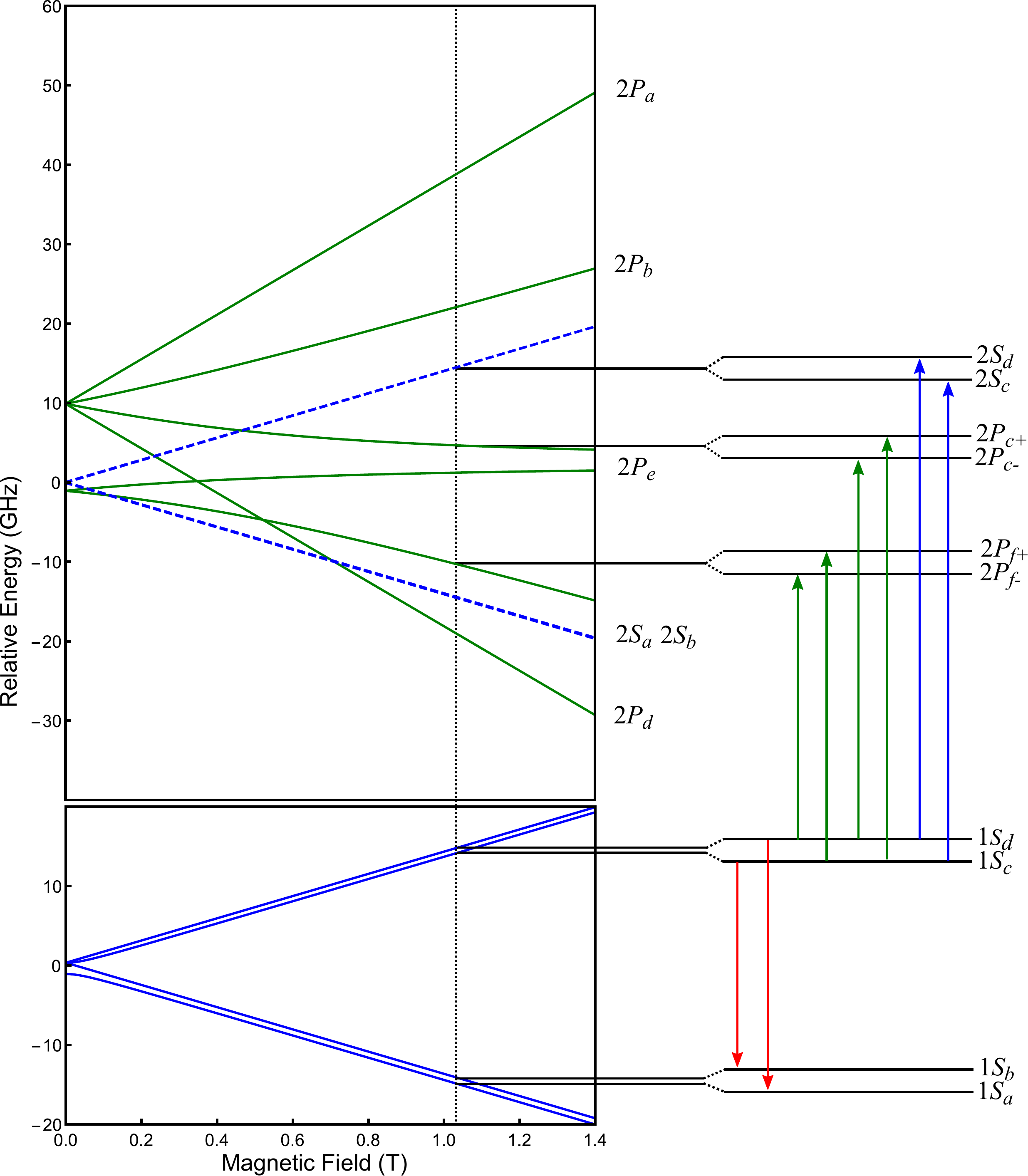}
\caption{The energy levels of the antihydrogen $1S$, $2S$  and $2P$ states in a uniform magnetic field, 
with the central magnetic field in the ALPHA apparatus indicated by a vertical dashed line. 
The origins of the vertical axis in the upper and lower panels are separated by the $1S$-$2S$ energy difference. 
Transitions discussed in the text are denoted by arrows from the initial state in experiments. 
Channels to final states in transitions excited during laser spectroscopy are omitted for clarity.}
\label{Fig 4}
\end{figure}

The corresponding energy levels are
\begin{align}
E_{d,b} ~&=~ \frac{1}{4} \mathcal{E}_{HF} \,\pm\, \frac{1}{2} g_e \m_B B (1-\e_p) \nonumber \\
E_{c,a} ~&=~ -\frac{1}{4} \mathcal{E}_{HF} \,\pm\, 
\frac{1}{2} \left[ \mathcal{E}_{HF}^2 \,+\, \left(g_e \m_B B (1+\e_p)\right)^2\,\right]^{1/2} \ ,
\label{d6a}
\end{align}
reproducing the well-known Breit-Rabi formula \cite{Rasmussen:2017pyn}\footnote{Here, 
we seek the leading order form of the states for calculating expectation values in the SME. 
We note that in order to work out precise energies and energy differences, such as those 
required for two-photon spectroscopy, corrections due to the difference in the magnetic moment 
and the diamagnetic term need to be taken into account.}.

These energy levels are shown for the $1S$ antihydrogen states in the lower panel of
Figure \ref{Fig 4}. In the ALPHA trap, only the two `low-field seeking' states
$|d\rangle$ and $|c\rangle$ are trapped, while the `high-field seeking' states
$|b\rangle$ and $|a\rangle$ escape the trap.

\vskip0.2cm
The ground state hyperfine energy levels receive corrections due to the Lorentz and $\mathsf{CPT}$ violating 
couplings at $O(1)$ in the fine structure constant. We therefore need only calculate the contributions 
of the $A$ and $B_k$ terms in (\ref{d2}) to these levels. For this, we need the expectation values of 
the antiproton and positron spin operators. These are most simply evaluated using the spin basis,
{\it e.g.}
\begin{align}
\langle c|\, S_z^e\,|c\rangle 
&= \left(\cos\theta_1 \langle\Uparrow\,\, \downarrow |
+ \sin\theta_1 \langle \Downarrow\,\, \uparrow|\right) S_z^e
\left(\cos\theta_1 |\Uparrow\,\, \downarrow\rangle
+ \sin\theta_1 | \Downarrow\,\, \uparrow \rangle \right) \nonumber \\
&= - \frac{1}{2}\, \cos 2\theta_1   \ ,
\label{d7}
\end{align}
using $S_z^e |\Uparrow\,\, \downarrow\rangle = - \tfrac{1}{2} |\Uparrow\,\, \downarrow\rangle$, etc.
The full set of SME corrections are then found to be
\begin{align}
E_{\rm SME}^d &= A^e + A^p - \left(B_3^e + B_3^p\right)   \nonumber \\
E_{\rm SME}^c &= A^e + A^p - \cos2\theta_1 \left(B_3^e - B_3^p\right)   \nonumber \\
E_{\rm SME}^b &= A^e + A^p + \left(B_3^e + B_3^p\right)   \nonumber \\
E_{\rm SME}^a &= A^e + A^p + \cos2\theta_1 \left(B_3^e - B_3^p\right) \ .
\label{d8}
\end{align}

At zero magnetic field, $\cos2\theta_1 \rta 0$, and the $|c\rangle$ and $|a\rangle$ states are
only shifted by the $A^e + A^p$ term common to all states. The $|c\rangle \rta |a\rangle$
transition is therefore insensitive to the Lorentz and $\mathsf{CPT}$ violating couplings in
the minimal SME and simply measures the hyperfine splitting $\mathcal{E}_{HF}$. 
In fact, this result holds even in the non-minimal theory \cite{Kostelecky:2015nma}.

The two transitions measured
by ALPHA \cite{Ahmadi17} are $|d\rangle \rta |a\rangle$ and $|c\rangle \rta |b\rangle$,
both from trapped to untrapped states (red arrows in Figure~\ref{Fig 4}).
The SME contributions to these transition frequencies are 
\begin{align}
\D E_{d\ltrta a}^{\bar{\rm H},\, {\rm SME}} &= \D E_{c\ltrta b}^{\bar{\rm H},\, {\rm SME}} \nonumber \\  
&= - \left(1 + \cos 2\theta_1\right) B_3^e
- \left(1 - \cos 2\theta_1\right) B_3^p \nonumber \\
&= - 2 \cos^2\theta_1 \, B_3^e  - 2 \sin^2\theta_1 \, B_3^p \ .
\label{d9}
\end{align}
Comparing with the corresponding transition for hydrogen, and recalling that the hydrogen
hyperfine states are spin-flipped relative to antihydrogen, we find
\begin{equation}
\D E_{d\ltrta a}^{\bar{\rm H},\, {\rm SME}}  -  \D E_{d\ltrta a}^{{\rm H},\, {\rm SME}}  
= 4 \cos^2\theta_1 \, b_3^e + 4 \sin^2\theta_1 \, b_3^p  \ ,
\label{d10}
\end{equation}
as first shown in \cite{Bluhm:1998rk}.

This shows that at leading order in $\a$, there is a predicted difference in the hydrogen
and antihydrogen $1S$ hyperfine transitions proportional at zero magnetic field to the
combination $b_3^e + b_3^p$ of SME couplings, while at high magnetic fields as in the
ALPHA trap this dependence is on the electron coupling $b_3^e$ alone.
Recalling that the $b_\m$ coupling in the SME Lagrangian is $\mathsf{CPT}$ odd, a difference 
between ${\rm H}$ and $\overline{\rm H}$ in these transitions could therefore be interpreted as a signal
of $\mathsf{CPT}$ violation. Assuming the energy levels described by (\ref{d6a}) are the same
for hydrogen and antihydrogen, and neglecting
the $O(\a^2)$ contributions from other SME couplings, a measurement in ALPHA of the individual
transition frequencies for $1S_d$\,-\,$1S_a$ or $1S_c$\,-\,$1S_b$ would therefore allow a
bound to be placed on the $\mathsf{CPT}$-violating electron coupling $b_3^e$.
A corresponding measurement at zero magnetic field, as proposed by ASACUSA 
\cite{Malbrunot:2017ncm}, would measure the combination $b_3^e + b_3^p$.

In fact, the published ALPHA measurement \cite{Ahmadi17},  with its quoted precision 
of $4\times 10^{-4}$, is of the {\it difference} of the transition frequencies 
$\D E_{d\ltrta a} - \D E_{c\ltrta b}$, {\it not} the individual transition frequencies.
The experimental procedure leading to this measurement is outlined above in section \ref{sect 3.2.1}.
This has the advantage that the result is independent of the magnetic field $B$, as is evident from
(\ref{d6a}), 
so any uncertainty in determining the magnetic field that affects both resonances equally is not relevant.
Unfortunately, from (\ref{d9}) we see that the dependence on the $\mathsf{CPT}$
violating parameters also cancels in this difference.  This would at first sight imply 
that the difference of the hyperfine transition frequencies does not lead to a determination of
the parameters $b_3^{e,p}$ and could not be considered as a test of $\mathsf{CPT}$.

However, what we have shown is that even in the Lorentz and $\mathsf{CPT}$ violating SME theory, 
the difference of the hyperfine transition frequencies is still
\begin{equation}
\D E_{d\ltrta a} - \D E_{c\ltrta b} ~=~ \mathcal{E}_{HF} \ ,
\label{d10aa}
\end{equation}
{\it i.e.}~the difference of the frequencies simply measures the hyperfine splitting
$\mathcal{E}_{HF}$. Now, from (\ref{d5c}), $\mathcal{E}_{HF}$ depends only on the mass and charge of the
positron and antiproton, together with the $g_e$ and $g_p$ factors.
As explained in section \ref{sect 2} and \ref{sect 3.1}, any difference in the mass or charge of a 
particle and its antiparticle leads to a violation of causality, and potentially unitarity or locality.
These properties are far more fundamental than $\mathsf{CPT}$ symmetry, and their preservation
is implicitly assumed in the formulation of the Lorentz and $\mathsf{CPT}$ violating SME effective
theory.  

This would, however, still leave the possibility that $g_e$ or $g_p$ could be different for the
electron/positron or for the proton/antiproton as an explanation for a difference in the hyperfine splitting
of antihydrogen and hydrogen. In this sense, the ALPHA measurement \cite{Ahmadi17} can be
viewed as a test of the identity of the anomalous magnetic moments for the particles and their antiparticles. 
Any difference would entail a violation of $\mathsf{CPT}$, but is not predicted
by the leading-order non-relativistic Hamiltonian derived from the SME.

\vskip0.5cm
\noindent $1S$\,-\,$2S$ {\it antihydrogen transitions}:
\vskip0.2cm
Next we consider the $1S$\,-\,$2S$ transition in antihydrogen (blue arrows in Figure~\ref{Fig 4}). 
This is forbidden by the usual dipole 
selection rules for a single-photon transition and instead takes place through the much slower
Doppler-free, two-photon transition. This gives rise to an extremely narrow spectral line
and should permit precise tests of Lorentz and $\mathsf{CPT}$ symmetry in antihydrogen spectroscopy.

The derivation of the relevant $1S$ and $2S$ hyperfine energy levels follows that given above
except that, as explained below, the leading correction to the $1S$\,-\,$2S$ transitions measured by 
ALPHA is $O(\a^2)$, so we need to consider both the $E_{ij}$ and $F_{kij}$ terms in the
Hamiltonian (\ref{d2}), along with the higher-dimensional operator contribution $H_{\rm SME}^{(5)}$
in (\ref{d5}). 

First note that at zero magnetic field, where $\cos 2\theta_n = 0$, the hyperfine energy levels
(\ref{d8}) can be written as
\begin{equation}
E_{\rm SME} = A^e + A^p + \left(B_3^e + B_3^p\right) m_f \ ,
\label{d10a}
\end{equation}
in the $|f\,\,m_f\rangle$ basis, independent of $n$. The two-photon selection rule 
$\D f = 0$, $\D m_f = 0$ therefore shows that these SME corrections give no contribution to the
frequency $\D E_{2S\ltrta 1S}$ of the $1S$\,-\,$2S$ transition.  This is not, however, true in a magnetic field due 
to the $n$-dependence of the mixing angle $\theta_n$ in (\ref{d6}) and footnote \ref{footnoteHYP}.
Only in the high-field limit, where $\cos2\theta_n \rta 1$, does the $1S$\,-\,$2S$ transition frequency
again become independent of the $A$ and $B_k$ terms in the SME Hamiltonian.
Now consider the $O(p_i p_j)$ terms in the Hamiltonian (\ref{d2}). 
To compute their contribution to the hyperfine energy levels, we need to calculate expectation
values of the form $\langle d| p_i p_j |d\rangle$, $\langle d| p_i p_j S_k^e|d\rangle$, {\it etc.}~for
all the states and for both the antiproton and positron spin operators. 

At this point, we draw attention to the extremely detailed and clear calculation of the energy levels
derived from $H_{\rm SME}$ for an arbitrary state $|n\, \ell\, j\, f\, m_f\rangle$ presented in 
\cite{Yoder:2012ks}.  This paper derives the necessary Clebsch-Gordan coefficients for the 
evaluation of the expectation values of all the required combinations of the momentum and spin 
operators in these states.  However, while this analysis is complete for calculating energy levels 
at zero magnetic field, we also require expectation values in the mixed hyperfine states,
especially $|c\rangle$, to compute energy levels in strong magnetic fields such as those in the
ALPHA trap. For these, we would also need {\it off-diagonal} matrix elements between states
with differing $f$, $m_f$ (see footnote \ref{footnoteHYP}) in addition to those quoted in 
\cite{Yoder:2012ks}. For states with arbitrary $\ell, j, f$ this is a substantial calculation.

Fortunately, the situation is enormously simplified for the $\ell = 0$ states, for which the
wave function is isotropic, and it is then again straightforward to use the spin basis.
Following \cite{Yoder:2012ks}, we first note
\begin{equation}
\langle n| p_i p_j |n \rangle = \frac{1}{3} \d_{ij} \langle n | p^2 |n\rangle 
= \frac{1}{3} \d_{ij} \frac{\a^2 m_e^2}{n^2} \ ,
\label{d11}
\end{equation}
which is independent of the spin of the state.  (Recall the Bohr energy levels
are $E_n = -\a^2 m_e/2n^2$.) 
The simplification is then clear. As an illustration, we can evaluate one of the expectation values 
contributing to $E_{\rm SME}^c$ in the same way as in (\ref{d7}), {\it viz.}
\begin{align}
\langle c| p_i p_j S_k^e|c\rangle 
&= - \frac{1}{2} \cos^2\theta_n \,\langle \Uparrow\,\, \downarrow | p_i p_j |\Uparrow\,\, \downarrow\rangle
+ \frac{1}{2} \sin^2\theta_n \,\langle \Downarrow\,\, \uparrow | p_i p_j |\Downarrow\,\, \uparrow\rangle
\nonumber \\
&= - \frac{1}{6} \d_{ij}  \left(
\cos^2\theta_n \,\langle \Uparrow\,\, \downarrow | p^2 |\Uparrow\,\, \downarrow\rangle
- \sin^2\theta_n \,\langle \Downarrow\,\, \uparrow | p^2|\Downarrow\,\, \uparrow\rangle \right)
\nonumber \\
&= - \frac{1}{6} \d_{ij} \cos 2\theta_n \,\, \frac{\a^2 m_e^2}{n^2} \ .
\label{d12}
\end{align}

Ultimately, we find the energy levels for the trapped hyperfine states:
\begin{align}
E_{\rm SME}^d~ &= ~\tilde{A}^e + \tilde{A}^p - \left(B_3^e + B_3^p\right)  \nonumber \\
&~~~~+ \frac{1}{3} \frac{\a^2}{n^2} \, \tr_{i,j} \left( \tilde E_{ij}^e + \e \tilde E_{ij}^p
- \left(F_{ij3}^e + \e F_{ij3}^p\right) \right)
\label{d13a}
\end{align}
and 
\begin{align}
E_{\rm SME}^c ~&=~ \tilde{A}^e + \tilde{A}^p - \cos 2\theta_n \left(B_3^e - B_3^p\right)  \nonumber \\
&~~~~+ \frac{1}{3} \frac{\a^2}{n^2} \, \tr_{i,j} \left( \tilde E_{ij}^e + \e \tilde E_{ij}^p
- \cos 2\theta_n \left(F_{ij3}^e - \e F_{ij3}^p\right) \right) \ ,
\label{d13b}
\end{align}
where $\tilde{A}= A - m^2 a^{(5)}$ and
$\tilde E_{ij} = E_{ij} - m^2 a_{ij}^{(5)}$.  
The $c_{ij}$ traces simplify,
since by suitable field redefinitions in the SME Lagrangian we are free to take
$ c^\m{}_\m = 0$.

The photon sector contributions are easily included using (\ref{d5a}). Note that $k_{AF}$ does not contribute 
here since we are considering $\ell = 0$ states. The $k_F$ contribution is readily evaluated for these 
states using (\ref{d11}) and is proportional to the trace $\kappa_0 = (k_F)_{0i0i}$.

Substituting in terms of the original SME couplings, we therefore find:
\begin{align} 
E_{\rm SME}^d 
= &-a_0^e - m_e^2\,a^{(5)\,e} - m_e\, c_{00}^e - a_0^p - m_p^2\,a^{(5)\,p} - m_p\, c_{00}^p  \nonumber \\
&+ b_3^e + m_e \, d_{30}^e +  H_{12}^e + b_3^p + m_e \, d_{30}^p + H_{12}^p \nonumber \\
&- \frac{1}{3} \frac{\a^2}{n^2}\,\bigg[ m_e^2 \left(a_{ii}^{(5)e} + a_{ii}^{(5)p} \right) 
+ \tfrac{5}{2} \left( m_e\, c_{00}^e + \e m_p\, c_{00}^p\right)   \nonumber \\
&~~~~~~~~~~~~~~~~~~~~+ 2 m_e \kappa_0  + \left( b_3^e - \tilde{d}_3^e  + \e \left(b_3^p 
-\tilde{d}_3^p\right) \right) \bigg] \ ,
\label{d14}
\end{align}
and
\begin{align}
E_{\rm SME}^c 
= &-a_0^e - m_e^2\,a^{(5)\,e} - m_e\, c_{00}^e - a_0^p - m_p^2\,a^{(5)\,p} - m_p\, c_{00}^p  \nonumber \\
&+ \cos 2\theta_n\left( b_3^e + m_e \, d_{30}^e +  H_{12}^e 
- b_3^p - m_e \, d_{30}^p - H_{12}^p \right)\nonumber \\
&- \frac{1}{3}\frac{\a^2}{n^2}\,\bigg[ m_e^2 \left(a_{ii}^{(5)e} + a_{ii}^{(5)p} \right) 
+ \tfrac{5}{2} \left( m_e\, c_{00}^e + \e m_p\, c_{00}^p\right) \nonumber \\
&~~~~~~~~~~~~~~~~~+2 m_e \kappa_0 + 
\cos 2\theta_n \left( b_3^e - \tilde{d}_3^e  - \e \left(b_3^p - \tilde{d}_3^p \right) \right) \bigg] \ ,
\label{d15}
\end{align}
where $\tilde{d}_3 = m d_{03} + \tfrac{1}{2} m d_{30}-\tfrac{1}{2} H_{12}$.\footnote{
To compare with \cite{Kostelecky:2013rta, Kostelecky:2015nma} 
(see also \cite{Vargas:2018isj}), these references rewrite the
SME tensor couplings in a spherical basis (after suitable field redefinitions) then
consider the isotropic components, which are sufficient to describe the $\ell = 0$ states. 
In their notation, with effective non-relativistic couplings $a^{\rm NR}_{njm}$,
the dictionary to compare with (\ref{d14})--(\ref{d16a}) is,
\begin{equation*}
\mathring{a}_0^{\rm NR} ~\equiv~ 
\tfrac{1}{\sqrt{4\pi}}\, a_{000}^{\rm NR}
~=~ a^{(3)}_0 + a^{(5)}_{000}\, m^2 ~=~a_0 + a^{(5)} \ ,
\end{equation*}
and 
\begin{equation*}
\mathring{a}_2^{\rm NR} ~\equiv~
\tfrac{1}{\sqrt{4\pi}}\, a_{200}^{\rm NR} ~=~ a^{(5)}_{000} + a^{(5)}_{0ii}  ~=~ 
\tfrac{1}{3} \tr\,a_{ij}^{(5)} \ ,
\end{equation*}
the first equalities giving the relation to the Lagrangian couplings in Cartesian coordinates 
and the second to our notation (see (\ref{d55})).
Similarly, 
\begin{equation*}
\tfrac{1}{\sqrt{4\pi}}\, c_{000}^{\rm NR} ~=~ m\,c_{00} \ ,
\end{equation*}
and
\begin{equation*}
\mathring{c}_2^{\rm NR} ~\equiv~
\tfrac{1}{\sqrt{4\pi}}\, c_{200}^{\rm NR} ~=~ 
\tfrac{1}{m}\,\left(\tfrac{1}{2} c_{00} + \tfrac{1}{3} c_{ii}\right)  ~=~ 
\tfrac{5}{6}\tfrac{1}{m}c_{00} \ ,
\end{equation*}
if we apply the condition $\tr\, c_{\m\n}=0$.  In this notation, the $c$ and $a$-dependent
contributions to the antihydrogen $1S$\,-\,$2S$ transitions (\ref{d16}), (\ref{d16a})
are written as \cite{Kostelecky:2015nma}, 
\begin{equation*}
\D E_{2S\ltrta 1S}^{\bar{\rm H}} ~=~ \frac{3}{4}\a^2 m_e^2 \,\frac{1}{\sqrt{4\pi}}\,
\sum_{\omega = e,p} \left( c_{200}^{{\rm NR}\omega} \,+\, a_{200}^{{\rm NR}\omega} \right) ~~+~~ 
O(b,d,H,\kappa) \ .
\end{equation*}
Later, when we consider
the $1S$\,-\,$2P$ transitions, we find a further dependence on the combination 
$a^{(5)}_Q = a^{(5)}_{11} + a^{(5)}_{22} -2 a^{(5)}_{33}$  of the $a^{(5)}_{ij}$,
which is proportional to the non-isotropic couplings $a_{220}^{\rm NR}$. 
Similarly for $c_Q$. \label{KVcouplings}}

This reproduces the relevant parts of eq.~(37) of \cite{Yoder:2012ks} for the special case of $\ell=0$
states in $\overline {\rm H}$, and also includes the magnetic field dependence of the hyperfine state $|c\rangle$.
Note that \cite{Yoder:2012ks} includes also the $e,f,g$ SME couplings which have been omitted here.

Finally, we can read off the SME contributions to the $1S_d$\,-\,$2S_d$
and $1S_c$\,-\,$2S_c$ transitions in antihydrogen, as measured by ALPHA \cite{ALPHA1S2S}
(see Figure \ref{Fig 4}) :
\begin{align}
\D E_{2S_d\ltrta 1S_d}^{\bar{\rm H} } &\,=\, \frac{1}{4} \a^2\,\bigg[
m_e^2 \left(a_{ii}^{(5)e} + a_{ii}^{(5)p} \right)  \,+\,2 m_e \kappa_0  \nonumber \\
&~~~~~~~~~~~~+\tfrac{5}{2}\left(m_e\, c_{00}^e + \e m_p \, c_{00}^p\right)
+ \left( b_3^e - \tilde{d}_3^e  + \e b_3^p -\e \tilde{d_3}^p  \right) \bigg]  \ ,
\label{d16}
\end{align}
and
\begin{align}
\D E_{2S_c\ltrta 1S_c}^{\bar{\rm H} } \,=\,
&\,(\cos 2\theta_2 - \cos 2\theta_1) \left(b_3^e + m_e \, d_{30}^e +  H_{12}^e 
- b_3^p - m_e \, d_{30}^p - H_{12}^p \right)\nonumber \\
&+ \frac{1}{4} \a^2\,\bigg[
m_e^2 \left(a_{ii}^{(5)e} + a_{ii}^{(5)p} \right)  \,+\,2 m_e \kappa_0  
+\frac{5}{2}\left(m_e\, c_{00}^e + \e m_p \, c_{00}^p\right) \nonumber \\
&~~~~~~~~~~~~~~~~~~~ - \frac{1}{3}\left(\cos 2\theta_2 - 4\cos 2\theta_1\right)
\left( b_3^e - \tilde{d}_3^e -\e b_3^p  + \e \tilde{d_3}^p  \right) \bigg] \ .
\label{d16a}
\end{align}

Although $\cos 2\theta_n \simeq 1$ at the high magnetic field in the ALPHA trap,
we have kept this dependence here. It shows an important feature, {\it viz.}~that unlike
the $|d\rangle \rta |d\rangle$ transition, the $|c\rangle \rta |c\rangle$ transition 
has a contribution at $O(1)$ in the fine structure constant, albeit highly suppressed by
a magnetic field factor. We come back to this below.
At zero field, the $\tilde{d}_3$ and $b_3$ terms do not contribute to the $E_{\rm SME}^c$ energy levels,
both then being proportional to $m_f$.
In either case, we see from (\ref{d16}) and (\ref{d16a}) that the SME contributions to the 
$c$ and $d$ transitions are different. 

To compare with the $1S$\,-\,$2S$ transitions in hydrogen, again recalling that the corresponding states
are spin-flipped, we make the $\mathsf{CPT}$ conjugation sign changes on the relevant SME couplings
as described above and find:
\begin{equation}
\D E_{2S_d\ltrta 1S_d}^{\bar{\rm H} } - \D E_{2S_d\ltrta 1S_d}^{{\rm H} }  
=~ \frac{1}{2} \a^2 \left[ m_e^2 \left( a_{ii}^{(5)\,e}  + a_{ii}^{(5)\,p}\right)
+ b_3^e + \e b_3^p \right] \ ,
\label{d17} 
\end{equation}
and
\begin{align}
\D E_{2S_c\ltrta 1S_c}^{\bar{\rm H} } &- \D E_{2S_c\ltrta 1S_c}^{{\rm H} } ~=~
2\left(\cos 2\theta_2 - \cos 2\theta_1\right) \left(b_3^e - b_3^p\right) \nonumber \\
&~~~~~~~~~~~~+ \frac{1}{2} \a^2 \bigg[ m_e^2 \left( a_{ii}^{(5)\,e}  + a_{ii}^{(5)\,p}\right) 
- \frac{1}{3} \left(\cos2\theta_2 - 4\cos2\theta_1\right) 
\left(b_3^e - \e b_3^p\right) \bigg] \ .
\label{d17a} 
\end{align}
Clearly, this only depends on the $\mathsf{CPT}$ odd couplings in the SME Lagrangian. 
Again note the $O(1)$ contribution in the $|c\rangle \rta |c\rangle$ transition only.
We comment on the significance of these results on $1S$-$2S$ spectroscopy for fundamental physics
below, after first considering other transitions accessible to the ALPHA programme.

\vskip0.4cm
\noindent $1S$\,-\,$2P$ {\it and other antihydrogen transitions}:
\vskip0.2cm

ALPHA have also recently measured the $1S$\,-\,$2P$ transition in antihydrogen 
\cite{ALPHA1S2P, Ahmadi:2020ael} (green arrows in Figure~\ref{Fig 4}),
the first involving a state with non-zero orbital angular momentum. As such, it has some extra
interest from a fundamental physics perspective since it is directly sensitive to the potential 
spin-independent $\mathsf{CPT}$ violation (\ref{d5a}) in the photon sector, parametrised in the SME 
by the effective coupling $k_{AF}$. 

In the absence of an external magnetic field, the $2P$ states are split by the spin-orbit coupling
into a $j=3/2$ quartet and a $j=1/2$ doublet, with energy difference $\mathcal{E}_{FS}$. 
With non-zero $B$, Zeeman splitting removes the remaining degeneracy with respect to $m_j$,
with the $m_j = 1/2$ states with $j=3/2,1/2$ being mixed and similarly for the $m_j=-1/2$ states.
The first step is therefore to determine these energy eigenstates and mixing angles for the
magnetic fields present in the ALPHA trap.

In this case, it is a good approximation to neglect the hyperfine splitting, which is relatively small
for the $2P$ states, and include the $m_I = \pm 1/2$ antiproton spin only after finding the 
positron eigenstates. The effective Hamiltonian, including the spin-orbit coupling, is then simply
\begin{align}
H_{SO} &= \frac{2}{3} \mathcal{E}_{FS} \, {\bf L}\,.\,{\bf S} - \m_B\left( {\bf L} + g_e {\bf S}\right)\,.\,{\bf B} 
\nonumber \\
&= \frac{1}{3} \mathcal{E}_{FS} \left( J^2 - L^2 - S^2\right) - \m_B \left(L_z + g_e S_z\right)\,B
\label{d18}
\end{align}
Neither $|n\, \ell\, s\, j\, m_j\rangle$ nor $|n\, \ell \, s\, m_{\ell}\, m_s\rangle$ states are eigenstates
of $H_{SO}$ and either basis can be used to describe the mixed states at non-zero $B$.
Since the magnetic field in ALPHA is relatively high, and with an eye to the inclusion of 
SME couplings, we find it more convenient to describe the states in the $|m_\ell\,\, m_s\rangle$ basis.
Note also that $m_j = m_\ell + m_s$ is a good quantum number for $H_{SO}$, since $[H_{SO},J_z] =0$,
but $j$ is not. This selects the mixed states as described above and, after diagonalising the Hamiltonian,
we find the following $2P$ eigenstates and corresponding energy eigenvalues in the 
$|m_\ell\,\,m_s\rangle$ basis:
\begin{align}
|a\rangle ~&=~ |-1\,\, -\thalf\rangle   \nonumber \\
|b\rangle ~&=~ \cos\psi\, |-1\,\, \thalf\rangle + \sin\psi\, |0\,\,-\thalf\rangle \nonumber \\
|c\rangle ~&=~ \sin\eta\, |1\,\, -\thalf\rangle + \cos\eta\, |0\,\, \thalf\rangle  \nonumber \\
|d\rangle ~&=~ |1\,\, \thalf\rangle  \nonumber \\
|e\rangle ~&=~ - \sin\psi\, |-1\,\, \thalf\rangle + \cos\psi\, |0\,\, -\thalf\rangle \nonumber \\
|f\rangle ~&=~ \cos\eta\, |1\,\, -\thalf\rangle - \sin\eta\, |0\,\, \thalf\rangle
\label{d19}
\end{align}
where 
\begin{equation}
\tan\psi = \frac{1}{2\sqrt{2} \mathcal{E}_{FS}} \,\left(\mathcal{E}_{FS} + 3 \m_B B + 6 \mathcal{E}_1(B) \right) \ ,
\label{d20}
\end{equation}
and
\begin{equation}
\tan\eta = \frac{1}{2\sqrt{2} \mathcal{E}_{FS}}\,\left(-\mathcal{E}_{FS} + 3 \m_B B + 6 \mathcal{E}_1(-B) \right) \ ,
\label{d21}
\end{equation}
with
\begin{equation}
\mathcal{E}_1(B) = 
\tfrac{1}{2} \left[\mathcal{E}_{FS}^2 + \tfrac{2}{3} \m_B B\, \mathcal{E}_{FS}  + (\m_BB)^2 \right]^{1/2}
\label{d22}
\end{equation}
and we have set $g_e=2$.\footnote{These results
agree with those in \cite{Rasmussen:2017pyn}, up to a different choice of phase (sign) for the $|e\rangle$ state 
and an alternative definition of the mixing angle for $|b\rangle,\,|e\rangle$. With our choice,
$|e\rangle$ becomes the $|j=1/2,\,m_j=1/2\rangle$ state at $B=0$.}

The corresponding energy levels of $H_{SO}$ are:
\begin{align}
E_a ~&=~   \frac{1}{3} \mathcal{E}_{FS} + 2 \m_B B    \nonumber \\
E_b ~&=~    -\frac{1}{6} \mathcal{E}_{FS} + \frac{1}{2} \m_B B + \mathcal{E}_1(B)  \nonumber \\
E_c ~&=~    -\frac{1}{6} \mathcal{E}_{FS} - \frac{1}{2} \m_B B + \mathcal{E}_1(-B)  \nonumber \\
E_d ~&=~    \frac{1}{3} \mathcal{E}_{FS} - 2 \m_b B   \nonumber \\
E_e ~&=~    -\frac{1}{6} \mathcal{E}_{FS} + \frac{1}{2} \m_B B - \mathcal{E}_1(B)  \nonumber \\
E_f ~&=~    -\frac{1}{6} \mathcal{E}_{FS} - \frac{1}{2} \m_B B - \mathcal{E}_1(B). 
\label{d23}
\end{align}
These are illustrated in green in the upper panel of Figure~\ref{Fig 4}, with reference to the $2S$ level 
including the Lamb shift. 
Clearly, for zero magnetic field, $E_{a,\ldots d} = \tfrac{1}{3}\mathcal{E}_{FS}$, and
$E_{e,f} = - \tfrac{2}{3} \mathcal{E}_{FS}$, consistent with their interpretation as the $j=3/2$ and
$j=1/2$ states respectively.  The corresponding limits for the mixing angles are
$\tan\psi = \sqrt{2}$ and $\tan\eta = 1/\sqrt{2}$, which reproduce the required Clebsch-Gordan factors 
in (\ref{d19}) to convert between $|j\,\, m_j\rangle$ and $|m_\ell\,\, m_s\rangle$ states. 

For very large fields, the mixing angles both go to the limit $\pi/2$, and the limiting form of the
states can be immediately read off from (\ref{d19}).  For these $2P$ states, however, the ALPHA
magnetic field $B= 1.03\,{\rm T}$ does not fully reach this limit. In fact, at this value of $B$, 
the mixing angles are $\tan\psi \simeq 3.76$ and $\tan\eta \simeq 2.49$. These imply the following
values which we need below to parametrise the contributions from Lorentz and CPT violating parameters,
$\cos 2\psi = -0.868$ and $\cos 2\eta = -0.721$.  

Finally, we include the antiproton spin by simply taking the direct product with each of these
states, with notation $|a+\rangle = |a\rangle\,|\Uparrow\rangle$, 
$|a-\rangle = |a\rangle\,|\Downarrow\rangle$, {\it etc}.

\vskip0.2cm

The transitions of interest to us here are $1S_d$\,-\,$2P_{c-}$,
$1S_d$\,-\,$2P_{f-}$, $1S_c$\,-\,$2P_{c+}$ and $1S_c$\,-\,$2P_{f+}$, with the notation 
in (\ref{d6}) for the $1S$ hyperfine states. 
Of these, the first two have recently been measured by ALPHA \cite{Ahmadi:2020ael},
with the hyperfine states resolved.
To find the contribution to these transition frequencies from the Lorentz and $\mathsf{CPT}$ couplings, 
we first need to find the expectation value of the effective SME Hamiltonian (\ref{d2}) in these
$2P$ states. Evidently, there is a contribution already at $O(1)$ in the fine structure constant, 
but we shall give a complete result up to $O(\a^2)$ including also the photon sector couplings. 

Technically, the new feature arising with the $2P$ states is that since they have a non-zero angular
momentum, the wave functions are no longer isotropic and we cannot use the simplification (\ref{d11})
for the expectation values $\langle \hat{p}_i \hat{p}_j\rangle$. To overcome this \cite{Yoder:2012ks},
we first express $\hat{p}_i \hat{p}_j$ in a spherical tensor basis, defining coefficients $C_{ij}^M$
from the expansion in spherical harmonics,
\begin{equation}
\hat{p}_i \hat{p}_j - \frac{1}{3} \d_{ij} = \sum_{M=-2}^2\, C_{ij}^M Y_2^M(\theta,\phi) \ .
\label{d24}
\end{equation}
Matrix elements in an $|n\,\,\ell\,\,m_\ell\rangle$ basis are then found using a well-known formula for the
product of three spherical harmonics in terms of Clebsch-Gordan coefficients. The result is,
\begin{align}
\langle n\,\,\ell\,\,m_\ell\,|\hat{p}_i \hat{p}_j| n\,\,\ell\,\,m'_\ell\rangle 
~=~  &\frac{1}{3} \,\d_{ij} \,\d_{m_\ell m'_\ell}  \nonumber \\
&+ \sqrt{\frac{5}{4\pi}}\, C_{ij}^{m_\ell - m'_\ell} \, \langle \ell\,\,0\,;\,2\,\,0\,|\,\ell\,\,0\rangle \,
\langle \ell\,\,m'_\ell\,; \,2\,\, m_\ell - m'_\ell \,|\,\ell\,\,m_\ell\rangle \ .
\label{d25}
\end{align}

For example, for the diagonal matrix elements in (\ref{d25}), specialising to $\ell=1$ and 
$m_\ell = m'_\ell$, we have
\begin{equation}
\langle 2\,1\,m_\ell\,|\hat{p}_i \hat{p}_j| 2\,1\,m_\ell\rangle 
~=~  \frac{1}{3} \,\d_{ij}  - \sqrt{\frac{1}{2\pi}}\, C_{ij}^0 \,
\langle 1\,\,m_\ell\,; \,2\,\, 0 \,|\,1\,\,m_\ell\rangle \ .
\label{d26}
\end{equation}
The relevant coefficients are $C_{ij}^0 = -\sqrt{\frac{4 \pi}{45}} \,\d_{ij}^Q$ with 
$\d_{ij}^Q = \d_{i1}\d_{j1} + \d_{i2}\d_{j2} - 2 \d_{i3}\d_{j3}$ \cite{Yoder:2012ks}, 
so we find the required expectation value,
\begin{equation}
\langle 2\,1\,m_\ell\,|\hat{p}_i \hat{p}_j| 2\,1\,m_\ell\rangle 
~=~  \frac{1}{3} \,\d_{ij}  + \frac{1}{3}\sqrt{\frac{2}{5}}\, 
\langle 1\,\,m_\ell\,; \,2\,\, 0 \,|\,1\,\,m_\ell\rangle \, \d_{ij}^Q \ .
\label{d27}
\end{equation}
Analogous results hold for the $m_\ell \neq m'_\ell$ matrix elements, with
$C_{ij}^{\pm 1} = \sqrt{\frac{2\pi}{15}} \, \d_{ij}^{P\pm}$, where
$\d_{ij}^{P\pm} = \left( \mp \d_{i1} \d_{j3} + i \d_{i2} \d_{j3} 
+ i \leftrightarrow j\right)$.

One of the simplifications of expressing the $2P$ Zeeman states (\ref{d19})
in the $|m_\ell\,\,m_s\rangle$ basis is that 
it is now straightforward to evaluate all the required matrix elements such as 
$\langle \hat{p}_i \hat{p}_j\rangle$, $\,\langle \hat{p}_i \hat{p}_j S_z^e\rangle$, {\it etc.} in these states. 
The photon sector contributions from (\ref{d5a}) are also readily incorporated using these
matrix elements. Here, we simply quote the final results for the transition frequencies 
for $1S_d$\,-\,$2P_{c_-} $ and $1S_d$\,-\,$2P_{f_-}$.  For
$1S_d$\,-\,$2P_{c_-} $,
we have\footnote{Note that for simplicity we have set 
$\cos2\theta_1 \rta 1$ here, since for the ALPHA magnetic field, $\cos 2\theta_1 \simeq 0.998$.}
\begin{align}
\D E^{\bar{H}}_{2P_{c-}\ltrta 1S_d} =
&- (1+ \cos 2\eta) \left(b_3^e + m_e d_{30}^e + H_{12}^e\right) \nonumber \\
&+ \frac{1}{4} \a^2\bigg[-m_e^2\left(a^{(5)e}_{ii} + a^{(5)p}_{ii}\right) 
- \frac{1}{30} m_e^2 \left(1 + 3 \cos 2\eta\right) \left(a^{(5)e}_Q + a^{(5)p}_Q\right)    \nonumber \\
&~~~~~~~~~~~+\frac{5}{2}\left(m_e c_{00}^e + \e m_p c_{00}^p\right) 
+ \frac{1}{30} \left(1 + 3 \cos2\eta\right) \left(m_e c_Q^e + \e m_p c_Q^p\right) \nonumber \\
&~~~~~~~~~~~+ \frac{1}{3}\left(4 + \cos2\eta\right) \left(b_3^e - \tilde{d}_3^e \right) 
+ \frac{1}{3}\e \left(b_3^p -\tilde{d}_3^p \right) \nonumber \\
&~~~~~~~~~~~-\frac{1}{30} \left(3 + \cos2\eta\right) \left(b_3^e + 2 \tilde{d}_3^e \right) 
+\frac{1}{30} \e \left(1 +3 \cos2\eta\right) \left(b_3^p +2 \tilde{d}_3^p\right)  \nonumber \\
&~~~~~~~~~~~~+ \frac{\sqrt{2}}{10}\,\sin2\eta \, \left(b_3^e + 2 \tilde{d}_3^e\right) \nonumber \\
&~~~~~~~~~~~+\,2 m_e \k_0 \,-\, \frac{1}{30} \left(1 +3\cos2\eta\right) m_e \k_Q 
\,+\,\frac{1}{2} \left(1 - \cos 2\eta\right) (k_{AF})_3 ~\bigg] \ , 
\label{d28}
\end{align}
where $a^{(5)}_Q = a^{(5)}_{ij}\d_{ij}^Q$, $c_Q = c_{ij}\d_{ij}^Q$ and $\k_Q = (k_F)_{0i0j}\d_{ij}^Q$. 
Recall (footnote \ref{KVcouplings}) that these non-isotropic couplings $a^{(5)}_Q$, $c_Q$ are proportional to the
couplings denoted $a^{\rm NR}_{220}$, $c^{\rm NR}_{220}$ respectively in \cite{Kostelecky:2013rta, Kostelecky:2015nma}.
An identical result holds for $\D E^{\bar{H},SME}_{2P_{f-}\ltrta1S_d}$
with the substitutions $\cos2\eta \rta - \cos2\eta$ and $\sin2\eta \rta - \sin2\eta$
throughout, and similar expressions can be derived for transitions involving $1S_c$. 
Experimentally, while inhomogenous broadening of the $1S$\,-\,$2P$ line initially obscured 
the hyperfine structure \cite{ALPHA1S2P}, 
state selectivity has since been achieved by ejecting the unwanted $1S_c$ hyperfine population from the trap 
before spectroscopy begins \cite{Ahmadi:2020ael}.

Notice that for the magnetic field in ALPHA, both transitions have $O(1)$ contributions from the SME couplings 
dependent on the positron spin, {\it viz.} $b_3^e$, $d_{30}^e$ and $H_{12}^e$, but not (in this approximation) 
from the corresponding antiproton couplings. At still larger magnetic fields, where $\cos2\eta \rta -1$,
these contributions would cancel out in the $1S_d$\,-\,$2P_{c_-}$ transition only. Also recall that at zero field,
$\cos2\eta = 1/3$.

The other qualitatively new feature of (\ref{d28}) compared to the $1S$\,-\,$2S$ transition is in the photon sector, 
where the $1S$\,-\,$2P$ transitions become sensitive to the $\mathsf{CPT}$ violating coupling $(k_{AF})_3$,
as well as an independent combination $\k_Q$ of the $\mathsf{CPT}$ even $(k_F)_{0i0j}$ couplings.
Note also the appearance of the new Lorentz violating combination $c_Q^e$ in the positron sector.

As always, to expose the difference with the corresponding transitions in hydrogen, we keep only the
$\mathsf{CPT}$ odd couplings, {\it viz.} $a^{(5) e,p}_{ij}$, $b_3^{e,p}$ and $(k_{AF})_3$. This leaves 
the much simpler formula (approximating $\e=0$ here),
\begin{align}
&\D E^{\bar{H}}_{2P_{c-}\ltrta 1S_d} - \D E^H_{2P_{c-}\ltrta 1S_d}  ~=~ -2 (1 + \cos2\eta) \,b_3^e \nonumber \\
&~~~~~~~~~~~~~~~~~~~+\frac{1}{2} \a^2 
\bigg[ -m_e^2\left(a^{(5)e}_{ii} + a^{(5)p}_{ii}\right) 
-\frac{1}{30}m_e^2\left(1 + 3 \cos 2\eta\right)\left(a^{(5)e}_Q + a^{(5)p}_Q\right)  \nonumber \\
&~~~~~~~~~~~~~~~~~~~~~~~~~~~~~~+\frac{1}{3} (4 + \cos2\eta)b_3^e
  -\frac{1}{30}(3+\cos2\eta) b_3^e  + \frac{\sqrt{2}}{10} \sin2\eta \, b_3^e \nonumber \\
&~~~~~~~~~~~~~~~~~~~~~~~~~~~~~~
+\frac{1}{2}(1-\cos2\eta) (k_{AF})_3 \bigg] \ ,
\label{d29}
\end{align}
with the corresponding result for $1S_d$\,-\,$2P_{f-}$.  Note that, unlike the $1S$\,-\,$2S$
transitions, the $1S$\,-\,$2P$ transition frequencies also involve the non-isotropic $a^{(5)}_Q$ 
SME couplings.

\vskip0.2cm
The ALPHA programme involves a detailed study of a variety of further transitions, outlined
in \cite{ALPHA2019}. Similar theoretical methods can be applied to determine the dependence on the
Lorentz and $\mathsf{CPT}$ violating couplings for all the relevant eigenstates, using the 
key formula (\ref{d25}) with the appropriate Clebsch-Gordan coefficients and taking account 
of the magnetic field dependence of the mixing angles amongst the Zeeman states.

\vskip0.4cm
\noindent {\it Testing Lorentz and $\mathsf{CPT}$ symmetry}:
\vskip0.2cm

We close this section with a general discussion of the implications of the ALPHA measurements of the
$1S$ hyperfine, $1S$\,-\,$2S$ and $1S$\,-\,$2P$ antihydrogen transitions for the effective theory
of Lorentz and $\mathsf{CPT}$ violation. Conversely, we comment on how the general features of this
theory may motivate future measurements in antihydrogen spectroscopy.

The first point to emphasise is that the precision reached experimentally in, for example, the $1S$\,-\,$2S$ 
transition frequency far exceeds that possible from a first principles QED calculation.
This means that to establish any violations of fundamental principles such as Lorentz and
$\mathsf{CPT}$ symmetry we need to {\it compare} measurements. In the case of $\mathsf{CPT}$, this 
means comparing the spectrum of ${\rm H}$ and $\overline{\rm H}$ in sufficiently similar environments that
we can control systematics to high precision. Such {\it instantaneous} comparisons of the ${\rm H}$ and 
$\overline{\rm H}$ spectra will be different if and only if $\mathsf{CPT}$ is broken. 
For Lorentz invariance, ${\rm H}$ (or $\overline{\rm H}$) transition frequencies should be compared at 
different times, to look for possible sidereal or annual variations. 

Second, in comparing sensitivities to the SME couplings from different transitions, we need to distinguish 
between the {\it relative} precision of the measurement and the {\it absolute} precision of the energy
sensitivity, which bounds both the dimensional SME couplings such as $b_3$ and the dimensionless
couplings such as $c_{00}$ (which arise in energy levels accompanied by factors of $m_e$). 

A further important point, which is evident from our explicit expressions for the atomic energy levels
and transitions, is that the Lorentz violating couplings always appear in combinations comprising
$\mathsf{CPT}$ even and $\mathsf{CPT}$ odd couplings. For example, the $1S$ hyperfine transitions
(\ref{d9}) involve the combination $B_3^e = -b_3^e - m_e d_{30}^e - \tfrac{1}{2} \e_{kij} H_{ij}^e$, arising 
directly from the Hamiltonian (\ref{d2}), where $b_3$ is $\mathsf{CPT}$ odd while $d_{30}$ and $H_{12}$ are 
$\mathsf{CPT}$ even. Similarly, the $1S$\,-\,$2S$ transition involves the $\mathsf{CPT}$ even $c_{00}$ 
together with the higher-dimensional $\mathsf{CPT}$ odd coupling $\tr\,a^{5}_{ij} \sim a_{200}^{\rm NR}$. 
As noted in \cite{Kostelecky:2015nma}, this is a very general feature of the SME effective theory. 
In practice, this means that a search for Lorentz violation from an experiment with matter alone
can only bound this combination. It cannot {\it on its own} test for $\mathsf{CPT}$ violation, since there
remains the possibility of a cancellation of the $\mathsf{CPT}$ violating couplings ({\it e.g.}~$b_3$)
and the $\mathsf{CPT}$ even couplings ({\it e.g.}~$d_{30}$ and $H_{12}$). Again, this shows that 
to identify unambiguously a signal for $\mathsf{CPT}$ violation, we need to compare experiments on
equivalent matter and antimatter systems.

In essence, this is the same idea we have already tried to exploit in section \ref{sect 2.6}, where we considered 
potential new background fields and the equivalence principle. Here, we similarly entertain the possibility
of a {\it cancellation} between the Lorentz violating couplings within a pure matter system, while 
allowing them to add to give an observable effect for the equivalent pure antimatter system.

Focusing first on $\mathsf{CPT}$, comparing the $1S$ hyperfine spectrum (\ref{d10}) of ${\rm H}$ and 
$\overline{\rm H}$ at zero magnetic field allows a bound to be placed on the combination $b_3^e + b_3^p$, while
at the ALPHA magnetic field the sensitivity is essentially to $b_3^e$ alone. Combining these, we could 
bound both $b_3^e$ and $b_3^p$ (always recalling that $b_3^p$ and related quantities is an effective
parameter for the QCD bound state proton, not actually a parameter in the SME Lagrangian itself).
This would, however, require individual measurements of the $1S_d$\,-\,$1S_a$ and 
$1S_c$\,-\,$1S_b$ transitions whereas the so-far published ALPHA results \cite{Ahmadi17},
with a quoted precision of $4\times 10^{-4}$, are for the difference alone 
for which the SME corrections cancel (recall (\ref{d9}) and (\ref{d10aa})).
If, for illustration, we assume a similar precision could be reached for the individual transition frequencies,
then since the frequency of the $1S_d$\,-\,$1S_a$ hyperfine transition is $29$\,GHz, this
would correspond to an absolute energy precision of $12$\,MHz or 
$4.8\times 10^{-8}$\,eV and would imply a bound on $|b_3^e| \lesssim 10^{-17}$\,GeV.

Now consider the $1S$\,-\,$2S$ transitions, for which ALPHA have published a high precision measurement
of the $1S_d$\,-\,$2S_d$ transition \cite{Ahmadi:2018eca}.
The SME contributions to the difference of the $1S_d$\,-\,$2S_d$ transitions for ${\rm H}$ and
$\overline{\rm H}$ are given in (\ref{d17}). Here, the 
leading contribution of the $\mathsf{CPT}$ violating SME couplings arises only at $O(\a^2)$, and depends 
on $b_3^{e,p}$ and the higher-dimensional $a^{(5)}$ couplings, whose contribution is theoretically expected 
to be relatively suppressed by $O(m_e^2/\L^2)$. The $b_3^p$ contribution is accompanied by the mass suppression factor 
$\e = m_e^2/m_p^2 \sim 10^{-6}$.
So if first we interpret (\ref{d17}) as bounding $b_3^e$, the ALPHA precision of $2\times 10^{-12}$ on the 
$1S_d$\,-\,$2S_d$ transition frequency of $2.466\times 10^{15}$\,Hz \cite{Ahmadi:2018eca} 
gives a bound $|b_3^e| \lesssim 7 \times 10^{-16}$\,GeV.
This illustrates the point raised above, that a higher relative precision measurement of a higher frequency 
spectral line can nevertheless result in a less stringent bound on the $\mathsf{CPT}$ violating 
coupling $b_3^e$.

On the other hand, if we impose the existing bounds on $b_3^e$ from other antimatter systems quoted in 
\cite{Kostelecky:2008ts}\footnote{Note that the 
formula quoted in \cite{Kostelecky:2015nma, Vargas:2018isj} for the
SME contribution to the $1S$\,-\,$2S$ transition omits the spin-dependent couplings such as $b_3$ in 
(\ref{d17}). This is because they assume prior to calculating the energy levels that these couplings 
are negligible compared to $a_{200}^{\rm NR}$ and $c_{00}$, on the basis of the existing experimental
constraints given in \cite{Kostelecky:2008ts}.}, 
then the $1S_d$\,-\,$2S_d$ measurement can be used to give an experimental bound on
the sum of the higher-dimensional couplings $a^{(5)}$ for $e$ and $p$,
{\it viz.} $\sum_{\omega = e,p}\tr\,a^{(5)\omega}_{ij} \sim 
\sum_{\omega=e,p}a_{200}^{{\rm NR} \,\omega} \lesssim 10^{-9}\,{\rm GeV}^{-1}$.

However, with high precision measurements of {\it both} the $1S_d$\,-\,$2S_d$ and $1S_c$\,-\,$2S_c$ transitions 
in ${\rm H}$ and $\overline{\rm H}$, we could use (\ref{d17}) and (\ref{d17a}) to obtain bounds 
on both the $b_3$ and $a^{(5)}$ couplings directly from antihydrogen, and {\it with no prior assumptions} on 
their magnitude.
Note first that the isotropic $\tr\,a^{(5)}_{ij}$ couplings cancel in the difference of
$1S_d$\,-\,$2S_d$ and $1S_c$\,-\,$2S_c$, so a difference measurement is sensitive only to the
$b_3$ coefficients. 
From (\ref{d17a}), the SME contributions for the $1S_c$\,-\,$2S_c$ transition has a term proportional 
to $b_3^e - b_3^p$ at $O(1)$, which is not present for $1S_d$\,-\,$2S_d$.
While this is suppressed by the magnetic field dependent factor 
$\left(\cos2\theta_2 - \cos2\theta_1\right) = 1.2 \times 10^{-3}$ 
at the ALPHA trap magnetic field of $1.03$\,T, this is greater than the $O(\a^2)$ $b_3$ contributions
discussed above. So an ALPHA measurement of the frequency difference for the 
$1S_d$\,-\,$2S_d$ and $1S_c$\,-\,$2S_c$ transitions, with a precision for $1S_c$\,-\,$2S_c$
matching that already achieved for $1S_d$\,-\,$2S_d$,
would give a bound on the combination 
$|b_3^e - b_3^p| \lesssim 10^{-17}\,{\rm GeV}$. This is comparable with the illustrative
bound given above for a potential determination of $b_3^e$ from the hyperfine transitions.
Then, assuming this applies generically to both $b_3$ coefficients separately, 
we can deduce the bound  
$\sum_{\omega=e,p}\tr\,a^{(5)\omega}_{ij} \lesssim 10^{-9} \,{\rm GeV}^{-1}$
from ${\rm H}$ and $\overline{\rm H}$ spectroscopic measurements alone.

The $b_3^e$ coupling can also be bounded from the $1S$\,-\,$2P$ transitions. Temporarily neglecting the 
photon coupling $(k_{AF})_3$ in (\ref{d29}), along with the $a^{(5)}_{ij}$ couplings, 
and using the ALPHA precision of 76\,MHz for 
the resolved $1S_d$\,-\,$2P_{c_-}$ transition \cite{Ahmadi:2020ael} gives a bound 
$|b_3^e| \lesssim 5 \times 10^{-16}$\,GeV. 

Conversely, imposing the bound \cite{Kostelecky:2008ts} on $b_3^e$, and assuming 
the $1S$\,-\,$2S$ bound on $a^{(5)}_{ii}$, would enable
a bound $a^{(5)}_Q \sim a_{220}^{\rm NR} \lesssim  10^{-3} \,{\rm GeV}^{-1}$
to be placed on the non-isotropic higher-dimensional couplings.

As emphasised above, however, a new feature of $1S$\,-\,$2P$, and any transition involving states
with non-zero orbital angular momentum, is its sensitivity to a potential spin-independent 
$\mathsf{CPT}$ violation arising in the photon sector. 
Accepting the quoted bounds on $b_3^e$ in \cite{Kostelecky:2008ts}, the result (\ref{d29})
bounds the $\mathsf{CPT}$  violating photon coupling $|(k_{AF})_3| \lesssim 10^{-11}$\,GeV.
This is many orders of magnitude below the bound 
$|k_{AF}| \lesssim 10^{-42}$\,GeV \cite{Kostelecky:2008ts}  deduced from astronomical observations 
from gamma ray bursts or the CMB. 
Nevertheless, the comparison of the ALPHA $1S$\,-\,$2P$ result with hydrogen gives an interesting illustration
of limiting $\mathsf{CPT}$ violation in the photon sector in an atomic physics experiment.

\vskip0.3cm

Next, we consider how these spectral transitions in ${\rm H}$ and $\overline{\rm H}$ may test for violations 
of Lorentz symmetry. Of course, all the SME couplings violate Lorentz invariance but we focus first on
the $c^{(4)}$ and $a^{(5)}$ couplings and the bounds arising from measurements looking for sidereal 
or annual variations in the $1S$\,-\,$2S$ transitions.\footnote{In principle, the independent role of all 
the Lorentz violating couplings, including the spin-dependent $b_3$, $d_{30}$ and $H_{12}$, may be 
extracted from the frequencies of the variety of transitions considered here, especially given their 
different magnetic field dependence.}
Currently, the most stringent bounds come from searches for annual variations in
high-precision hydrogen spectroscopy \cite{Matveev:2013orb}, 
which now allows the $1S$\,-\,$2S$ frequency to be measured with a precision of $4.5 \times 10^{-15}$. 

A detailed presentation of the formalism required to compare measurements made in the laboratory frame with 
a standard Sun-centred frame, as used in the SME `Data Tables' \cite{Kostelecky:2008ts},
is given in a form relevant to ${\rm H}$ and $\overline{\rm H}$ spectroscopy in \cite{Kostelecky:2015nma}. 
Briefly, the Sun-centred frame comprises a pseudo-orthonormal basis $(\hat{T}, \hat{X},\hat{Y},\hat{Z})$ 
with the $\hat{Z}$-axis aligned with the Earth's rotation axis and the $\hat{X}$-axis with the vernal
equinox, and with the origin of the time coordinate $T$ taken as the vernal equinox in 2000.
We are concerned with the Lorentz boosts arising 
from the Earth's rotation and annual orbital motion around the Sun. (Note that for the {\it isotropic} 
$c^{(4)}$ and $a^{(5)}$ coefficients measured in $1S$\,-\,$2S$, there is no contribution from the rotation
between the laboratory and Sun-centred frames.) The Earth's orbital velocity in these coordinates 
is \cite{Kostelecky:2015nma},
\begin{equation}
{\bf v}_\oplus ~=~ v_\oplus\,\left(\sin\Omega_\oplus T\, \hat{X} ~-~ 
\cos\Omega_\oplus T \left(\cos\eta\,\hat{Y} \,+\, \sin\eta\, \hat{Z} \right)\,\right) \ ,
\label{dL1}
\end{equation}
where $\Omega_\oplus$ is the orbital frequency and $\eta \simeq 23.4^\circ$ is the angle between the
$XY$-plane and the orbital plane. The Earth's rotational velocity is simply
\begin{equation}
{\bf v}_L ~=~ v_L\,\left(-\sin\omega_\oplus T_\oplus\,\hat{X} ~+~ \cos\omega_\oplus T_\oplus \, \hat{Y}\right )\ ,
\label{dL2}
\end{equation}
where $T_\oplus$ is the local Earth sidereal time and $\omega_\oplus$ is the rotation frequency.

For annual variations, we are interested in the change of frequency as measured in the laboratory 
due to its orbital motion 
relative to the inertial Sun-centred frame (the formalism for sidereal variations is identical). 
For this, we need the effect of a Lorentz boost with velocity ${\bf v}_\oplus$ 
on the $(t,x,y,z)$ components of the isotropic $c^{(4)}$ and $a^{(5)}$ SME couplings entering in the
formulae for the $1S$\,-\,$2S$ transitions (see footnote \ref{KVcouplings} for notation).
To first order in $v_\oplus$, the boost gives:
\begin{equation}
c_{00} ~=~ c_{TT} \,+\, 2 v_\oplus^J\, c_{JT} \ , ~~~~~~~~~~
c_{ii} ~=~ c_{KK} \,+\, 2 v_\oplus^J\, c_{JT} \ ,
\label{dL3}
\end{equation}
where the tracelessness of $c_{\m\n}$ ensures these are identical,
and
\begin{equation}
a^{(5)}_{000} ~=~ a^{(5)}_{TTT} \,+\, 3 v_\oplus^J\, a^{(5)}_{JTT} \ , ~~~~~~~~~~
a^{(5)}_{0ii} ~=~ a^{(5)}_{TKK} \,+\, v_\oplus^J \left( a^{(5)}_{JKK} \,+\, 2 a^{(5)}_{JTT}\right) \ .
\label{dL4}
\end{equation}

The difference in frequency of the $1S_d$\,-\,$2S_d$ transition measured at two points on the Earth's
orbit with velocity difference $\d {\bf v}_\oplus$ is therefore given by:
\begin{equation}
\d\left(\D E_{2S_d\ltrta 1S_d}^{\bar{\rm H}}\right) ~=~ \frac{3}{4} \a^2\,m_e \,\sum_{\omega = e,p} 
\d v^J_\oplus \left( \frac{5}{3}\, c^\omega_{JT}  \,+\, m_e \left(
a^{(5)\omega}_{JKK} \,+\, 5\, a^{(5)\omega}_{JTT}\right)  \,\right) \ ,
\label{dL5}
\end{equation}
neglecting the anisotropic spin couplings, with the same result with $a^{(5)} \rightarrow - a^{(5)}$ for
hydrogen. As usual, complementary measurements of the hydrogen and antihydrogen transitions would
enable us to determine {\it both} the $\mathsf{CPT}$ odd $a^{(5)}$ couplings, from the difference of
${\rm H}$ and $\overline{\rm H}$, and the $\mathsf{CPT}$ even $c^{(4)}$ couplings.  Bounds for the latter,
expressed as combinations of the $c_{JT}$, were quoted in \cite{Matveev:2013orb} from limits on
annual variations of the $1S$\,-\,$2S$ transitions in hydrogen alone by neglecting the contribution
to the transition frequencies from the $a^{(5)}$ couplings.

Importantly, note that measurements constraining the annual variations of the transition frequency
are sensitive to {\it different} combinations of the SME couplings, though suppressed by
the boost factor $v_\oplus \sim O(10^{-4})$. For example, annual-averaged
measurements allow a bound to be placed on  
$\tfrac{1}{\sqrt{4\pi}}a^{\rm NR}_{200}\,=\,a^{(5)}_{TTT} + a^{(5)}_{TKK}$ as discussed above,
while annual variations constrain a combination of the couplings $a^{(5)}_{JTT}$ and $a^{(5)}_{JKK}$
weighted according to the laboratory velocity components $\d v^J$ in the Sun-centred frame. 

Unlike the case of these isotropic couplings, for transitions involving the vector coupling 
$b_\m$ we also need to consider the rotation between the laboratory and Sun-centred frames. 
For simplicity, we consider here a `standard laboratory frame', with the $z$-axis pointing
vertically to the local zenith and the $x$-axis directed due south. We assume the applied 
magnetic field is in the $z$ ($i=3$) direction.\footnote{This is not the case for the ALPHA 
experiment, where the magnetic field is horizontal, at an angle $60\deg$ E of N. 
This requires an additional rotation matrix to convert the experimental frame to the 
standard laboratory frame.}
The relevant components of the rotation matrix $R_{i J}$
are then:
\begin{equation}
R_{3X} \,=\, \sin\chi \,\cos(\omega_\oplus T_\oplus + \phi) \ , ~~~~~~
R_{3Y} \,=\, \sin\chi \,\sin(\omega_\oplus T_\oplus + \phi) \ , ~~~~~~
R_{3Z} \,=\, \cos\chi \ ,
\label{dL6}
\end{equation}
where the colatitude $\chi$ is the angle between the $z$-axis and the Earth's rotation axis, 
while the phase $\phi$ is the angle between the $x$ and $X$-axes at the origin of the chosen
sidereal time $T_\oplus = 0$ \cite{Kostelecky:2015nma}. Including the Lorentz boosts with both
the Earth's rotational and orbital velocities, it is straightforward to see that the component
$b_3$ is expressed in the Sun-centred frame as,
\begin{equation}
b_3 ~=~ R_{3J}\, b_J  ~+~ R_{3J}\, (v_\oplus^J \,+\, v_L^J )\,b_T \ .
\label{dL7}
\end{equation}

For example, if we were to isolate the $b_3$-dependence experimentally by taking the difference 
of the $1S_c$\,-\,$2S_c$ and $1S_d$\,-\,$2S_d$ transition energies for antihydrogen compared 
to the same difference for hydrogen, we find from (\ref{d17}) and (\ref{d17a}),
\begin{align}
\bigl(\D E_{2S_c\ltrta 1S_c}^{\bar{\rm H}} \,-\, \D E_{2S_d\ltrta 1S_d}^{\bar{\rm H}} \bigr) ~-~
&\bigl(\D E_{2S_c\ltrta 1S_c}^{\rm H} \,-\, \D E_{2S_d\ltrta 1S_d}^{\rm H} \bigr) \nonumber \\
&=~  2\left(\cos 2\theta_2 \,-\,\cos 2\theta_1 \right)\, \left(b_3^e - b_3^p\right) ~~+~~ O(\a^2) \ ,
\label{dL8}
\end{align}
with $b_3^{e,p}$ given by (\ref{dL7}). Averaging over the sidereal variations so that only
$R_{3Z}$ is non-zero, but keeping the dependence on annual variations through ${\bf v}_\oplus$, 
we find simply using (\ref{dL7}) with (\ref{dL1}) and (\ref{dL2}),
\begin{align}
\bigl(\D E_{2S_c\ltrta 1S_c}^{\bar{\rm H}} \,&-\, \D E_{2S_d\ltrta 1S_d}^{\bar{\rm H}} \bigr) ~-~
\bigl(\D E_{2S_c\ltrta 1S_c}^{\rm H} \,-\, \D E_{2S_d\ltrta 1S_d}^{\rm H} \bigr)  \nonumber \\
&~=~  2\left(\cos 2\theta_2 \,-\,\cos 2\theta_1 \right)\,\, 
\cos\chi \Bigl[ \left(b_Z^e - b_Z^p\right) 
- \,v_\oplus\,\cos\Omega_{\oplus}T \,\sin\eta \, \left(b_T^e - b_T^p\right) \Bigr] \ . 
\label{dL9}
\end{align}
Notice again how the annual variations give access to a new component of the SME coupling,
in this case $b_T$.

On the other hand, if we consider sidereal variations, the dominant contribution is from the
time-dependent rotation term $R_{3J}b_J$ in (\ref{dL7}). (In fact the $O(v_L)$ contribution is
time-independent, as is readily checked from (\ref{dL2}) and (\ref{dL6})).  So in this case,
\begin{align}
&\bigl(\D E_{2S_c\ltrta 1S_c}^{\bar{\rm H}} \,-\, \D E_{2S_d\ltrta 1S_d}^{\bar{\rm H}} \bigr) ~-~
\bigl(\D E_{2S_c\ltrta 1S_c}^{\rm H} \,-\, \D E_{2S_d\ltrta 1S_d}^{\rm H} \bigr)  \nonumber \\
&=\,   2\left(\cos 2\theta_2 \,-\,\cos 2\theta_1 \right)\Big[
\sin\chi \,\Big(\cos(\omega_\oplus T_\oplus + \phi) \,b_X  
\,+\, \sin(\omega_\oplus T_\oplus + \phi) \,b_Y \Big) \,+\,\cos\chi \, b_Z \Big] \ ,
\label{dL10}
\end{align}
showing how sidereal variations are sensitive to the couplings $b_X$ and $b_Y$ as well as $b_Z$.

\vskip0.3cm

Finally, to compare the potential sensitivity of the antihydrogen bounds on $b_3^e$ with existing
results, we note again that measurements on purely matter systems, {\it e.g.}~on sidereal
variations in the spin precession frequency of electrons in a Penning trap (see \cite{Ding:2016lwt} for a 
summary), can only bound the combination of $B_3^c$ (the charge conjugate of $B_3$ as defined here).
These bounds (see under $|\tilde{b}_{X,Y,Z}|$ in \cite{Kostelecky:2008ts}) are typically 
of $O(10^{-23})$\, though can be significantly lower from torsion pendulum experiments.
However, the only bound quoted in \cite{Kostelecky:2008ts} on $|\underline{b}^e|$ itself 
comes from the comparison of spin precession frequencies of electrons and positrons 
\cite{Dehmelt:1999jh}, with the result $|\underline{b}^e| \lesssim 10$\,Hz, 
equivalent to $4 \times 10^{-23}$\,GeV.

It is interesting to compare the interpretation of these measurements of the anomalous magnetic 
moment for the electron/positron \cite{Dehmelt:1999jh} and proton/antiproton 
\cite{Nagahama17, Smorra17, Smorra:2019qfx, Smorra:2020nko} in a Penning trap with
the corresponding determination from the hyperfine splitting in H/$\overline{\rm H}$.
Penning trap measurements compare the spin precession (Larmor) and cyclotron frequencies
$\w_s$ and $\w_c$ in a background magnetic field, the difference being the `anomalous'
frequency $\w_a = \w_s - \w_c$. In a conventional theory, this measures the $g-2$ factor
for the test particle.\footnote{The cyclotron frequency is $\w_c = eB/m$ while the 
spin precession frequency, which depends on the magnetic moment, is $\w_s = g\mu_B B$.
The difference therefore gives
\begin{equation*}
\frac{\w_a}{\w_c} ~\equiv~ \frac{\w_s - \w_c}{\w_c} ~=~ \frac{1}{2}(g-2) \ .
\end{equation*} 
However, in the SME, the spin-dependent $b_3$ term in the Hamiltonian (\ref{d2}) gives an 
extra contribution to the spin precession frequency alone, so that $\w_s = g\mu_B B - 2 b_3$.
In this theory, the ratio $\w_a/\w_c$ therefore has the additional, $B$-dependent, 
factor shown in (\ref{d300}). }
In the SME, however, the $\mathsf{CPT}$ odd $b_3$ coefficient in the Hamiltonian (\ref{d2})
modifies the spin precession frequency \cite{Bluhm:1997qb} and to leading order,
\begin{equation}
\frac{\w_a}{\w_c} ~=~ \frac{1}{2}(g-2) \,-\, \frac{2m}{eB}\, b_3 \ .
\label{d300}
\end{equation}
A possible $\mathsf{CPT}$-violating difference in the ratio $\w_a/\w_c$ for, say, the electron 
and positron could then be attributable either to a difference in the $g$ factors for the 
particle and antiparticle, which depends on quantum loop corrections in QED, or to the 
direct Lorentz and $\mathsf{CPT}$ violating $b_3$ coefficient in the SME.
The bounds on $b_3^{e,p}$ quoted in \cite{Dehmelt:1999jh} and 
\cite{Nagahama17, Smorra17, Smorra:2019qfx, Smorra:2020nko}
are subject to the assumption that $g^e$ and $g^p$ are the same for the particle and
antiparticle. 

In contrast, while the individual (anti)hydrogen hyperfine transitions depend at leading order
on the SME coefficients $b_3^{e,p}$, as we showed in (\ref{d10aa}) this dependence cancels out
in the difference of the $1S_d$\,-\,$1S_a$ and $1S_c$\,-\,$1S_b$ transitions. 
The ALPHA \cite{Ahmadi17} measurement of the hyperfine splitting $\mathcal{E}_{HF}$
depends purely on the anomalous magnetic moments of the positron and antiproton
given by the corresponding $g$ factors alone. Comparison of (\ref{d300}) with (\ref{d5c})
for $\mathcal{E}_{HF}$ then makes it clear that the $e^-/e^+$ and $p/\overline{p}$ Penning trap
and H$/\overline{\rm H}$ hyperfine splitting measurements provide {\it complementary}
tests of $\mathsf{CPT}$ invariance.

To summarise, setting aside the details of the SME parametrisation, the results above make it 
clear that Lorentz and $\mathsf{CPT}$ violation can arise in subtly different ways in all the 
antihydrogen spectral transitions and, in the event of a non-null discovery, many measurements 
may be necessary to pin down the origin of $\mathsf{CPT}$ breaking. Sidereal and annual variations
can also place competitive bounds on Lorentz violation.  Moreover, in terms of looking for
radically new physics, we should not lose sight of the fact that the SME is itself in some sense
conservative, being a conventional effective quantum field theory built in the standard way 
from causal fields in representations of the Lorentz group. 
All this further motivates the most extensive and high precision analysis of
the whole antihydrogen spectrum, including the search for sidereal and annual variations.

\subsubsection{New background fields}

Until now, we have considered the possibility of new, long-range background fields (`fifth forces')
from the perspective of their gravity-like effects on weak equivalence principle experiments
(see section \ref{sect 2.6}).
Here, we point out an interesting effect of a long-range $U(1)_{B-L}$ interaction on atomic 
spectroscopy and show how this can limit the allowed coupling strength to high precision.

The idea is that in a $U(1)_{B-L}$ gauge theory in which the gauge boson is essentially massless
(strictly, with a sufficiently small mass $m_{Z'} <  10^{-14}$\,eV that the force has a range greater 
the Earth's radius), the Earth itself acts as a source creating the $U(1)_{B-L}$ analogue of an electric field 
at the surface of magnitude $\mathcal{E}_{B-L} = Q_{B-L}^{Earth} g'/4\pi R_E^2$.
Here, $Q_{B-L}^{Earth}$ is the $B-L$ charge of the Earth (the number of neutrons) and $R_E$ is its radius.
A hydrogen atom placed in this field acts as a $U(1)_{B-L}$ dipole since the proton has 
$Q_{B-L}^p = 1$ while the electron has $Q_{B-L}^e = -1$.

The atom therefore experiences a $U(1)_{B-L}$ analogue of the Stark effect, which occurs when
an atom is placed in a conventional electric field. This produces a $U(1)_{B-L}$ linear Stark shift in the
$n=2$ energy levels given by adapting the usual formula, {\it viz}.
\begin{equation}
\D E ~\simeq~ \pm g' \mathcal{E}_{B-L} a_0 \ ,
\label{d30}
\end{equation}
where $a_0$ is the Bohr radius. That is,
\begin{equation}
\D E ~\simeq~ Q_{B-L}^{Earth}\, \a' \,a_0/R_E^2  
~\simeq~ 10^{21} \,\a' \,\,{\rm eV}\ .
\label{d31}
\end{equation}

This Stark shift would be opposite in sign for antihydrogen because of the opposite $B-L$ charge
of the antiparticles, so in the absence of any other non-standard model interactions, a comparison of the 
${\rm H}$ and $\overline{\rm H}$ spectra would reveal the shift (\ref{d31}). 
From the current absolute energy sensitivity of the antihydrogen $1S$\,-\,$2S$ transition of 
$2 \times 10^{-20}$\,GeV \cite{Ahmadi:2018eca} 
we can therefore place a bound $\a' \lesssim 10^{-33}$ on the $U(1)_{B-L}$ coupling.

This is a remarkable level of precision at which to bound a new gauge coupling. Nevertheless,
as noted in section \ref{sect 2.6}, existing equivalence principle experiments already constrain
the coupling to the far lower value $\a' < 10^{-49}$, the extremely small value of course reflecting 
the huge $B-L$ charge of the Earth. Despite the high precision of atomic spectroscopy, it therefore
seems that gravitational tests remain a better test of new long-range `fifth force' interactions.

\subsection{Antihydrogen and Gravity}\label{sect 3.3}

There are currently three initiatives aimed at direct investigations of antimatter gravity based upon free fall 
measurements \cite{Testera15,Perez15,Bertsche18}. We briefly describe these below, noting that the ALPHA-g 
experiment \cite{Bertsche18} has developed using experience and knowledge gained from the trajectory analyses 
described in section \ref{sect 3.1}. This was also to the fore in the gravity investigation described by Amole and
coworkers \cite{Amole13} in which, in essence, the equation of motion  (\ref{d1}) was used, 
but assuming the antihydrogen ``charge anomaly'' $Q=0$. The ALPHA
apparatus \cite{ALPHAApp} is a horizontal antihydrogen trapping device (which as described in \cite{Zhmoginov} is not 
optimum for gravity investigations), and the annihilations of the anti-atoms escaping from the trap as the magnetic holding
fields were lowered were used to deduce rough limits, mostly as a demonstration of proof-of-principle, on the 
aforementioned parameter $F$. Values of $-65 < F > 110$ were excluded at a level of 95\% statistical significance.

The detailed work of Zhmoginov {\em et al.} \cite{Zhmoginov} (see also \cite{Zhong}) has informed the design of a new,
vertically orientated apparatus named ALPHA-g \cite{Bertsche18}, whose initial aim is to improve the limit of $|F|$ to
around unity ({\it i.e.}, a so-called ``up-down" determination for the gravitational behaviour of antimatter). 
ALPHA-g consists of two symmetrically located atom and Penning trap arrangements at either end of the apparatus, 
with a high precision trapping region in the centre, with the magnetic potential in the vertical direction controlled 
by a series of coils. These coils will allow a bias field to be added to the bottom of the trap that can compensate 
for the gravitational potential difference across the trap.
A determination of  the gravitational acceleration of antihydrogen will be carried out by varying the compensation 
fields and monitoring the relative populations of the anti-atoms leaving the top and bottom of the trap \cite{ALPHA2019}.
A measurement accuracy of $|F| \sim 1$ can be approached using a few hundred trapped
anti-atoms, as verified using the aforementioned simulations, and something that can be achieved in a single 
8-hour antiproton shift at the AD. Bertsche \cite{Bertsche18} has also argued that with augmentation of ALPHA-g using 
various techniques such as in-situ magnetometry and antihydrogen laser cooling, it should be possible to lower the 
systematic errors on the measurement of $|F|$ to the 1\% level.

An extension of capability for the vertical ALPHA-g has been suggested by M\"uller and co-workers \cite{Mueller:2013ysa}
who envisage using a novel light-pulse (anti)matter-wave interferometer (see {\it e.g.}~\cite{Chung:1999}) with trapped 
and cooled antihydrogen atoms, which are then released into the device. Without going into further details here, 
M\"uller {\em et al.} envisage a so-called basic scenario, capable of probing $|F|$ to around 1\%, with further advances 
possibly allowing measurements approaching the ppm-level.

The AEgIS collaboration ({\it e.g.}~\cite{Testera15}) is planning to use a beam of antihydrogen atoms formed 
via the reaction of very cold ($\sim$ a few mK) antiprotons with excited positronium atoms 
({\it i.e.}~$\overline{p} + {\rm Ps}^* \rightarrow \overline{\rm H} + e^-$, a reaction first suggested as a useful 
source of $\overline{\rm H}$ some time ago \cite{Charlton90}) to perform a moir\'e deflectrometry experiment. 
The relative sensitivity to the gravitational acceleration $g$ is expected to be around 1\%, and a demonstration of 
the technique using a flux of fast $\overline{p}$\,s has recently been reported \cite{Aghion14}.

GBAR is planning to perform free fall experiments on ultra-cold antihydrogen atoms formed via 
photoionisation of cold antihydrogen positive ions, $\overline{\rm H}^+$. The latter is to be produced 
following the $\overline{p}$\,-\,${\rm Ps}^*$ reaction (see above), and a further charge exchange as 
$\overline{\rm H} + {\rm Ps} \rightarrow \overline{\rm H}^+ + e^-$, most probably with ground state Ps. 
The $\overline{\rm H}^+$ 
will likely be formed at kinetic energies in the keV range, so will be decelerated and then individually 
sympathetically cooled in a Paul-type charged particle trap (using a laser-cooled Be$^+$ ion) into the mK regime. 
It will then be photo-ionised, allowing the resulting antihydrogen atom to undergo free fall: its time of flight 
between the pulsed laser used for ionisation and the subsequent annihilation on the chamber wall will be used 
to determine $g$. 

A further free fall approach has been proposed by Voronin and co-workers \cite{Voronin11,Voronin14} which relies
upon the interaction of cold antihydrogen with a material surface. Low energy antihydrogen will, due to 
quantum effects arising from the Casimir-Polder interaction, be efficiently quantum reflected from a surface and, 
in the presence of the Earth's gravitational field, will form long-lived quantum states. It has been 
shown \cite{Voronin11,Voronin14} that measuring the difference in the energy of the states using atom 
interferometry can yield a value for $g$ for antihydrogen. With the anti-atoms at temperatures 
of $\sim 100$\,mK, it is claimed that a flux of a few antihydrogen atoms per second can yield a precision 
of around $10^{-3}$.

Finally, in addition to these free-fall experiments, there is the possibility in future of performing
a gravitational redshift experiment of the Pound-Rebka type directly on antihydrogen. Some basic theoretical
considerations are discussed in section \ref{sect 3.3.2}.

\subsubsection{Antihydrogen free fall}\label{sect 3.3.1}

We now consider some of the theoretical ideas introduced in sections \ref{sect 2.5} and \ref{sect 2.6}
on how GR may be modified, or extended, to predict violations of the weak equivalence
principle (WEPff) in experiments on antimatter, specifically antihydrogen.

The difficulty of course is that existing experiments already place extremely small limits on any 
possible WEPff violation in matter systems. If there is to be any chance of measuring deviations 
from WEPff in the forthcoming antihydrogen free-fall experiments, we therefore need to find some
mechanism which is effectively shielded in matter interactions but leaves a residual effect on
antimatter.

\newpage

\noindent {\it Strong equivalence principle violation}:
\vskip0.2cm
As described earlier, the simplest modification of GR is to break the 
strong equivalence principle (SEP) by introducing 
direct couplings of the elementary particle fields to the local curvature, for example generalising the
Dirac action to:
\begin{equation}
S = \int d^4 x \sqrt{-g} \biggl(\frac{R}{16\pi G} + 
\bar\psi\left(i \c^\m D_\m - m\right) \psi + a\,\partial_\m R\, \bar\psi \c^\m \psi
+ c\,R_{\m\n}\,\bar\psi i \c^\m \overleftrightarrow{D}^\n \psi   + \ldots \biggr) \ .
\label{d32}
\end{equation}
These extra terms  (see (\ref{b42})) may be viewed either as new couplings in a fundamental theory, or as 
an effective theory where they are generated by quantum loop corrections. In either case, they modify 
the fermion dispersion relation and the equation of motion for free-fall, which is then no longer the 
geodesic equation. Moreover, defined in a local inertial frame, the operator $\bar\psi \c^a\psi$
is $\mathsf{CPT}$ odd and its coupling to $\partial_\m R$ modifies the geodesic equation 
{\it differently} for fermions and antifermions. 

Unfortunately, it seems this mechanism cannot be exploited for free-fall experiments on Earth,
since the Schwarzschild metric describing the gravitational field outside the source region 
is Ricci flat -- only the Riemann curvature $R_{\m\r\n\s}$ is non-zero, while $R_{\m\n}$ and $R$ vanish.
As noted in \cite{Shore:2004sh}, it is not possible to construct a term bilinear in the Dirac fields 
in (\ref{d32}) involving $R_{\m\r\n\s}$ which cannot be reduced at linear order in curvature to those
expressible in terms of the Ricci tensor alone.

We conclude that in general relativity itself, even allowing for the SEP-violating interactions
in the effective field theory, no observable WEPff violations are predicted in free-fall experiments in 
the Earth's gravitational field.

\vskip0.5cm
\noindent {\it Lorentz and $\mathsf{CPT}$ violation}:
\vskip0.2cm

A more radical alternative is to take the local Lorentz and $\mathsf{CPT}$ violating effective field 
theory discussed at length above and couple it to gravity. Incorporating spontaneous Lorentz violation 
into GR is not without subtlety, however, and the resulting theory involves many 
features requiring an extensive theoretical analysis \cite{Kostelecky:2003fs, Kostelecky:2008in,
Kostelecky:2010ze}. 
Here, we just present a simplified account of how this could affect antihydrogen free-fall.

We consider for simplicity only two of the possible couplings, with the action:
\begin{equation}
S = \int d^4 x \sqrt{-g} \biggl(\frac{R}{16\pi G} + 
\bar\psi\left(i \c^\m D_\m - m\right) \psi 
+ a_\m\, \bar\psi \c^\m \psi
+ c_{\m\n}\,\bar\psi i \c^\m {D}^\n \psi   + \ldots \biggr) \ .
\label{d33}
\end{equation}
The analogy with (\ref{d32}) is obvious. Here, however, $a^\m$ and $c_{\m\n}$ are entirely new
background fields. If they take fixed background values (or VEVs if they are considered as 
quantum fields) this selects a preferred direction in the local orthonormal frame at each point in 
spacetime, thereby breaking local Lorentz invariance. 

To find the classical, single-particle equation of motion originating from the theory (\ref{d33}),
we follow the method described in section \ref{sect 2.4} where it led to the geodesic equation.
First, we write the single-particle action in curved spacetime including the two new
background fields as,
\begin{equation}
S = - m \int d\l\, \left( \sqrt{\left(g_{\m\n} + c_{\m\n}\right) u^\m u^\n} \,+\,  a_\m u ^\m \right) \ ,
\label{d34}
\end{equation}
with $u^\m = dx^\m/d\l$. Here, $x^\m(\l)$ is the particle trajectory, with $\l$ an affine parameter
which we could choose directly as proper time.
To motivate this, note that the field $c_{\m\n}$ effectively enters (\ref{d33}) as a modification to the
metric, while $a^\m$ is analogous to an external electromagnetic field. It follows that we should not
take $a_\m$ of the `pure gauge' form $\partial_\m \phi$, otherwise it could be absorbed into a
phase redefinition of the fermion field in (\ref{d33}). 

The modified geodesic equation is found as before by extremising this action with respect to
$x^\m(\l)$. Under various technical assumptions, in particular that $c_{\m\n}$ and $a^\m$
are slowly varying, and using the fact that we can take $c_{\m\n}$ 
to be traceless, a short calculation yields the following equation of motion 
in the weak-field, non-relativistic ($u^i \ll u^0$) limit,\footnote{An 
essentially identical equation follows from a careful treatment of the VEVs and fluctuations
of $c_{\m\n}$ and $a_\m$ in the framework of the gravitationally coupled SME, as detailed
in \cite{Kostelecky:2003fs, Kostelecky:2008in, Kostelecky:2010ze, Kostelecky:2015nma}}
\begin{equation}
m_i\, \frac{d^2 x^i}{dt^2} + m_g \, \partial_i U = 0 \ ,
\label{d35}
\end{equation}
where \begin{align}
m_i &= m\Bigl( 1 + \frac{5}{6} c_{00}\Bigr) \nonumber \\
m_g &= m\Bigl(1 + \frac{1}{2} c_{00} +  a^0 \Bigr) \ .
\label{d36}
\end{align}
As in section \ref{sect 2.4}, we have used $\Gamma^i_{00} \simeq -\tfrac{1}{2}  h_{00,i}$
with the gravitational potential $U= - \tfrac{1}{2} h_{00} = - GM/r$ for the Schwarzschild metric.

The key point here is that the new background field values $c_{00}$ and $a^0$ appear with
{\it different} coefficients in front of the ``acceleration'' and ``gravity'' terms in this
modified geodesic equation. See the discussion following (\ref{b32b}). 
In the simple weak-field, non-relativistic limit, these can be interpreted as modifications to
(or better, definitions of) the ``inertial mass'' and ``gravitational mass'' respectively.
This is clearly a violation of WEPff. 

Moreover, for the corresponding antimatter test particle, we replace the $\mathsf{C}$ and 
$\mathsf{CPT}$ odd field $a^0$ in (\ref{d36}) by $-a^0$. This means we have a {\it different} 
ratio of $m_g/m_i$ for matter and antimatter. 

Now, in order to satisfy the extremely strong limits $\D g/g = m_g/m_i - 1 \lesssim 10^{-15}$
on WEPff for matter \cite{Touboul:2017grn}, which of course are possible 
because we can perform matter experiments with large test bodies, we have to assume a specific
relation between $c_{00}$ and $a^0$ in this model. Imposing $m_g = m_i$ for matter fixes 
$a^0 = c_{00}/3$, so then for antimatter $\D g/g \simeq - 2 c_{00}/3$.

This is reminiscent of the mechanism introduced for new `fifth force' background fields in 
section \ref{sect 2.6}. There, we exploited the fact that a vector field couples oppositely 
to particles and their antiparticles to arrange a cancellation of the effect of new background fields 
on matter while leaving a residual unconstrained effect on antimatter. The idea is similar here.

Of course, this particular choice of parameters 
is otherwise unmotivated and requires fine-tuning to evade existing equivalence principle bounds.
Moreover, going beyond the leading non-relativistic, Newtonian-like approximation (\ref{d35})
to the equation of motion will introduce corrections at least of $O(v^2/c^2)$, and the precise
cancellation required to shield the effect of the new $a^\m$ and $c_{\m\n}$ fields on matter
cannot be maintained for the whole range of WEPff tests involving different velocities. 
Depending on the test considered (compare section \ref{sect 2.6}),  the size of potential
WEPff violations in antimatter would then be limited to around $\D g/g \lesssim 10^{-7}$.
Identifying $c_{00}$ in (\ref{d36}) with the parameter bounded by the absence of observed Lorentz
violation through annual variations in the hydrogen spectrum \cite{Matveev:2013orb} would
also limit $\D g/g$ for antihydrogen in this model.

\vskip0.7cm
\noindent {\it New background fields}:
\vskip0.2cm

In section \ref{sect 2.6}, we discussed at some length the implications of new long-range scalar,
vector or tensor ($S,V,T$) fields for WEPff violations. In particular, we discussed the conditions under 
which it may be possible to limit WEPff violations in matter experiments to satisfy the current
experimental bounds while retaining a possible observable signal with antimatter.
This discussion, and the associated bounds, apply directly to the forthcoming antihydrogen free-fall
experiments.

Various other far more speculative models and suggestions have been advanced to try and justify
an $O(1)$ WEPff violating effect with antimatter, none of which in our view is theoretically 
well-founded or can be made compatible with the huge body of experimental and observational
evidence supporting GR and the standard model. 
Straightforward phenomenologies, such as the parametrisation $(g_{00})_{\rm eff} = 1 - \a_g 2GM/r$
of the metric at the Earth's surface with $\a_g$ chosen independently for antimatter,  give rise 
in the obvious way to a non-vanishing $\D g/g$, given by
\begin{equation}
\frac{\D g}{g} \,=\, \a_g - 1 \ .
\label{b100}
\end{equation}
However, as already noted in section \ref{sect 2.5}, this is unmotivated in terms of fundamental theory 
and apparently inconsistent since it would imply a direct dependence on the absolute gravitational 
potential and not, more realistically, only on a potential difference.\footnote{A {\it caveat} here 
is that we can devise experiments analogous to the 
Bohm-Aharanov effect in gauge theories in which a redshift may be affected by the gravitational 
potential without a direct gravitational force acting \cite{Hohensee:2011yt}. 
However, as in the Bohm-Aharanov effect,
this requires an element of non-trivial topology which is not the case for the simple free-fall
experiments described here.}
Even so, as we note below, redshift experiments involving antimatter interpreted with this 
parametrisation typically bound $|\a_g -1| \lesssim 10^{-6}$.  Certainly, there is no indication of 
the extreme ``antigravity'' value $\a_g = -1$.

The overall conclusion from theory is therefore that while possible violations of WEPff in antihydrogen
free-fall can be envisaged, every viable model suggests that they are extremely small, almost 
certainly already constrained at the $\D g/g \lesssim 10^{-7}$ level.
In this light, the current generation of antihydrogen gravity experiments should be regarded as an
important first step, with the ultimate goal of reaching this realistic, but challenging, 
level of precision.

\subsubsection{Antihydrogen spectrum and gravitational redshift}\label{sect 3.3.2}

Throughout the discussion so far, we have seen many ways in which the free-fall equivalence principle
(WEPff) may be violated, even in ways which differ depending on whether the test particle is 
matter or antimatter.

On the other hand, none of these models has challenged the fundamental assumption of the
`universality of clocks' equivalence principle (WEPc), which asserts that all ideal clocks measure
the same gravitational time dilation. This is ensured in general relativity by the fact that time measurements 
are determined by the metric component $g_{00}$ and all matter (and antimatter) couples universally
to the same metric.

Of course, the identification of an ``ideal'' clock is not straightforward and we have discussed
how atomic spectral frequencies in atoms (and anti-atoms) may mimic WEPc violations due to
unconventional new interactions or Lorentz or $\mathsf{CPT}$ violation. These effects must be
identified and eliminated before we can attribute any anomalies to a genuine gravitational
WEPc violation. For example, as described in the discussion near the end of section \ref{subsubsection:LVCPTV},
studying possible annual variations in the spectra of hydrogen and antihydrogen during the 
Earth's elliptical orbit round the Sun would allow both the SME Lorentz violating parameter $c_{200}^{\rm NR}$ 
and the Lorentz and $\mathsf{CPT}$ violating parameter $a_{200}^{\rm NR}$ to be determined separately, 
and isolated from any WEPc violating effect dependent on the gravitational field.

Theoretically, one way to achieve a genuine WEPc violation is to construct a multi-metric theory
(see {\it e.g.}~\cite{Drummond:2013ida, Drummond:2016ukf}) in which different particle species
impose different metric structures on the underlying spacetime manifold. A discussion of these theories
is beyond the scope of this article but, even so, no causal QFT of this type has
been established for which particles and their antiparticles couple to different metrics.

In this section, therefore, we consider tests of gravitational time dilation and redshift with 
antimatter, specifically antihydrogen, in the standard framework of GR.
The essential results have already been presented in section \ref{sect 2.4}. 
To make contact with existing literature, we also comment on the phenomenological 
parametrisation of WEPc violations described, and criticised, at the end of section \ref{sect 2.5}.

To begin, we reflect on how in principle a frequency measurement of an atomic transition such as 
the $1S$\,-\,$2S$ antihydrogen line is made, taking GR into consideration.
Essentially, the desired transition is induced using a laser with a tuned frequency which is in 
turn locked onto a reference time standard such as
a co-located Cs atomic clock. The key point is that ultimately any frequency or time measurement 
is in fact simply a {\it ratio} of the frequency/time interval to be measured with another
atomic transition frequency characterising the reference clock. 

Now, as we have seen in section \ref{sect 2.4}, an atomic transition frequency in a gravitational
field will be redshifted proportionally to the local gravitational potential.
However, the {\it same} redshift applies to the co-located reference clock.
So, to take a specific example, the $1S$\,-\,$2S$ antihydrogen frequency measured by a 
co-located (and co-moving) Cs atomic clock will remain the same, independent of the local
gravitational potential. Colloquially, though imprecisely, we may say that time is running slowly in 
the gravitational field for both the measured atom and the reference clock, but crucially -- according 
to GR -- at the same rate.

Similarly, if we consider measurements taken through the annual cycle of the Earth's elliptical 
orbit of the Sun, so that the atom and reference clock are co-moving through the varying
gravitational field of the Sun, the measured atomic transition frequency remains the same. 

The gravitational redshift can however be detected if the atom and reference clock are not
co-located, or not co-moving. In this case we would compare the atomic transition frequencies
with the atom at different points with differing gravitational potentials as measured
by a reference clock which remains fixed. The analysis of this situation in the idealised 
context of the Pound-Rebka experiment has been described in section \ref{sect 2.4}. 
Here, two atoms at different heights in the Earth's gravitational field compare frequencies
through the direct exchange of a photon. In practice, this could be performed with an 
extended ALPHA-g apparatus with an upper and lower trap. The antihydrogen $1S$\,-\,$2S$ 
frequencies in these upper and lower traps would then be measured by lasers with frequencies
calibrated to a single {\it fixed} reference Cs clock. This allows a measurement of the ratios 
of the frequencies of the upper and lower atoms for which the GR prediction is
\begin{equation}
\frac{\D \n}{\n} \,=\, - \frac{G M_E h}{R_E^2} \ .
\label{d60}
\end{equation}
Here, $\D \n$ is the difference of the upper and lower frequencies, while the height difference 
is $h$.  The key point of course is that this is dependent only on the {\it difference} of the
gravitational potentials in the upper and lower trap. This gives
\begin{equation}
\frac{\D \n}{\n} \,=\, - 1.1 \times 10^{-16}\, \frac{h}{1\,{\rm m}} \ .
\label{d61}
\end{equation}
With the frequency precision currently attained with hydrogen, this gravitational redshift
effect could in principle be measured with a height difference of order $h \simeq 10\,$m.
Of course, a practical realisation of this experiment would nevertheless present many challenges.
Note also that, according to GR, this redshift is universal and should be the same for antihydrogen
and hydrogen, the matter-antimatter distinction being irrelevant.

Naturally, any deviations from these predictions would constitute an effective violation of WEPc
and, if all non-gravitational origins could be excluded, would be in direct conflict with general
relativity. 

At this point, we would ideally have a well-defined extension of GR against which
to compare any anomalous measurements and constrain new parameters. The gravitational
extension of the Lorentz and $\mathsf{CPT}$-violating SME is one such model and its
implications for a variety of types of clock measurements have been extensively discussed
in \cite{Kostelecky:2008in, Kostelecky:2010ze, Kostelecky:2018fmc}. Without invoking
Lorentz violation, a popular phenomenological parametrisation of possible beyond-GR effects
was introduced by Hughes and Holzscheiter \cite{Hughes:1990ay}. It essentially violates the 
weak equivalence principle by asserting that different particle species couple to different
spacetime metrics, as described in section \ref{sect 2.5}. Equivalently, it conjectures that the
coupling of matter to gravity, in the weak field approximation, is given by the Lagrangian,
\begin{equation}
L \,=\, \frac{1}{2}\,\a_g\, h_{\m\n} T^{\m\n} \ ,
\label{d62}
\end{equation}
with the deviation from the flat spacetime metric given by $h_{\m\n}$ with
$h_{00} = h_{rr} = -2U$ for the Schwarzschild metric,
where $U= -GM/r$ is the local gravitational potential. 

The violation of GR comes in allowing $\a_g$ to differ for different particle species, 
and for particles and their antiparticles, removing the universality of matter couplings to gravity 
embodied in the equivalence principle. It has to be immediately recognised, however,
that (\ref{d62}) cannot be embedded in a relativistic QFT of the type used
so successfully in the standard model, and indeed the Lorentz-violating SME.
In these theories, based on causal fields lying in representations of the Lorentz group 
(see section \ref{sect 2.1}), the energy-momentum tensor is necessarily built from 
fields describing both particles and antiparticles. 

Nevertheless, in the absence of an alternative well-motivated theory, we can still parametrise 
the scenarios described above in this model.
The simplest case is to compare the antihydrogen and hydrogen $1S$\,-\,$2S$ transition frequency
at the same place, measured through the intermediary of a reference Cs atomic clock.  
The key conceptual point here is that if the period of the antihydrogen transition being observed 
is $dT$ in flat spacetime, then in a gravitational field this is the period {\it in the atom's
proper time}, {\it i.e.}~$d\t = dT$, where $d\t = \sqrt{-g_{00}(\a_g^\Hbar)}\,\, dt$ with
$g_{00}(\a_g^\Hbar) = - \left(1 - \a_g^\Hbar\, 2GM/r\right)$. It follows that the period
{\it in coordinate time} is $dt = dT/\sqrt{-g_{00}(\a_g^\Hbar)}$. 
This is measured by the Cs reference clock in terms of its own proper time, which in this model
is determined by a metric with a potentially different coefficient $\a_g^{\rm Cs}$.
So, in terms of the flat spacetime frequency $\n = 1/dT$, the frequency of the antihydrogen 
transition {\it as measured by the {\rm Cs} atomic clock}}, satisfies
\begin{equation}
\frac{\n^\Hbar}{\n} \,=\, \frac{\sqrt{-g_{00}\left(\a_g^\Hbar\right)}}
{\sqrt{-g_{00}\left(\a_g^{\rm Cs}\right)}}  ~~
\simeq\, 1 + \left(\a_g^\Hbar - \a_g^{\rm Cs}\right)\, U  \ .
\label{d63}
\end{equation}
Note though that since we cannot have a physical measurement of the atomic frequency
in the absence of any gravitational field, the flat spacetime frequency $\n$ is at best a theoretical,
not a measured, quantity. 

A more physically direct result follows immediately by comparing the frequencies of 
hydrogen and antihydrogen {\it as measured by the same reference clock}
in a gravitational field. Details of the clock cancel from this ratio and we are left with
the conceptually clear prediction,
\begin{equation}
\frac{\D \n^{\Hbar - {\rm H}}}{\n^{\rm H}} \,\equiv\,  \frac{\n^\Hbar - \n^{\rm H}}{\n^{\rm H}} 
= \left(\a_g^\Hbar - \a_g^{\rm H} \right) \, U \ .
\label{d64}
\end{equation}
We have assumed here that the flat spacetime frequencies of $\Hbar$ and H are the same
so, as elsewhere in this subsection, this formula applies if the possibility of $\mathsf{CPT}$
violation has been excluded {\it a priori}. Then, unless the coupling $\a_g$ is the same for 
$\Hbar$ and H, there will be a difference in the transition frequencies, allowing a bound
to be placed on $\D\a_g^{\bar{\rm H} - {\rm H}}$.
Here, however, we are confronted most starkly with a basic problem of this phenomenological 
model. By violating WEPc, this model predicts that a physical quantity is dependent on the local
value of the gravitational potential itself, not a difference in potentials. So what value of the 
potential is relevant here? At first sight, one might suppose that we should use the Earth's gravitational
potential at the surface, $|U| \simeq GM_E/R_E \simeq  7\times 10^{-10}$. However, the Sun's potential 
is bigger, $|U| \simeq 1\times 10^{-8}$. In fact, the potential becomes greater for more distant 
gravitational structures. For the galaxy, we may estimate $|U| \simeq 10^{-6}$ \cite{Karshenboim:2008zj}
with even higher values for the local galactic cluster. The limit placed on $\D\a_g^{\bar{\rm H} - {\rm H}}$
entirely depends on this choice, with the most stringent bound coming from the highest potential.
If we take the current precision of antihydrogen spectroscopy, and in the absence of annual variations,
we would find $\D\a_g^{\bar{\rm H} - {\rm H}} \lesssim 10^{-6}$. However, this has little credence
and we would find an even smaller bound if, for example, we chose the gravitational potential of
the Virgo cluster.

At first sight more reasonably, though disguising the same fundamental difficulty with the model,
we could consider frequency measurements which depend only on differences of the gravitational 
potential. So next we consider the bounds on $\a_g^\Hbar$ that would arise from the
non-observation of annual variations in the antihydrogen spectrum during the Earth's orbit in the 
Sun's gravitational field. 

In the same way, we find the difference in antihydrogen frequencies measured by the Cs reference clock 
at two different distances $r_1$ and $r_2$ from the Sun is
\begin{equation}
\frac{\D \n^\Hbar(r_1|r_2)}{\n^\Hbar} \,\equiv\, 
\frac{\n^\Hbar(r_1) - \n^\Hbar(r_2)}{\n^\Hbar(r_2)}  ~=~
\left(\a_g^\Hbar - \a_g^{\rm Cs}\right) \, \D U_S(r_1|r_2) \ ,
\label{d65}
\end{equation}
where $\D U_S(r_1|r_2)$ is the difference in the Sun's potential at $r_1$ and $r_2$, and as usual
we quote the result to $O(U)$ only. An annual variation would therefore indicate a difference between
the parameters $\a_g^\Hbar$ and $\a_g^{\rm Cs}$ characterising the antimatter antihydrogen atom
and the matter Cs atom respectively. 

As before, we can cancel out the characteristics of the reference clock and compare the 
measured $\Hbar$ and H frequencies directly. We then find,
\begin{equation}
\frac{\D \n^{\Hbar-{\rm H}}(r_1|r_2)}{\n^{\rm H}} \,\equiv\, 
\frac{\n^\Hbar - \n^{\rm H}}{\n^{\rm H}} \Big|_{r_1}  \,-\,
 \frac{\n^\Hbar - \n^{\rm H}}{\n^{\rm H}} \Big|_{r_2}  ~=~
\left(\a_g^\Hbar - \a_g^{\rm H}\right) \, \D U_S(r_1|r_2) \ ,
\label{d66}
\end{equation} 
giving a very direct antimatter-matter comparison depending only on a difference of 
gravitational potentials.

To quantify this, at aphelion $r_1 = 1.52 \times 10^{11}\,$m while at perihelion 
$r_2 = 1.47\times 10^{11}\,$m, so for the Earth's orbit we find,
\begin{equation}
\D U_S(r_1|r_2) \,\simeq\, \frac{G M_S}{r_2^2} \, \left(r_1 - r_2\right) \,=\, 3.3\times 10^{-10} \ .
\label{d67}
\end{equation}
So with the current antihydrogen $1S$\,-\,$2S$ precision of $10^{-12}$, the non-observation of
annual variations in $\overline{\rm H}$ and ${\rm H}$ would place a bound 
$\D \a_g^{\Hbar - {\rm H}} \lesssim 3 \times 10^{-3}$, while if the antihydrogen precision 
would match that currently available with hydrogen, this bound could be improved to
$\D \a_g^{\Hbar - {\rm H}}  \lesssim 10^{-6}$.  A search for annual variations in the antihydrogen
spectrum could, if this model is adopted, provide a very competitive test of WEPc.

The analysis of an antihydrogen Pound-Rebka experiment in this model is a straightforward
extension of the usual GR treatment given in section \ref{sect 2.4}.  From (\ref{b41}) and 
(\ref{d60}), the difference of the frequencies of the `emitter' and `observer' antihydrogen 
atoms is,
\begin{equation}
\frac{\D\n^\Hbar}{\n^\Hbar} \,\equiv\, \frac{\n_O - \n_E}{\n_E} \,=\, - \a_g^\Hbar\, \frac{GM_E h}{r_E^2} \ .
\label{d68}
\end{equation}
As noted above, measurement of this redshift factor is at the limit of what may be attained
even if the antihydrogen $1S$\,-\,$2S$ precision could match that of hydrogen. So this experiment,
while in principle an excellent test of WEPc, would only be sensitive to $O(1)$ deviations from the 
GR value $\a_g^\Hbar = 1$.

Finally, the gravitational redshift measurement using atom matter-wave interferometry is
modified in the same way in the phenomenological WEPc violating model. All that changes
in (\ref{b41a}) -- (\ref{b41d}) is the inclusion of the relevant $\a_g$ factor modifying the
gravitational potential. In particular, the final formula for the phase shift (\ref{b41d}) becomes
\begin{equation}
\D \phi_{redshift}\,=\, \a_g^\Hbar\, \omega_C\, \int_0^T dt\, g\, \D r(t) \ ,
\label{d69}
\end{equation}
for an antihydrogen interferometry experiment. Again, this allows a bound to be placed on the antimatter WEPc
violation factor $|\a_g^\Hbar - 1|$, and the proposal \cite{Mueller:2013ysa} suggests that precisions of
order $|\a_g^\Hbar - 1| \lesssim 10^{-6}$ are possible.\footnote{In an experiment of this type
with Cs atoms, the gravitational redshift is observed to agree with GR for 
trajectory separations of $O(0.1{\rm mm})$, giving a bound $|\a_g^{\rm Cs} - 1| < 7\times 10^{-9}$.
Assuming no WEPc violation, this experiment can also bound the local gravitational acceleration,
finding $\D g/g < 3 \times 10^{-9}$.
See \cite{Chung:1999, Chung:2001, Muller:2010zzb, Hohensee:2011wt} 
for further details and references to the literature on atom matter-wave interferometry.}

\vskip0.5cm
\noindent{\it $U(1)_{B-L}$ and gravitational redshift}:
\vskip0.2cm
In previous sections, we have considered the {\it direct} effects of a possible long-range $U(1)_{B-L}$
field sourced by the Earth. Here, we consider briefly the {\it indirect} effects arising from the
modifications to the Schwarzschild metric induced by the Earth's $U(1)_{B-L}$ charge. 
We focus on gravitational redshift. 

The picture is straightforward. A massless $U(1)_{B-L}$ gauge field sourced by the large 
$Q_{B-L}$ charge of the Earth is entirely analogous to the electromagnetic field around a charged object.
Its gravitational effects are therefore described by the analogue of the Reissner-Nordstr\"om
metric, usually used to describe the spacetime around a static, charged black hole.
This metric is therefore:
\begin{align}
ds^2 &= - \left(1 - \frac{2GM}{r} + \a' \,\frac{G Q_{B-L}^2}{r^2} \right) dt^2 \nonumber \\
&~~~~~~~~~+ \left(1 - \frac{2GM}{r} + \a' \,\frac{G Q_{B-L}^2}{r^2}\right)^{-1} dr^2 
+ r^2 \sin^2(\theta) d\phi^2 \ ,
\label{d50}
\end{align}
where $\a' = g^{\prime 2}/4\pi$ is the fine structure constant corresponding to the fundamental 
$U(1)_{B-L}$ coupling, and $Q_{B-L}$ is the numerical $U(1)_{B-L}$ charge of the source, of mass $M$.

The standard analysis of gravitational redshift now follows from our earlier discussion,
with the new metric component
\begin{equation}
- g_{00}(r) = 1 - \frac{2GM}{r} + \a' \,\frac{G Q_{B-L}^2}{r^2} \ .
\label{d51}
\end{equation}
Since this new metric is still a function of $r$ only, the redshift derivation presented in 
section \ref{sect 2.4} applies directly.
With an emitter at radius $r_E$ above the Earth's centre and a receiver at $r_O = r_E + h$, 
the ratio of frequencies measured at $r_O$ and $r_E$ is (see \ref{b41}),
\begin{equation}
\frac{\n_O}{\n_E} = \frac{\sqrt{-g_{00}(r_E)}}{\sqrt{-g_{00}(r_O)}} \ ,
\label{d52}
\end{equation}
and we find the leading contributions,
\begin{equation}
\frac{\n_O}{\n_E} = 1 - \frac{G M_E}{R_E^2} \,h + \a'\, \frac{G Q_{B-L}^2}{R_E^3} \,h + \ldots 
\label{d53}
\end{equation}
with $M_E, R_E$ the mass and radius of the Earth. 

To establish the relative size of this new contribution to the redshift, which is of course the same
for a matter or antimatter clock, we take the ratio of the final two terms in (\ref{d53}), 
which is approximately $10^{28}\a'$. Verification of the GR prediction for the redshift
would therefore constrain $\a' \lesssim 10^{-28}$. Small though this is, it is still however many
orders of magnitude bigger than the experimental limit already imposed by conventional
equivalence principle tests (section \ref{sect 2.6}) which require $\a' \lesssim 10^{-49}$.

Alternatively, we could consider the redshift effect due to the eccentricity of the Earth's orbit
around the sun, for which the corresponding ratio is still comparable, roughly  $10^{30}\a'$.
Despite the theoretical elegance of this theory, it therefore seems that there is no
realistic possibility to detect a long-range $U(1)_{B-L}$ field even with the most sensitive
gravitational redshift experiments.

Finally, we should note that the modification (\ref{d51}) of the metric also alters the
equation of motion for free-fall, replacing the effective gravitational potential 
$U = - \tfrac{1}{2}h_{00} = - GM/r$ with
\begin{equation}
V = - \frac{GM}{r} + \frac{\a'}{2} \frac{G Q_{B-L}^2}{r^2} \ .
\label{d54}
\end{equation}
This induces an extra $1/r^3$ component in the gravitational force, though again the
order of magnitude is far too small to be detectable given the existing constraints on 
the coupling $\a'$.

\section{Other Antimatter Species}\label{sect 4}

In this section, we discuss briefly some possibilities for testing fundamental physics principles
using antimatter species other than antihydrogen.   In Table \ref{TableTests} we have 
provided a summary of some of the antimatter species that have been, or may be in the not-too-distant
future, subject to investigation and have indicated the types of test that may be performed, 
organised according to the discussion in section \ref{sect 2}.  The types of experiments that permit
such tests are also shown. Note that we confine 
ourselves to laboratory tests involving low ($\approx$ eV or below) kinetic energies,
and do not consider high energy measurements such as the special storage ring experiments 
aimed at the muon $g-2$ value \cite{Muong-2}.

Most of the systems listed are stable against decay or self-annihilation. 
The notable exceptions are the bound-states muonium Mu ($\mu^+\,e^-$) 
\cite{MuonRev,MuonCPT}, positronium Ps ($e^+\,e^-$) \cite{Cassidy:2018tgq,KarshenboimRev}
and antiprotonic helium He$^+\overline{p}$, 
the metastable bound state of an antiproton and a helium ion \cite{Hayano07,Hori13}).
All these have, however,  been studied spectroscopically and are the subject of ongoing investigations 
relevant to the types of fundamental physics tests described here. 

\begin{table}[h!]
\centering
\begin{tabular} {c  l rl r l r l c l c }
\\
\hline\hline\\
 \multicolumn{1}{c} {\raisebox{1.5ex} {Species} } &
  \multicolumn{1}{c} {\raisebox{1.5ex} {  }  }  &
\multicolumn{1}{c} {\raisebox{1.5ex} {$Q,\,B-L$}  }  &
   \multicolumn{1}{c} {\raisebox{1.5ex} {  }  }  &
 \multicolumn{1}{c} {\raisebox{1.5ex} { Tests } } &
   \multicolumn{1}{c} {\raisebox{1.5ex} {  }  } &
\multicolumn{1}{c} {\raisebox{1.5ex} { Experiments}  } 
\\
\hline 
\\
$\overline{p}$ &&$-1,\,\,-1$  && $\mathsf{CPT}$, WEPc, Lorentz && Traps	\\
$\overline{d}$ &&$-1,\,\,-2$ && $\mathsf{CPT}$, WEPc, Lorentz && Traps	\\
$~~e^+$ &&$1,\,\,\,\,\,\,\,1$&&$\mathsf{CPT}$,  WEPc, Lorentz && Traps\\
\\
\hline
\\
$\overline{\rm H}$ &&~~~$0,\,\,\,\,\,\,0$ && $\mathsf{CPT}$, WEPc, WEPff, Lorentz && Spectroscopy, AI, free fall \\
$\overline{\rm D}$ &&~~~$0,\,\,-1$ && $\mathsf{CPT}$, WEPc, WEPff, Lorentz && Spectroscopy, free fall \\
$~~\overline{\rm H}^+$ &&~~~$1,\,\,\,\,\,\,1$ && $\mathsf{CPT}$, WEPc, Lorentz &&Traps\\
$~~\overline{{\rm H}}_2^-$&&$-1,\,-1$&& $\mathsf{CPT}$, WEPc, Lorentz&& Traps, Spectroscopy\\
Mu&&~~~$0,\,\,\,\,\,0$&& WEPc, WEPff, Lorentz  &&Spectroscopy, free fall	\\
Ps&&~~~$0,\,\,\,\,\,0$&& WEPc, WEPff, Lorentz &&Spectroscopy, free fall\\
He$^+\overline{p}$&&~~~$0,\,\,\,\,\,2$&&$\mathsf{CPT}$, WEPc, Lorentz&&Spectroscopy\\
	\\
\hline\hline

\end{tabular}
\caption{The antimatter particles and bound states discussed in section \ref{sect 4}, 
together with their electric charge and $B-L$ quantum number, the type of fundamental
principles which they enable to be tested, and the types of experiments possible. WEPff and WEPc, as defined 
in section \ref{sect 1}, refer to the universality of free-fall and the universality of clocks respectively.
AI denotes atomic matter-wave interferometry. We have only shown this in the table for the neutral 
antihydrogen, although AI experiments with other species may also be feasible.}

\label{TableTests}
\end{table}

As indicated in Table \ref{TableTests}, the antiparticle species we consider are the positron, 
the antiproton and the antideuteron $\overline{d}$. 
Though heavier antibaryons have been created, {\it e.g.}~anti-$^3$He 
\cite{antiHe3} and the anti-alpha particle \cite{antialpha}, their yields are currently 
too small to permit precision experimentation. The $\overline{d}$ has also not so far been 
subjected to such study, but it can be produced at $\approx 10^{-3}$ of the $\overline{p}$ 
flux and may be amenable to capture and manipulation.
The possibility of doing was briefly discussed some time ago \cite{dbar1,dbar2}. 
The antineutron \cite{nbar} is not amenable to capture and, 
like the neutron, is expected to be unstable as a free particle \cite{Tanabashi:2018oca}.

The antiproton and positron can be held for experimentation for extended periods  
(several months or longer, if required \cite{vanDyck87,Gabrielse90,Sellner17}) in 
charged particle traps. The latter are typically so-called Penning traps (or variants 
thereof) \cite{Ghosh,BG} in which an harmonic electrical potential is used, together 
with a uniform magnetic field, to confine charged species. Measurement of the motion 
of the particles, and perhaps the spin-flip (with respect to the direction of the applied 
magnetic field), can determine properties such as the charge-to-mass ratio 
(often interpreted as a direct mass measurement under the assumption that the fundamental charge 
is quantised and is equal and opposite for particles and antiparticles) and the $g-2$ ratio.
 
 Recent highlights have included the work of the BASE collaboration that has provided systematic 
improvements in $\overline{p}$ storage, manipulation and interrogation to yield values of the 
charge-to-mass ratio and magnetic moment to unprecedented accuracies 
\cite{Ulmer15,Nagahama17,Smorra17}. For example, in \cite{Ulmer15}, the ratio of the
charge-to-mass ratios for the antiproton and proton was determined through cyclotron frequency 
comparisons as
\begin{equation}
\frac{q/m\,(\overline{p})}{q/m\,(p)} - 1 \,< \,  10^{-12} \ ,
\label{d80}
\end{equation}
improving on a previous precision of $< 9\times 10^{-11}$ \cite{Gabrielse:1999kc}.
While loosely interpreted as a high-precision test of $\mathsf{CPT}$, as we have discussed in section
\ref{sect 2} this is really a test of more fundamental principles, particularly causality.
The magnetic moment measurements in \cite{Nagahama17,Smorra17} can be interpreted in 
the SME as a bound on the coefficient $|b_3^p|\lesssim 10^{-24}\,{\rm GeV}$ \cite{Kostelecky:2008ts},
and also place stringent bounds on Lorentz violation through the absence of sidereal variations. 
On the other hand, if we assume no other non-standard physics, (\ref{d80}) could be viewed 
as a WEPc equivalence principle test. Interpreted in terms of the phenomenological model
(\ref{d62}), and using the local galaxy supercluster gravitational potential $U \simeq 3\times 10^{-5}$, 
the equality of the charge-to-mass ratios places a bound of
$|\a_g^{\bar{p}} - 1| < 8.7 \times 10^{-7}$ (where again we assume $\a_g^p = 1$) \cite{Ulmer15}.

Antihydrogen is of course the archetypal neutral antimatter bound state and is capable of 
investigation both via spectroscopy and in free fall. As described throughout this paper, this
offers the means of testing $\mathsf{CPT}$ and Lorentz symmetries with high precision
as well as exploring the gravitational properties of antimatter with WEPc and WEPff tests. 

Anti-deuterium ($\overline{\rm D}$) should also become available for experiment
in future, and offers further opportunities for complementary tests \cite{ALPHA2019}.
In terms of Lorentz and $\mathsf{CPT}$ violation as parametrised by the SME, its spectrum
would be sensitive to SME couplings involving the antineutron as well as the antiproton
\cite{Kostelecky:2015nma},
with the corresponding additions to the transition frequency calculations in section 3.
In principle, since unlike antihydrogen it has a non-zero $B-L$ charge, it would experience 
a violation in WEPff if there were a long-range $B-L$ interaction with the Earth. 
However, as with any antimatter species carrying a $B-L$ charge, since it necessarily follows 
that the corresponding matter system also has non-vanishing $B-L$, such interactions 
are already constrained to greater precision from studies of the equivalence principle with bulk matter.

Other current experimental work at CERN includes antiprotonic helium, He$^+\overline{p}$ \cite{Yamazaki02}.  
This unique bound state is formed by stopping $\overline{p}$\,s in dilute He gas, with around 3\% of the states 
formed being metastable against annihilation with lifetimes in the $\mu$s range. It has been the subject of a 
sustained programme of spectroscopic investigations (see {\it e.g.}~\cite{Hayano07,Hori13} for reviews). 
Recent highlights have included quasi-two-photon spectroscopy \cite{Hori11} and buffer gas cooling of 
the He$^+\overline{p}$~\cite{Hori16}, which have allowed, for instance, the determination of  some 
$\overline{p}$ properties to high precision.  
Possible types of fundamental physics test with He$^+\overline{p}$ 
are given in Table \ref{TableTests}.  Note that since there is no matter counterpart of He$^+\overline{p}$ 
available for experimental study, such tests require comparisons with detailed few body atomic structure 
calculations (see {\it e.g.}~\cite{Korobov14}). Laser spectroscopy of pionic helium $\pi{\rm He}^+$ has
also been proposed and could allow a measurement of the $\pi^-$ mass with a fractional
precision $10^{-6}$\,-\,$10^{-8}$ \cite{Hori:2014uta}.

The antihydrogen positive ion $\overline{\rm H}^+$ ($\overline{p}\,e^+\,e^+$)
and the antihydrogen molecular anion $\overline{\rm H}_2^-$ ($\overline{p}\,\overline{p}\,e^+$)
offer exciting possibilities for new tests of fundamental physics.
These states have yet to be observed in the laboratory, but both have well studied and 
important matter counterparts and may be produced using  interactions of trapped antihydrogen, 
or beams of the anti-atom. Possible mechanisms for $\overline{\rm H}^+$ include radiative 
$\overline{\rm H}/e^+$ combination and charge exchange in $\overline{\rm H}$/Ps collisions 
\cite{Keating14,Keating16,Perez12} and for $\overline{\rm H}_2^-$, radiative $\overline{\rm H}/\overline{p}$ 
association and 
$\overline{\rm H}$-$\overline{\rm H}$ associative attachment \cite{Myers18, Zammit:2019etx} . 
The production of $\overline{\rm H}^+$ is envisaged within the GBAR programme \cite{Perez12,vdWerf14} 
currently underway at the AD at CERN.

$\overline{\rm H}^+$ is expected, as is its matter counterpart H$^-$, to be bound by around 
$0.75\,{\rm eV}$ and to have only a single bound state. As such, it is not amenable to spectroscopic 
investigation although it should be able to be stored for long periods in Penning traps for interrogation. 
$\overline{\rm H}_2^-$ is a bound state expected to have a rich vibrational and rotational spectrum 
and H$_2^+$ has already been suggested as a possible ultra-precise optical clock \cite{Schiller14}. 
Myers \cite{Myers18} has described how measurements of analogous clock transitions with 
$\overline{\rm H}_2^-$ may be used as $\mathsf{CPT}$ tests with the potential for greater precision 
than is possible with antihydrogen. This is due primarily to the very long lifetimes (of the order of years, 
rather than seconds for antihydrogen) of some of the excited states.

\section{Summary and Outlook}\label{sect 5}

In the last three years, the long-standing ALPHA programme of producing and trapping
cold antihydrogen atoms in sufficient numbers and with the required control to permit 
precision spectroscopy has been achieved. This enables the anti-atom to be used 
for state-of-the-art precision tests with antimatter of the most fundamental principles of 
modern theories of particle physics and gravity, notably local Lorentz and $\mathsf{CPT}$ invariance,
causality and, eventually, the various forms of the Equivalence Principle.

In section \ref{sect 1} we briefly summarised how the development of sources of low (eV) energy positrons 
and antiprotons, and schemes for the accumulation and manipulation of the antiparticles culminated in the 
controlled creation of cold antihydrogen at the AD \cite{ATHENA,ATRAP1}. Extending the techniques that 
made this possible, and superimposing neutral atom trapping technology in the form of a magnetic minimum
atom trap onto the antihydrogen creation region, allowed the trapping of the anti-atom to be achieved
some years later \cite{ALPHATrap1,ATRAPTrap}.

This advance was key, particularly as long confinement times were quickly forthcoming \cite{ALPHATrap3}, 
ensuring that the antihydrogen (initially produced in highly excited states, as discussed in
section \ref{sect 1}) would  decay to the ground state in readiness for experimentation. 
More recently, lifetimes in the trap in excess of 60 hours have been achieved, allowing of
the order of a thousand antihydrogen atoms to be accumulated on a shot-by-shot basis, thereby 
ensuring  optimum use of the antiproton flux from the AD.

We have also summarised the main achievements in physics with antihydrogen, principally the work 
of the ALPHA collaboration. This has included: the setting of a limit on the charge neutrality of antihydrogen; 
the development of a method of investigating the behaviour of the anti-atom in the earth's gravitational field 
and the observation and characterisation of hyperfine and positronic transitions, including the seminal observation
of the $1S$\,-\,$2S$ transition and a first determination of the Lamb shift. 
The pace of progress, with the $1S$\,-\,$2S$ transition already known to a few parts in
$10^{12}$, bodes well for future improvements in precision and the concomitant limits set upon fundamental 
physics principles. 

On the theory side, we have discussed how each of these experiments tests the fundamental
principles on which our current understanding of particle physics is based. We highlighted
how these principles are interwoven in the structure of relativistic quantum field theories
and their extension to include gravity through general relativity.

Particular emphasis was placed on the role of causality, and we reviewed how the existence
of antiparticles with properties {\it exactly} matching those of the corresponding particles is 
required in a Lorentz invariant theory to preserve causality. The quantum field theory is then 
built from causal fields transforming in representations of the Lorentz group; these causal fields 
necessarily contain particles {\it and} antiparticles. A key feature of such local field theories is that 
they exhibit $\mathsf{CPT}$ symmetry -- breaking $\mathsf{CPT}$ necessarily implies
breaking Lorentz invariance and would undermine the fundamental principles on which 
our theories of particle physics are built, unlike the individual symmetries $\mathsf{C}$, 
$\mathsf{P}$, $\mathsf{T}$, $\mathsf{CP}$ {\it etc.}~which may be trivially broken in the
standard model.

Gravity is introduced classically through GR; for the atomic physics experiments
described here, the scales and weak gravitational fields involved mean that quantum gravity
is irrelevant. The key feature of general relativity is that it describes gravitational forces by
formulating the theory on a Riemannian curved spacetime. This is locally flat and it is
in this local tangent space that  relativistic QFT is formulated (so for example $\mathsf{CPT}$ symmetry
has nothing to do with the curvature of the spacetime). Insisting on the strong equivalence
principle, as defined here, imposes the requirement that the local causal fields couple
only to the spacetime metric connection, {\it not} the curvature. All this realises a universality
of matter-gravity couplings which implies the WEPff and WEPc equivalence principles.

This raises the question of how we could imagine breaking Lorentz or $\mathsf{CPT}$ symmetry 
or the EEP.  Two minimal extensions of the standard model were discussed. 
First, an effective field theory in which additional local but non-Lorentz invariant (and 
possibly $\mathsf{CPT}$ odd) operators are added to the standard model was described.
The new couplings in this SME model \cite{Colladay:1996iz, Colladay:1998fq, Kostelecky:2015nma}
may be included in calculations of atomic physics
properties, especially in spectroscopy but also in modifying `gravitational' and `inertial' masses
in free-fall experiments, and bounds placed on them. Here, we have carried this out explicitly 
for the frequencies associated with the particular transitions measured by ALPHA with the
magnetic field in their confining trap. This allows quantitative bounds to be placed on Lorentz 
and $\mathsf{CPT}$ breaking and compared between quite different experiments.

Second, an effective field theory in which additional local operators coupling directly to the 
spacetime curvature \cite{Shore:2004sh} was described. This implies that in a local inertial frame, 
physical measurements are sensitive to the gravitational field strength in violation of SEP, 
and also WEPff. However, although this has interesting applications in cosmological 
spacetimes where it can give rise to matter-antimatter asymmetry through gravitational leptogenesis 
\cite{McDonald:2015ooa, McDonald:2016ehm}, it was pointed out in section \ref{sect 2.5}
that this would not affect atomic physics experiments in the Earth's gravity.
This is because above the Earth's surface, the gravitational field is described by
a Ricci-flat spacetime and at leading order there are no fermion bilinear couplings to the
Riemann tensor itself. 

It was also emphasised that these theories must be regarded as low-energy {\it effective} 
field theories only. Although they break Lorentz invariance or SEP only minimally and are 
still built from local causal Lorentz fields, they do in general break causality. 
This was discussed in some detail here in section \ref{sect 2.3}. Causality is, however, a property 
of high-momentum propagation and it is possible for a theory which would be non-causal
in itself to be the low-energy limit of a causal theory valid also at high energies. 
The existence of a causal UV completion (see section \ref{sect 2.3})
can impose special relations amongst the couplings in its low-energy approximation.
Here, in the absence of any knowledge of this UV theory (whether a Planck-scale QFT,
string theory or other formulation of quantum gravity) we have to consider the SME 
or SEP-violating couplings as arbitrary parameters to be fixed by experiment.

We also followed the literature by analysing certain equivalence principle experiments
in terms of an entirely phenomenological model \cite{Hughes:1990ay} in which matter and
antimatter were supposed to couple to different metrics, immediately violating WEPc
and WEPff. This model was criticised, however, on two counts -- first that it cannot be
realised in terms of a quantum field theory with causal fields, and therefore cannot 
be related in a plausible way to the standard model, and second that they have the 
apparently unphysical feature of making local frequency comparisons dependent on the
absolute value of the local gravitational potential, rather than on potential differences or 
the field strength. 

The introduction of new `fifth-force' interactions can also produce physical effects which
mimic a genuine gravitational violation of the equivalence principle, SEP and WEPff in particular.
We considered several potential new interactions, especially $U(1)_{B-L}$ gauge theories,
supergravity inspired models and general features of gravitational-strength theories with 
new scalar, vector or tensor ($S,V,T$) fields. Of these, only vector-mediated interactions 
act with opposite  sign on matter and antimatter (analogously to electromagnetism). 
A tensor interaction such as conventional gravity acts attractively on both matter
and antimatter, coupling through the energy-momentum tensor and not some 
`gravitational charge'. In early work, it was speculated that this could be exploited to hide
WEPff violations for matter, for example by arranging a cancellation between new $V$ and $S$
forces for matter, while they would appear with double strength for antimatter where the sign
of the $V$ interaction would be reversed. We analysed these ideas in section \ref{sect 2.5}
with the conclusion that, partly due to the different velocity-dependence of $S,V,T$ interactions,
such cancellations cannot be exact. Existing bulk matter experiments therefore
already place severe constraints on possible WEPff violations with antimatter systems, indeed
several orders of magnitude below those accessible to the first-generation direct WEPff tests
planned with antihydrogen.

This work began by looking at the ways in which individual experiments on antimatter, 
especially antihydrogen, test specific fundamental principles. What soon becomes apparent,
however, is that it is generally impossible to associate a particular experiment with an
unambiguous test of a particular principle, such as WEPc or $\mathsf{CPT}$. 
In terms of theory, these fundamental properties are all part of a single, tight theoretical 
structure and individual elements cannot be violated without impacting the whole
construction. The most obvious example is the $\mathsf{CPT}$ theorem, where with our current
understanding of local  relativistic QFT, $\mathsf{CPT}$ violation necessarily involves breaking 
Lorentz symmetry.

Experimentally, we saw how, for example, each transition frequency in the antihydrogen 
spectrum depends in a different way on the many couplings in the SME, each of which could be 
the place where Lorentz or $\mathsf{CPT}$ violation is hiding. Experiments such as the search for 
annual variations in the antihydrogen spectrum could be an indication either of Lorentz
violation or WEPc violation, or indeed both. The possible existence of new fifth forces violating
WEPff but not WEPc means that a null experimental finding limiting WEPc violation, 
such as the equality of $q/m$ for protons and antiprotons, does not necessarily imply
a null result in an antihydrogen free-fall experiment looking for WEPff violations.

In general, the moral of this is that in searching for presumably tiny violations of the 
fundamental principles of particle physics, {\it all} possible high precision experiments are
valuable and well-motivated. In the event of an unexpected result in antimatter physics,
many complementary measurements would probably be required to pin down its implications
for fundamental theory.

The outlook for experimental antihydrogen physics is currently excellent, and we can look forward to many 
new endeavours and potentially dramatic improvements of the precision of spectroscopic measurements. 
It is hoped that the new Extra Low ENergy Antiproton facility (ELENA) \cite{Maury14} will soon be fully operational at 
the AD. This should allow low energy antiproton capture efficiencies to increase by up to two orders of magnitude. 
Coupled with delivery to experiments using easily switchable electrostatic beamlines, this will result in much 
optimised use of the antiproton flux. One can envisage antihydrogen experiments working continuously, 
with much shorter frequency scan times, for instance, leading to much enhanced capabilities for the exploration 
and understanding of systematics and to measurement campaigns over extended periods of time.
This promises a bright future for high precision tests of fundamental physics with antihydrogen.

\restoresymbol{yyy}{o}
\restoresymbol{zzz}{O}

\vskip2cm

\noindent {\bf Acknowledgements}
\vskip0.3cm

MC and SE are members of the ALPHA collaboration. Their work is
supported by EPSRC grant EP/P024734/1.
The work of GMS is supported in part by the STFC theoretical particle physics grant
ST/P00055X/1. We would like to thank M.~Fujiwara, N.~Madsen,
F.~Robicheaux, S. Jonsell and D.~P.~van der Werf 
for helpful discussions and C.~{\O}.~Rasmussen for valuable comments on the manuscript.

\newpage


\end{document}